%% file: article.tex
\pdfoutput=1  % for arxiv
\documentclass[acmtog,screen]{acmart}
\setcitestyle{square}

\copyrightyear{2018}
\setcopyright{none}
\acmConference{}
\acmIsbn{}
\acmVolume{}
\acmPrice{}
\acmDOI{}
\acmYear{2018}

\renewcommand\footnotetextcopyrightpermission[1]{} % removes footnote with conference information in first column
\title{End-to-end Sampling Patterns}

\author{Thomas Leimk\"uhler}
\affiliation{MPI Informatik}

\author{Gurprit Singh}
\affiliation{MPI Informatik}

\author{Karol Myszkowski}
\affiliation{MPI Informatik}

\author{Hans-Peter Seidel}
\affiliation{MPI Informatik}

\author{Tobias Ritschel}
\affiliation{University College London}

\usepackage{soul}
\usepackage{microtype}
\usepackage{amssymb}
\usepackage{amsmath}
\usepackage{booktabs}
\usepackage{hyperref}
\usepackage{multirow}
\usepackage{graphicx}
\usepackage{dsfont}
\usepackage{pifont}
\usepackage{xcolor}
\usepackage{tikz}
\usepackage{wrapfig}

\newcommand{\eg}{e.\,g.,\ }
\newcommand{\ie}{i.\,e.,\ }
\newcommand{\etal}{et~al.\ }

\newcommand{\refSec}[1]{Sec.~\ref{sec:#1}}
\newcommand{\refFig}[1]{Fig.~\ref{fig:#1}}

\def\figurePath{images/}

\def\myfigure#1#2{\begin{figure}[h]\centering\includegraphics*[width = \linewidth]{\figurePath#1}\vspace{-.2cm}\caption{#2}\label{fig:#1}\end{figure}}

\def\mycfigure#1#2{\begin{figure*}[t]\centering\includegraphics*[clip, width = \linewidth]{\figurePath#1}\vspace{-.2cm}\caption{#2}\label{fig:#1}\end{figure*}}

\def\mysection#1#2{\section{#1}\label{sec:#2}}
\def\mysubsection#1#2{\subsection{#1}\label{sec:#2}}
\def\mysubsubsection#1#2{\subsubsection{#1}\label{sec:#2}}

\definecolor{fixedcolor}{rgb}{.8,1,.7}
\definecolor{fixedncolor}{rgb}{.8,.2,.1}

\newcommand{\unsure}[1]{#1}

\soulregister\ref7
\soulregister\cite7
\soulregister\refFig7
\soulregister\cite7
\soulregister\ref7
\soulregister\pageref7
\soulregister\shortcite7
\soulregister\eg0
\soulregister\ie0
\soulregister\etal0

\DeclareGraphicsExtensions{.png,.jpg,.pdf,.ai,.psd}
\begin{document}


\begin{teaserfigure}
   \includegraphics[width=\textwidth]{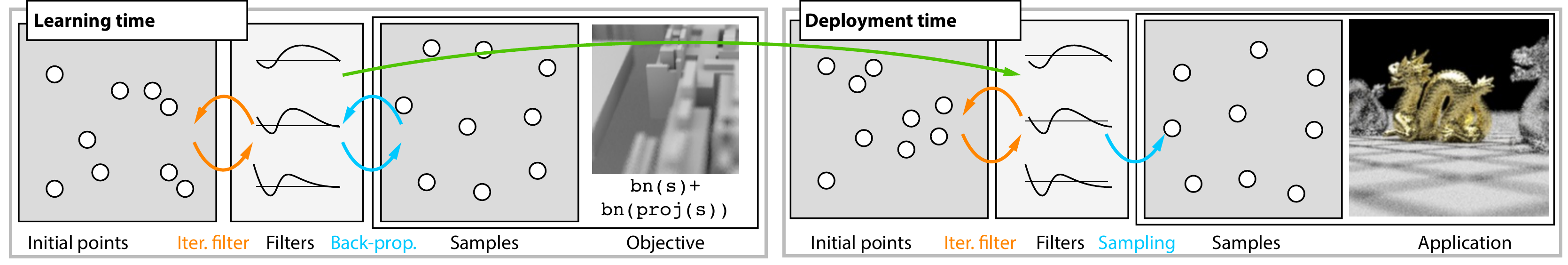}
   \vspace{-.4cm}
   \caption{
We use back-propagation in common Deep Learning frameworks to optimize for recursive unstructured filters that realize an objective, expressed as a small \emph{sample program} such as \texttt{bn(s)+proj(bn(s))}: a blue noise spectrum in 4D and in 2D \textbf{(left)}.
The resulting filters can be deployed to efficiently convert any random sample patterns into a pattern with properties suitable to solve the programmed task \textbf{(right)}.
   }
   \label{fig:Teaser}
\end{teaserfigure}

\begin{abstract}
Sample patterns have many uses in Computer Graphics, ranging from procedural object placement over Monte Carlo image synthesis to non-photorealistic depiction.
Their properties such as discrepancy, spectra, anisotropy, or progressiveness have been analyzed extensively.
However, designing methods to produce sampling patterns with certain properties can require substantial hand-crafting effort, both in coding, mathematical derivation and compute time.
In particular, there is no systematic way to derive the best sampling algorithm for a specific end-task.

Tackling this issue, we suggest another level of abstraction:
a toolkit to end-to-end optimize over all sampling methods to find the one producing user-prescribed properties such as discrepancy or a spectrum that best fit the end-task.
A user simply implements the forward losses and the sampling method is found automatically -- without coding or mathematical derivation -- by making use of back-propagation abilities of modern deep learning frameworks.
While this optimization takes long, at deployment time the sampling method is quick to execute as iterated unstructured non-linear filtering using radial basis functions (RBFs) to represent high-dimensional kernels.
Several important previous methods are special cases of this approach, which we compare to previous work and demonstrate its usefulness in several typical Computer Graphics applications.
Finally, we propose sampling patterns with properties not shown before, such as high-dimensional blue noise with projective properties.
\end{abstract}

\maketitle

% This is for arXiv only
\thispagestyle{empty}

\mysection{Introduction}{Introduction}
Sample patterns have many important uses in Computer Graphics, linking apparently disparate topics such as placement of procedural trees, casting shadows from an area light or choosing visually pleasing artistic stippling patterns.
Many algorithms have been proposed to generate sampling patterns and the instruments available for their analysis is large and growing.
Famous examples are Lloyd's \shortcite{lloyd1982least} relaxation algorithm or dart throwing \cite{mccool1992hierarchical} as well as deterministic combinatorial patterns \cite{shirley1991discrepancy,kuipers2012uniform} where Halton or Hammersley can serve as examples.
Analysis is typically done with respect to spectral properties, \ie the colors-of-noise or their discrepancy.
  
Choosing the sampling pattern most suited for an application is typically done by analyzing the presumed properties of a task with the alleged properties of a sample pattern.
As an example, we might want to choose blue noise for its spectral properties that shift the error into a less visible frequency band, yet we want to retain the favorable integration error of a low-discrepancy pattern.
Achieving both is an active topic of research, requiring involved mathematical derivations and algorithmic effort.
Regrettably, there is no systematic or automated way to produce such sampling patterns that are best for a specific task.
Consequently, substantial effort has to be invested into hand-crafting purpose-fitted solutions, a procedure expensive both in terms of mathematical derivation, implementation effort and finally compute time.

In this paper we leverage modern deep learning to introduce a new level of abstraction at which we end-to-end optimize for a sampling pattern algorithm that fits a specific property or task.
Instead of mathematical derivation, a user of our system provides a straight-forward implementation of the desired properties in form of a \emph{sample program} (\refFig{Teaser}, left) \ie a low integration error, and the system uses the sample program as a loss to find an algorithm that converts random patterns into patterns with the desired properties (\refFig{Teaser}, right).
This is in analogy to a CNN \cite{LeCun} that optimizes image filters to map an image to the likelihood of containing \eg a cat. %Karol: This is a well cite paper of LeCun but rather on handwriting recognition
The architecture comprises of iterated unstructured non-linear filtering that can efficiently be optimized in respect to a combination of losses.
Several important previous sampling patterns algorithms are special cases of ours.
The method works in high dimensions as our multi-dimensional analysis shows, is quick and simple to evaluate, and allows arbitrary combinations of losses which we demonstrate in rendering and object placement applications.

\mysection{Sampling in Computer Graphics}{Background}
Randomness can be quite useful in many computational sampling tasks, in particular in Computer Graphics.
What form of randomness is desirable, however, is an active topic of research.
We will here quickly review the two main properties (spectral and discrepancy) before relating to the  machine learning and domain-specific language background relevant for this work.
A survey of spectral properties from 2015 is provided by Yan and co-workers~\shortcite{yan2015survey} and formal details about discrepancy can be found in a textbook by Kuipers \shortcite{kuipers2012uniform}.

\mysubsection{Spectra}{Spectra}
Yellot \shortcite{yellott1983spectral} first noted, how the receptors on the retina are neither regular nor random but follow very specific patterns.
These patterns, as any sampling pattern, are now routinely characterized by  their \emph{spectra}.
A sampling pattern spectrum (or periodogram) is computed by averaging over many Fourier transforms of many instantiations of that sample pattern.
For two or more dimensions, the full spectrum is often further \emph{radially averaged} to a projected spectrum.
The variance of this estimate is called the radial \emph{anisotropy} which is low for radially symmetric patterns and large for others.
All these quantities will directly appear as losses of our formulation.

A \emph{blue noise} (BN) pattern is a pattern with a power spectrum with no power in low-frequency range.
Blue noise was first used in graphics for  dithering \cite{ulichney1988dithering} and stippling \cite{secord2002weighted,deussen2001floating}.
Classic ways to produce BN patterns are dart throwing \cite{mccool1992hierarchical} and Lloyd relaxation \cite{lloyd1982least}.
The first can be slow, while the latter often suffers from regularity artifacts, that need extra effort to be overcome \cite{balzer2009capacity,de2012blue}.
As in the context of dithering, models of human perception can be made used to improve quality \cite{mulligan1992principled}. %, we support perceptual losses in our framework. 
BN patterns are also used for Monte Carlo integration-based image synthesis, as they shift the error into the high-frequency bands, to which humans are less sensitive \cite{cook1986stochastic}.
However, there is no guarantee on the magnitude of the integration error.
Besides blue noise, other colors of noise are useful in tasks such as procedural primitive placement.
Some recent methods allow to produce patterns from the specified target spectrum in two dimensions \cite{wachtel2014fast,wei2011differential,ahmed2015aa,kailkhura2016stair}.
We take it a step further and allow to work in any dimension, also prescribe projective properties, combined with discrepancy or histograms.

A concept very similar to BN is the \emph{Poisson disk} \cite{mccool1992hierarchical} or max-min distance.
In such a pattern, the minimal distance from one point to the others is maximized over all points \ie all samples keep a minimal distance.
Alternative to a Fourier analysis, we allow losses using histograms of distances.
Histograms of points distances (the \emph{differential domain}) \cite{bowers2010parallel,wei2011differential} or point correlation \cite{oztireli2012analysis} are a more flexible tool to analyze sampling patterns.
In particular, they allow working on non-uniform samples and anisotropic spectra \cite{pilleboue2015variance,Singh:2017:CAA,Singh:2017:VCALS}.
We use both differential representations (pair correlations and spectra) in our network and define losses that enforce desired differential properties.

Many technical alternatives have been considered to produce blue noise patterns such as variational \cite{chen2012variational}, optimal transport~\cite{Qin:2017:WBN,DeGoes:2012:BNOT}, tiling \cite{ostromoukhov2004fast,wachtel2014fast}, Wang tiles \cite{kopf2006recursive}, kernel-density estimation \cite{fattal2011blue}, smooth particle hydro-dynamics \cite{jiang2015blue} or electro-statics \cite{schmaltz2010electrostatic}.
All these methods include involved mathematical derivations, can only realize a subset of the  properties and are limited in dimensionality and/or speed. 

Multi-class \cite{wei2010multi} blue noise is an extension where samples belong to different classes and get arranged, such that within each class the pattern is blue noise, as well as the union of all patterns.

In summary, we see that a lot of methods exist, but all  need a fairly involved mathematical derivation, support only a subset of properties and can be slow to compute.

\mysubsection{Discrepancy}{Discrepancy}
A concept orthogonal to the spectrum is the \emph{discrepancy} of a sample point set.
It is mainly relevant, if the set is to be used for Monte Carlo (MC) integration, such as in rendering \cite{cook1986stochastic}, or image reconstruction.
Loosely speaking, discrepancy computes the difference between an area and the number of points in a sub-domain (\eg a rectangle) \cite{shirley1991discrepancy,kuipers2012uniform}.
In a low-discrepancy pattern, this ratio is constant, \ie the same everywhere.
Typically, discrepancy measures the maximal deviation across all possible subareas.
The shape of the sub-area leads to other definitions of discrepancy (stars, boxes, etc).
Commonly, discrepancy is believed to be an indicator for a low integration error.

In particular, patterns that produce a low integration error, are often not random but structured in very specific ways \cite{kollig2002efficient}.
Regrettably, the low quasi-Monte Carlo (QMC) errors, come at the expense of structured patterns.
While the structured patterns can be reduced by randomized quasi-Monte Carlo (RQMC) that randomly shifts the pattern using so-called Cranely-Paterson rotation, no guarantees on the resulting spectrum can be given.

Another related property is $N$-rooks, which assures that subspaces are filled exactly once, or more general, a Latin-hypercube distribution.
Our approach includes a loss to address such projective properties.

As the integrands encountered in light transport are often very high-dimensional, the patterns need to scale to such dimensions.
While arguments exist, that under certain conditions, certain properties of some patterns are better than others, the relation between those properties and the ultimate perceived quality remain difficult to capture.

Discrepancy is routinely used as a measure to predict how useful the pattern would be for MC integration.
The notion of \emph{equidistribution} \cite{kuipers2012uniform} is much more immediately linked to MC: A sample pattern is said to be equidistributed, if and only if, it produces a low maximal error across all integrable functions (Eq.~1.2 in Kuipers and Niderreiter~\shortcite{kuipers2012uniform}).
In this work, we will rather optimize towards equidistribution than towards discrepancy, as the latter does not allow for effective back-propagation.
More specifically, we will investigate directly optimizing patterns for integrating a subset of functions, namely  those encountered in signals we actually wish to integrate: 2D images, 4D light fields, 5D temporal light fields etc.

\mycfigure{Method}{
Overview of our method.
The learning part starts from an initial point set shown left in 2D and as a stacked vector with $x$ and $y$ values.
We here show computation of the green point that depends on the blue and orange point.
The filter is translation-invariant, working on the offsets $d_1$ and $d_2$ that combine pairs of points.
The offsets are fed into the RBF-based non-linear filters \textbf{(b)} in multiple iterations that assign a weight to each offset.
In each iteration, a different filter is used (here we show three iterations), which typically  gets more spatially compact.  % Karol for subsequent iterations?
After the iteration, the resulting points are assessed by a combination of losses and improvement are back-propagated to the RBF filters.}

\mysubsection{Mixed}{Mixed}
Recent methods try to explicitly combine spectral and discrepancy properties.
Reinert~\etal\shortcite{reinert2016projective} produce blue noise patterns that share the Latin hypercube properties of typical low-discrepancy patterns: their patterns are blue noise also when projected to subspaces.
This results in a typical cross-like spectrum.
More general low-discrepancy was introduced by Ahmed~\etal\shortcite{ahmed2016low}.
The relation of blue noise and discrepancy is not fully clear as discussed by Subr and Kautz~\shortcite{subr2013fourier} as well as Georgiev and Fajardo~\shortcite{georgiev2016dithered}: It is evident, that there are methods that produce a low error, yet produce a more suspicious artifact pattern and that there are other approaches that produce visually pleasing patterns but a high error.
As losses can simply be added, we can combine spectral and integration desiderata.

\mysubsection{Learning}{LearningPW}
Computer graphics, and in particular filtering recently sees a push towards a learning-based paradigm, where, instead of implementing algorithms from first principles and mathematical derivations, data is used to optimize a general architecture to perform a task.
In particular for inverse problems, this idea has led to ground-breaking achievements \cite{Krizhevsky2012}. %\cite{krivniewzk2009}.
But also in graphics, learning of filters to solve tasks had been suggested before: Fattal~\etal\shortcite{fattal2011blue} optimize for hierarchical filters to solve tasks like Poisson integration.
Rendering is a key application of sampling patterns, where deep learning has been used for relighting \cite{Ren2015}, screen-space shading \cite{Nalbach2017}, volume rendering \cite{Chaitanya2017} or denoising \cite{Kallweit2017}. Recently, reinforcement learning is used for directing the samples to reduce error during light transport~\cite{Dahm:MLE:2017,Dahm:LTRW:2017}.

In this work, we apply the idea to learn filters to solve the task of creating sampling patterns.
We make use of unstructured convolutions, as pioneered by PointNet \cite{qi2017pointnet}, but suggest a simpler, fully convolutional radial basis function-based kernel representation.
Instead of inferring labels or per-pixel or per-point attributes such as normals, we optimize for filters that transform sets of random points into a sets of points with the desired properties.
Furthermore, our network is recursive, catering to the need of changing the point positions which are typically kept constant in PointNet and follow-up work.

\mysubsection{Domain-specific languages}{DSLs}
Our system introduces \emph{sample programs}, a notation to define sample pattern requirements using programmatic expressions.
This is a simple instance of a domain specific language, such as recently proposed for image synthesis \cite{anderson2017aether}, non-linear image optimization \cite{devito2017domain,heide2016proximal} or physics \cite{bernstein2016ebb}.
Instead of deriving our own parser, we provide functions in TensorFlow \cite{Abadi2016} that are parsed and evaluated efficiently during training and testing thanks to TensorFlow's symbolic analysis and GPU evaluation support.

\mysection{Sampling End-to-end}{OurApproach}
We refer to our approach as end-to-end sampling, as we do not construct a forward algorithm to produce a sampling pattern.
Instead we suggest a general sampling scheme that can be optimized in respect to a desired end-goal.

We now give an overview (\refSec{Overview}), followed by the details on how to achieve this: learnable non-linear recursive filters that work on unstructured data (\refSec{Filters}), an architecture to train those filters (\refSec{Architecture}) and, most importantly, the losses describing the goals (\refSec{Losses}).

\mysubsection{Overview}{Overview}
Our architecture comprises of two parts: A \emph{learning} stage and a \emph{test} stage (\refFig{Method}).
We will publicly provide both pre-trained filters that are readily applicable to produce the desired sample patterns, as well as the full architecture, including the losses, to construct new sample patterns.

At the learning stage, many training point sets are fed into the architecture that comprises of unstructured, non-linear recursive filters.
The output is analyzed in respect to the configurable losses.
This error is back-propagated to the filters such that their result improves.
Note, that we need to reformulate concepts like discrepancy, blue noise or progressiveness in order to become back-propagatable.
Providing these components is the key technical contribution of this paper.

It is important to see, that we are not given pairs of ``bad'' and ``good'' absolute sample patterns and learn how to transfer one into the other, which appears a daunting task.
Instead, we learn a much simpler task of adjusting a pattern such that its statistics follow prescribed goals.

At the test stage, a user provides a new set of points that can be converted to have the desired properties by applying the filters learned before.
The filters are compact and quick to apply.
While many ways exist to train the network in TensorFlow for different losses, there is only one resulting network structure, independent of the loss.
This structure is easily and efficiently implemented.
%(will provide a PBRT~\cite{Pharr:2016:PBRT} \texttt{Sampler} in C++ and a GLSL GPU implementation).
The particular instance is parametrized by a compact latent coding with only a handful of degrees of freedom.

\mysubsection{Tunable convolution on sample patterns}{Filters}
We start by defining our tunable filters, before we go into the details on how to optimize over their parameters.
These filters are required to work on unstructured data, \ie a list of $n$-d points, and that also recursively; and shall be expressive enough to perform non-linear operations. We go over those three aspects in the following paragraphs.

% X point set
% n dimensionality
% m RBF kernels
% o number of points
% n_s iterations

\paragraph{Convolution}
As our data comprises of $o$ unstructured points $X=\mathbf x_1,\ldots,\mathbf x_o$ in $n$-D, we first have to introduce a convolution on such data.
PointNet \cite{qi2017pointnet} and following papers have made use of symmetric functions and rotational transformers, but their tasks like segmentation and classification are different.
We found a much more straightforward extension from a structured to the unstructured domain to be effective: We parametrize our kernels as a weighted sum of $m$ RBFs (we use Gaussians) $\mathcal N$ with fixed $n$-d position $\mathbf\mu_i$ and variance $\sigma_{\mathcal N}^2$, where the weights $w_i$ are tunable and correspond to classic filter mask entries.
In other words, a convolution $\mathcal C$ of a sample point $\mathbf x_i$ with a kernel parametrized by a weight vector $\mathbf w$ is defined as:
\[
\mathcal C(
\mathbf x_i
|
\mathbf w)
=
\frac{
\mathbf x_i
+
\sum_{j\neq i}^o
\sum_{k=1}^m
w_k
\mathcal N(\mathbf x_i\ominus\mathbf x_j|\mathbf\mu_k,\sigma_{\mathcal N})
\mathbf x_j
}{
|
1+
\sum_{j\neq i}^o
\sum_{k=1}^m
w_k
\mathcal N(\mathbf x_i\ominus\mathbf x_j|\mathbf\mu_k,\sigma_{\mathcal N})
|
},
\]
where $\ominus$ denotes the torroidal vector difference.
In a slight abuse of notation, we will refer to the (overloaded) convolution of all points $X$ as $\mathcal C(X)$ as well.
Note, that the weights $\mathbf w$ can also be negative.
Typically we use \unsure{$m=20$}.
The means $\mathbf\mu$ are placed following a low-discrepancy pattern -- Hammersley -- to allow covering higher dimensions easily.
The variances $\sigma_\mathcal N$ are all chosen the same as \unsure{$0.4$}, the size of domain.
Please note, that the division produces a partition of unity.

\paragraph{Non-linearity}
The above filters are non-linear by construction.
We also experimented with introducing explicit non-linearities such as ReLUs but did not observe an improvement.

We found the residual approach in the above formulation to improve the results compared to a filter iterating over all points. 
The RBFs have to be able to represent the identity upon convergence, \ie a Dirac that maps the point to itself and nothing else.
This cannot be done when using a straightforward non-residual formulation for numeric reasons (a single extremely high Gaussian in combination with many zeros).
As a solution, we rather learn an update and always keep $\mathbf x_i$.
Now identity is easily produced using RBF weighs of zero.

\paragraph{Receptive field}
To avoid computing interaction between all samples in the convolution, which would imply quadratic time complexity, we limit the convolution to a constantly-sized neighborhood of a receptive field $\sigma$, that is typically chosen to be a fraction of the domain, such as $0.4$.
The variances of the Gaussians forming the RBFs are to be scaled accordingly, \ie their effective size in this example is $.1\times.4=.04$, \ie four percent of the entire domain.
We also found this critical for learning of more complex losses to converge.

\paragraph{Iteration}
The above can be applied to a point cloud directly to produce a new one.
Applying the filter is similar to an update in a Jacobi or Gauss-Seidel-type optimization or a generalized step of Lloyd relaxation.
Repeatedly applying the filter would further improve the result, given the filter is optimized for such an operation.
We call each repeated application of the filter an \emph{iteration}.
As our training set comprises of continuous random vectors, the concept of epochs is not applicable.
We will refer to learning effort in units of batch counts, \ie how many sample patterns were produced, divided by the number of patterns per batch, that needs to be larger than one for effective training.
After experimenting with sharing the filter weights $\mathbf w$ across iterations, we found better convergence by using different filters $\mathbf w^{(l)}$ in different iterations, so the overall forward expression for the network is
\[
\mathcal C(\mathcal C(\mathcal C(\ldots |\mathbf w^{(l-2)})|\mathbf w^{(l-1)})|\mathbf w^{(l)})
\]

\paragraph{Learning}
Subsequently, we simply stack all the filter weights $\mathbf w^{(1)},\ldots,\mathbf w^{(n_\mathrm s)}$ for all RBF kernels in all iterations into a parameter vector $\Theta\in\mathbb R^{n_\mathrm s\times m}$ to optimize over.

\mysubsection{Architecture}{Architecture}
After having established tunable, iterative filters that run on unstructured data, we can now optimize a parameter vector that produces the desired results.
Such an optimization comprises of an outer loop across many sets of random points and an inner loop of three steps: initialization, filtering and back-propagation in respect to a loss.

% \paragraph{Initialization}
% Different initializations are possible leading to different filters.
% Here, we denote the three options.

% A simple solution is to start from \texttt{uniform} random noise.
% However, it can happen that two or more points fall so close to each other, that applying an unstructured convolution runs into numerical difficulties.
% Consider two identical points and a loss that tries to enforce a non-zero distance: There would not be a unique and learnable response.
% Instead we require an input where no two points are closer than $\epsilon$, in our case \unsure{$.01$} where the domain is from 0 to 1.

% An alternative solution is to start \texttt{jittered}, a pattern that naturally fulfills the $\epsilon$ criterion.
% However, producing high-dimensional jittered patterns is not possible, other solutions are required.

% A third option is to start from \texttt{lowDiscrepancy} already.
% There are methods that easily produce well-placed samples in high dimensions.
% Starting from such a pattern, we can optimize for better discrepancy, other notions of discrepancy, \eg that adapt to the task, and most importantly for blue noise or progressiveness.

\paragraph{Filtering}
We can initialize the network with any known sampling pattern (random, jittered or low discrepancy). After this initialization, the point set is filtered many times ($n_\mathrm s$) using the iterative filter.
The resulting point set is then submitted to the loss in the next step.

\paragraph{Gridding}
We support both non-gridded and gridded sample patterns.
In a \emph{non-gridded} sample pattern, all dimensions are filtered.
This is the default.
A \emph{gridded} sampling pattern comprises of a $n'<n$ dimensions which are fixed and their values are not changed by the filters.
Always, all $n$ dimensions are input to the filters, but the filter only outputs the $n'$ dimensions to update.
An example of a gridded pattern with $n=3, n'=1$ is a pattern where the first two fixed dimensions are the pixel centers, and the third dimension is the wavelength.

\paragraph{Backpropagation and Losses}
The loss can be a linear combination of many sub-losses, which we will detail in \refSec{Losses}.
It maps the point set to a scalar value, that is low if the point set well-fulfills the requirements.
For now, it is enough to assume that the loss is back-propagatable in respect to the choice of filter parameters $\Theta$.

\mysubsection{Sample programs}{SamplePrograms}
Core of our method is to define the  desirable random sample pattern properties as a back-propagatable \emph{point sample program}.
The point sample program is an expression such as \texttt{bn(s) + bn(proj(1D,s)) + discrepancy(s)}: A pattern that is blue noise per-se, that has a blue noise projection and also is low-discrepancy.
A sample program expression is formed of three parts: The sample pattern \texttt{s}, losses that quantify the quality of s and operators that map a sample pattern into a new sample pattern, potentially of a different dimension and sample count. 
The name of the losses and operators  are denoted in \texttt{teletype} font.
The family relation of all losses is shown in \refFig{Family}.
Next, we discuss these losses (\refSec{Losses}), before explaining the operators (\refSec{Operators}).

\myfigure{Family}{Family tree of our operators \emph{(blue)} and losses \emph{(orange)}.}

\mysubsection{Losses}{Losses}
A loss maps a point sample set to a scalar.
It is the last (outermost) expression in a sample program and defines properties such as point correlation or discrepancy in several most relevant variations.

\mysubsubsection{Fourier}{FourierLoss}
The \texttt{spectral} loss measures the frequencies found in the pattern.
These are defined on correlation of pairs of sample points.
The spectral loss is defined as the $\mathcal L_2$ distance between a desired $n$-dimensional spectrum, given either as a table or an analytic function, and the spectrum of the current point set.
When \eg a blue noise pattern is desired, a user simply provides a 1D table where the blue frequencies are enhanced and others are suppressed.
The target spectrum is provided by means of a 1D table or 2D image.
In higher dimensions, implicit descriptions of the desired spectrum are used.

We have found the use of mini-batches essential to make a Fourier loss converge faster: Producing a single spectrum of a single realization is typically noisy, which is why, even for visualization, many spectra need to be averaged.
As the stochastic gradient descent is computing gradients in respect to this noisy spectrum, they are even more noisy, leading to low convergence or even divergence (especially, for small sample count $N$).
We observed, that with a mini-batch size of $4$, spectra are converged enough to be used for gradient computation.

We also found, that excluding the DC term (that is just a sum of all points) from the loss helps convergence, especially for sampling algorithms (e.g., \cite{wachtel2014fast}) that generates varying sample count over different realizations. Samplers for which the number of points doesn't change (e.g., for jittered) over multiple realizations, this over-constraints the problem artificially. 
 
\mysubsubsection{Differential}{DifferentialLoss}
The \texttt{differential} loss is another pair correlation loss.
Here, the relation between two points is reduced to a scalar distance that is inserted into a histogram using kernel density estimation (KDE) with a Parzen filter instead of hard counting.
Doing so, the construction becomes smooth and hence, back-propagatable.
We typically use histograms of $128$ bins.

\mysubsubsection{Anisotropy}{Anisotropy}
Anisotropy is special in that it only works in combination with a projective loss.
Whenever a projection is performed, a higher-dimensional distribution $F$ of values is replaced by its mean $\mathbb E[F]$.
Anisotropy is the variance $\mathbb V[F]$ of this distribution.

When performing a radial projection of a 2D pattern to 1D for example, this loss can be use used to encourage the pattern to have circular uniformity.
While in 2D, this could also be achieved directly by a \texttt{spectral} loss with a radially smooth 2D image, in higher dimensions, we typically do not work with a full spectrum, but rather its radial averages or projections.

\mysubsubsection{Discrepancy}{Discrepancy}
Discrepancy is a measure that compares if the number of points in a sub-domain is proportional to the area of that sub-domain for all possible sub-domains.
Typically, the sub-domains are quads or nested sequences of quads in different sizes.
Regrettably, this notion is not back-propagatable due to the piecewise constant box function.
We therefore suggest to use a slightly generalized notion of \texttt{discrepancy}: We call discrepancy the difference $d=F-\bar F$ between the analytic result $F$ of an integration and the result $\bar F$ found when using the sample pattern.
In other words, discrepancy of a pattern is low, if MC integration was successful.
Note, how box discrepancy is a special case of this, but this notion allows a generalization that includes smooth functions, which in turn become back-propagatable.
In particular, we make use of Gaussians again: To compute discrepancy, we simply sample a number of random Gaussians in the domain.
We know their analytic integral (care is to be taken to handle the boundary).
This analytic integral is easy to compare to the MC estimate of this integral when using the points.
Comparing the two provides the smooth discrepancy that is back-propagatable.

%We will show evidence, how this notion of discrepancy actually is a better predictor for performance in many Computer Graphics tasks then classic box discrepancy is.
%Even more importantly, it allows to extent the end of the optimization chain by another step: To include the task into the loss as detailed next.

\mysubsubsection{Task-discrepancy}{TaskDiscrepancyLoss}
Our discrepancy now allows to produce patterns that have a low discrepancy for a particular task.
Instead of looking at the sample pattern $S$ or its statistics, we look at a sampling-dependent signal $F(S)$ and a sample-independent reference signal $\bar F$.
If we were, say, to use the sample pattern to compute integrals of the product of natural illumination and BRDF, we can now inject this data at learning time, and optimize for filters that will turn point patterns into point patterns that are suitable for this particular task.
The same can be repeated for anti-aliasing of fonts, super-sampling of vector graphics, stippling etc.
Our current implementation allows for all tasks that can be expressed as fetching an image at a location and computing the sum.
This allows applications such as MC integration of environment lights or super-sampling of fonts.
Future work will seek to make the set of tasks larger, but several difficulties out of the scope of this work would need to be overcome to make a ray-tracer including visibility back-propagatable.

\mysubsection{Operators}{Operators}
Operators map point sample sets to point sample sets of a different dimension and sample number.

\mysubsubsection{Projective}{ProjectiveLoss}
It has recently become of interest to not only achieve spectral properties in the full space but also in one or multiple projections of that space onto planes \cite{reinert2016projective,Singh:2017:VCALS,ahmed2016low}.
Such \texttt{projective} operations are seamlessly integrated by performing a projection and creating a spectrum later, as all intermediate operations are back-propagatable.
We support both Cartesian \texttt{projection} and \texttt{radialProjection}.

\mysubsubsection{Progressiveness}{Progressiveness}
Orthogonal to the above we can optimize for \texttt{progressiveness}: Here, instead of projecting all points, we just take random subsets for which we enforce the above constraints.
This incentivizes a pattern where not only the first $n$ points are blue noise but also $1,\ldots,n/2$ and $n/2+1,\ldots,n$ are on their own.

\mysubsection{Implementation}{Implementation}

%\paragraph{Sample count}
%We train the architecture using a fixed number of samples $o_1$. To apply the filters to a different number of input $n$-d points, we simply need to scale their support by a factor \unsure{$d=(o_1/o_2)^{-n}$}. %Karol: was it always true?

\paragraph{Training}
Training is implemented in TensorFlow. We train the architecture using a fixed number of samples $N$. As our training data set is infinite, it does not make sense to talk about epochs in our context. 
Instead, we state training effort as batch counts.
A batch comprises of multiple sample patterns that contribute to one optimization step.
We found a batch size of $4$ to work well in practice for better convergence.
In particular it helps to reduce the inevitable noise in a periodogram of any finite realization of a pattern by averaging many iterations. However, we observe that mini-batch size of $1$ (for large $N \geq 1024$) also does the job, but at the cost of longer training time. We typically train for a batch count of \unsure{10,000} batches.

For training the recursive filters, we simply unroll them.
For an iteration count of, \eg 30, we simply create a network of depth 30.
We train using the ADAM optimizer, using an exponentially decreasing learning rate that is initialized by $10^{-6}$. Learning a typical filter requires around 45 minutes of training time on an Nvidia Tesla V. The trained kernels can then be directly used to generate points on the fly (the cost involves only the recursive convolutional operations).

\mysection{Results}{Results}
Here we perform a quantitative analysis for our approach, including comparison to previous work (\refSec{Analysis}), before showing applications to rendering and object placement (\refSec{Applications}).

\mysubsection{Analysis}{Analysis}
\input{\figurePath/fig-analysis-samplers}
\paragraph{Spectra}
We start by performing a multi-dimensional analysis in~\refFig{analysis-samplers} of different state-of-the-art sampling patterns that ranges from 1D to 5D dimensions. We also look at their (radial) projections. All target spectra are computed for $N=1024$ point samples and are appropriately scaled for different dimensions for better comparison of the low frequency region (frequency axis scaled by $N^{1/d}$ for $d$-th dimension in radial plots).
%\refFig{analysis-samplers} shows our multi-dimensional analysis.
%It looks into the spectra of the state of the art of sampling patterns, as well as their projections, including dimensions up to 5D.
%We will now discuss different properties of different methods.

Starting from the top in~\refFig{analysis-samplers}, a random pattern shows no pronounced spectra. A simple way to improve error in integration is via stratification, which can be easily achieved by Latinhypercube (LHC) or N-rooks sampling. This method is easily scalable to higher dimensions and has dark anisotropic cross along the canonical axes due to dense stratification (column 2, row 2). 
However, this doesn't help improve Monte Carlo (MC) variance convergence (unless the integrand variations are aligned to the canonical axes, as illustrated by Singh and Jarosz~\shortcite{Singh:2017:CAA}).
%This is desirable for all methods, but LH lacks blue noise off the canonical axes.

The third row shows jittered sampling spectrum (second column) that has significant low energy region in the low frequency region around the DC (which is at the center of each spectrum 2D image). This leads to good convergence properties~\cite{pilleboue2015variance}. %However, as shown in the 2D projections (columns 4,5,6) this dark region shrinks with increase in dimensionality leading to deterioration in convergence. Jittering also suffers from the curse of dimensionality due to the underlying grid construction.
Regrettably this low-power area shrinks in higher dimensions as seen in the 2D projections from 3D, 4D and 5D (the dark region shrinks), which is the other reason (besides the curse of dimensionality of a dense grid), that jittered patterns are unattractive for higher dimensions.

In the fourth row, BNOT~\cite{DeGoes:2012:BNOT}---generated using tiling approach by Wachtel et al.~\shortcite{wachtel2014fast}---produces a large blue noise region around DC, but is only available in 2D (therefore, any further columns are missing). Many other variants exist but have limitations over the dimensions.
%also produces some undesired anisotropy. 
Non-deterministic samplers like Halton (shown in the fifth row) are easily extended to higher dimensions, have anisotropic dark regions in different projections that is known to lower the variance during MC integration. 
%can produce an even larger dark area around the canonical axis, also in higher dimensions, but lacks BN.

Recently, Perrier et al.~\shortcite{perrier18eg} proposed a high-dimensional sampler 
that can combine both properties (following the work by Ahmed et al.~\shortcite{ahmed2016low}), but at the expense of high anisotropy and a reduced amount of blue noise in higher-dimensional projections. This algorithm also suffers in the mixed projections (e.g., blue noise in \texttt{(x, y)} \& \texttt{(u,v)} projections doesn't imply blue noise in the mixed \texttt{(x,u)} or \texttt{(y,v)} ones).
Our proposed framework (last two rows in~\refFig{analysis-samplers}) is expressive enough to control different properties along different projections (as demonstrated later in~\refFig{halftoning}). 

%We show results for two different target spectra (jittered and BNOT). 
In the last two rows, first we show (penultimate row) ours when trained with a loss encouraging a 2D jittered spectrum in all dimensions (see the third row %column 
to recall that this does not happen when running jittered sampling). 
Since training for a full dimensional power spectra has diminishing returns (due to shrinking power spectra as we go in higher dimensions), we consider a 2D jittered target spectrum and define our loss as a sum of the L2 losses over each 2D projection (wrt the target). 
As a result, the low-power area (dark) in our jittered samples remains relatively large, even in higher dimensions and their corresponding 2D projections. This is important to gain convergence improvements. %, independent of dimension.

When using BNOT's spectrum (last row) in our loss across all projections, we find an even larger BN area in 2D and a consistently large BN area in all projections. Note, that for both of our variants, the anisotropy is low, also in all projections.
Our method does not directly produce the acclaimed cross \cite{reinert2016projective,ahmed2016low} (albeit we could optimize for it), but instead tries to have a good spectrum in all subspaces, not only axis-aligned: In other words, the cross might just be a sign for partially successful attempt that has found a way to produce very good BN along some (canonical) directions, but remains poor along almost all others (diagonals).
%
%\mycfigure{HistogramResult}{
%Differential histogram analysis when using our approach with a \texttt{histogram} loss.
%From left to right the pattern dimensionality increases.
%The blue curve denotes the target, the green curve the histogram our approach produces.}
\input{\figurePath/fig-histogram-loss-poissondisk-all}
\paragraph{Discussion}
With our approach, we manage to increase the dark regions in higher dimensions without worrying about curse of dimensionality (well known for jittered samples). We further showed that our approach preserves some blue noise characteristics in all the projections. However, several questions on spectral properties remain to be investigated.
We do not see our patterns, while they have some unique properties, as the best pattern ever, but would like to recall they were produced without diving into any intricate mathematical details or implementation maneuvering.
%mathematical derivation or implementations.
Ultimately, we hope this ease of implementation to foster construction of new patterns that bring forward their overall understanding.

\paragraph{Histograms}
To show the versatility of our approach, we also injected differential histogram as a loss function and trained our network 
to get Poisson disk samples. We demonstrate our results in~\refFig{histogram-loss-poissondisk} for dimensions upto $5D$ for $N=1024$. Similar to the spectral loss, we consider a 2D Poisson Disk target PCF and define our loss as a sum of the L2 losses over each 2D projection (wrt the target).  
%Alternative to spectra, we look into patterns produced by optimizing for a certain differential histogram in \refFig{HistogramResult}.
For 2D, the histogram loss does a pretty good job but in higher dimensions the known Poisson disk bumps in the histogram are not preserved. However, we managed to preserve the histogram shape at small distances. 
%see that our method is able to closely match the target histograms. 

%\paragraph{Computational efficiency}
%The compute time for varying numbers of input points is show in \refFig{ComputeTime}.

\begin{wrapfigure}{r}{5cm}
\includegraphics[clip, width = 4.9cm]{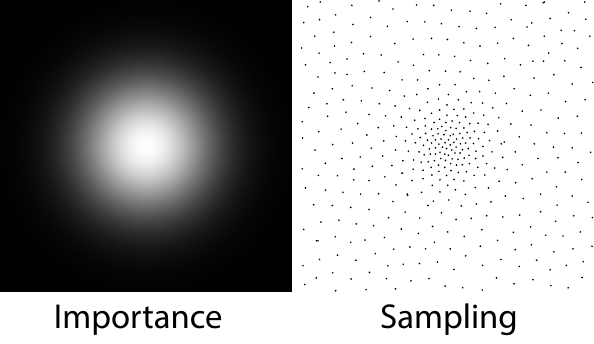}
\end{wrapfigure}
\paragraph{Adaptivity}
The ability to adjust for the number and therefore the density of samples naturally provides a means to perform adaptive sampling.
We show such a result in the figure to the right, where the left shows the importance map and the right our resulting pattern, that is \texttt{bn(s)} and follows the importance.

\input{\figurePath/fig-halftoning}

\paragraph{Gridding}
The ability to compute gridding masks (dithering patterns for rendering \cite{georgiev2016dithered}) is demonstrated in \refFig{halftoning}.
To achieve this, our framework doesn't require additional coding besides adding an enclosing \texttt{grid} operator that extracts the $x$ dimensions from a 3D pattern keeping, $y$ and $z$ fixed.
Explicit constructions of such masks can take considerable implementation effort  (simulated annealing). This also demonstrates our framework's ability to handle different target spectra along different projections (in this case, blue noise along 1D and uniform for the rest).
%and take long (Georgiev and Fajardo~\shortcite{georgiev2016dithered} report XX.X hours for patterns of similar size, while ours is computed in milliseconds).

\mysubsection{Applications}{Applications}
Our trained filters can efficiently be applied to problems such as rendering or object placement.
\paragraph{Rendering}
In~\refFig{renderings}, we render different scenes using PBRT~\cite{Pharr:2016:PBRT} with dimensionality varying from 3D to 5D and compare our jittered and blue noise (BNOT target) with Halton and classical jittered sampling. 
For fair comparison we implement 3D, 4D and 5D classical jittered 
sampling in the PBRT source code. All the scenes are rendered with a point light source to control the dimensionality of the underlying MC integration. First row shows 3D integrand for which the samples are generated over 2D pixel locations and along the 1D time axis to introduce motion blur. Visual inspection (insets) shows our jittered and BNOT manages to reduce the noise level (even if MSE values are not significantly improved). 
In the second row, 4D depth of field integrand (2D pixels + 2D lens) is computed using $N=256$ samples followed by a 5D integrand (2D pixels, 2D lens and 1D time) in the third row with $N=1024$ samples. 
We believe that our samplers could show  improvements in convergence compared to naive jittering due to the relatively large low-power (dark) region in higher dimensions (see~\refFig{analysis-samplers}). However, this requires a more focused convergence analysis that we leave for future work.

\paragraph{Object placement}
We further demonstrate the capability of our framework to handle different target spectra. In~\refFig{halftoning}, we show point set with a flower placed on it and the corresponding spectra for green and pink noises which is obtained from our trained filters.
%\myfigure{ObjectPlacement}{Object placement according to different spectra (shown as insets): blue, green and pink}
\input{\figurePath/fig-color-noise-spectralloss}
\mysection{Conclusion}{Conclusion}
We have proposed the first framework to end-to-end optimize for  filters that turn random points without properties into sample patterns with properties relevant for Computer Graphics tasks. 
Other than previous work that  requires mathematical derivation and implementation effort, we simply state the forward model as a loss and rely on modern back-propagation software to come up with a sampling method.
The methods resulting from our approach are very versatile: As we have shown several previous patterns can be emulated using our approach and in some cases even surpassed in terms of quality and/or computation speed.
We share execution efficiency with classic CNNs that require only a few passes across the input with constant time complexity and complete data-parallelism.  

Still many questions remain to be answered.
While we state the optimization and hope for modern optimizers to find good solution, at the one hand, we lack any theoretical guarantees.
On the other hand, most mathematical derivations also do not provide proofs, such as we are unaware of proofs that Lloyd relation converges in high dimensions.
Future work will need to investigate a detailed convergence analysis for different combination of losses.
Ultimately we would want to ask if any sample pattern can be learned as we here have only shown a small, but important, subset.

We think our approach to some extend is machine learning (ML), but with an indirection: a classic ML approach learns the mapping from input to output \eg a color image to a depth image.
Our task is slightly more indirect.
We do not provide supervision in form of pairs of input and output that sample a mapping.
Instead, we ``learn'' filters, that, when applied to originally random data are free to do to those points what they please, as long as they introduce structure in the form of the statistical properties.
This methodology might be applicable to other scientific questions, also beyond Computer Graphics.
% Tobias sees the interior of a bird bone.

Ultimately, we hope that our approach will support exploration of new sampling patterns, and both make their application easier in practical tasks as well as to move forward their theoretical understanding.
\input{\figurePath/fig-renderings}

\bibliographystyle{ACM-Reference-Format}
\bibliography{article}

\end{document}

%% file: images/fig-analysis-samplers.tex
%!TEX root = ../main.tex
%
\definecolor{mymagenta}{rgb}{1, 0, 1}
\definecolor{skyblue}{RGB}{135, 206, 255}
\begin{figure*}[t!]
\centering
\footnotesize
\hspace*{-1em}
\begin{tabular}{c@{\;}c@{\;}c@{\;}c@{\;}c@{\;}c@{\;}c@{\;}c@{\;}c@{\;}c@{}}
& \multicolumn{3}{c}{
\begin{tikzpicture}
\draw[magenta,thick] (0.1,0.125) -- (0.5,0.125);
\filldraw[black] (0.45,0.125) circle (0pt) node[anchor=west] {\tiny 5D};
\draw[green,thick] (0.8,0.125) -- (1.2,0.125);
\filldraw[black] (1.195,0.125) circle (0pt) node[anchor=west] {\tiny 4D};
\draw[skyblue,thick] (1.5,0.125) -- (1.9,0.125);
\filldraw[black] (1.85,0.125) circle (0pt) node[anchor=west] {\tiny 3D};
\draw[yellow,thick] (2.2,0.125) -- (2.6,0.125);
\filldraw[black] (2.55,0.125) circle (0pt) node[anchor=west] {\tiny 2D};
\draw[black,thick] (2.9,0.125) -- (3.2,0.125);
\filldraw[black] (3.15,0.125) circle (0pt) node[anchor=west] {\tiny 1D};
\draw[black,thick] (0,0) -- (3.8,0) -- (3.8,0.25) -- (0,0.25) -- cycle;
\end{tikzpicture}
} & 
\multicolumn{6}{c}{
\begin{tikzpicture}
\draw[<-] (0,0.05) -- (4.5,0.05);
\draw[->] (6,0.05) -- (11,0.05);
\filldraw[black] (4.5,0.05) circle (0pt) node[anchor=west] {Projections};
\end{tikzpicture}} 
\\
& 2D Samples & 2D Spectrum & Radial & to 1D &
\multicolumn{3}{c}{
\begin{tikzpicture}
\draw[<-] (0,0.25) -- (2.5,0.25);
\draw[->] (3.2,0.25) -- (5.5,0.25);
\filldraw[black] (2.5,0.25) circle (0pt) node[anchor=west] {to 2D};
\end{tikzpicture}} & to 3D & to 4D
\\
%& Samples & 2D Spectrum & Radial & to 1D & & to 2D & & to 3D & to 4D
%\\
\rotatebox{90}{\qquad Random}
&
\begin{tikzpicture}
  \node[anchor=south west,inner sep=0] (image) at (0,0)
  {
    \pdfliteral{ 1 w}\includegraphics[width=0.75in,page=1]{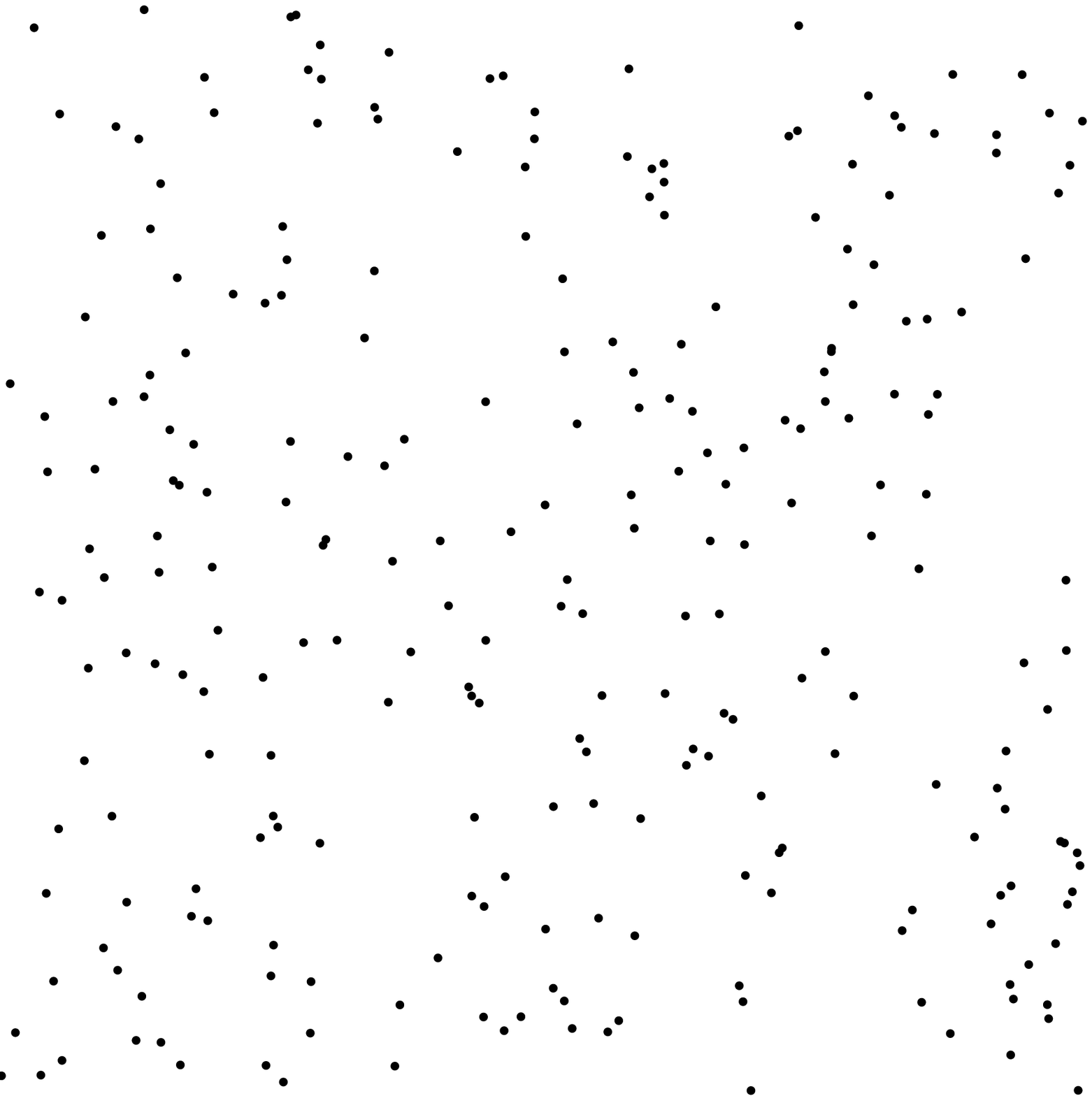}
  };
    \draw[black,thick] (0,0) -- (1.9,0) -- (1.9,1.9) -- (0,1.9) -- cycle;
\end{tikzpicture} 
&
\begin{tikzpicture}
  \node[anchor=south west,inner sep=0] (image) at (0,0)
  {
    \pdfliteral{ 1 w}\includegraphics[width=0.75in,page=1]{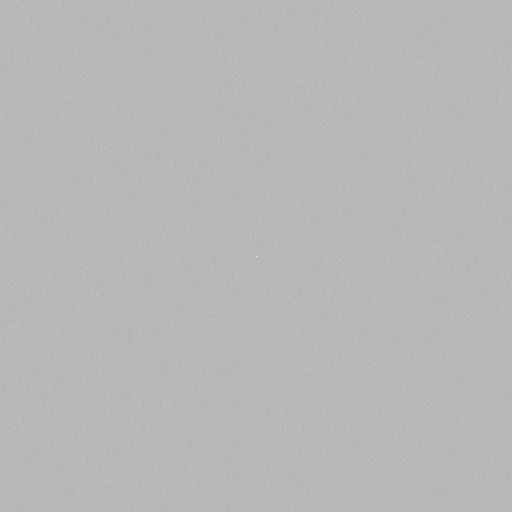}
  };
\end{tikzpicture} 
&
\begin{tikzpicture}
\node[anchor=south west,inner sep=0] (A) at (0,0)
  {
    \pdfliteral{ 1 w}\includegraphics[width=0.75in,height=0.35in,page=1] {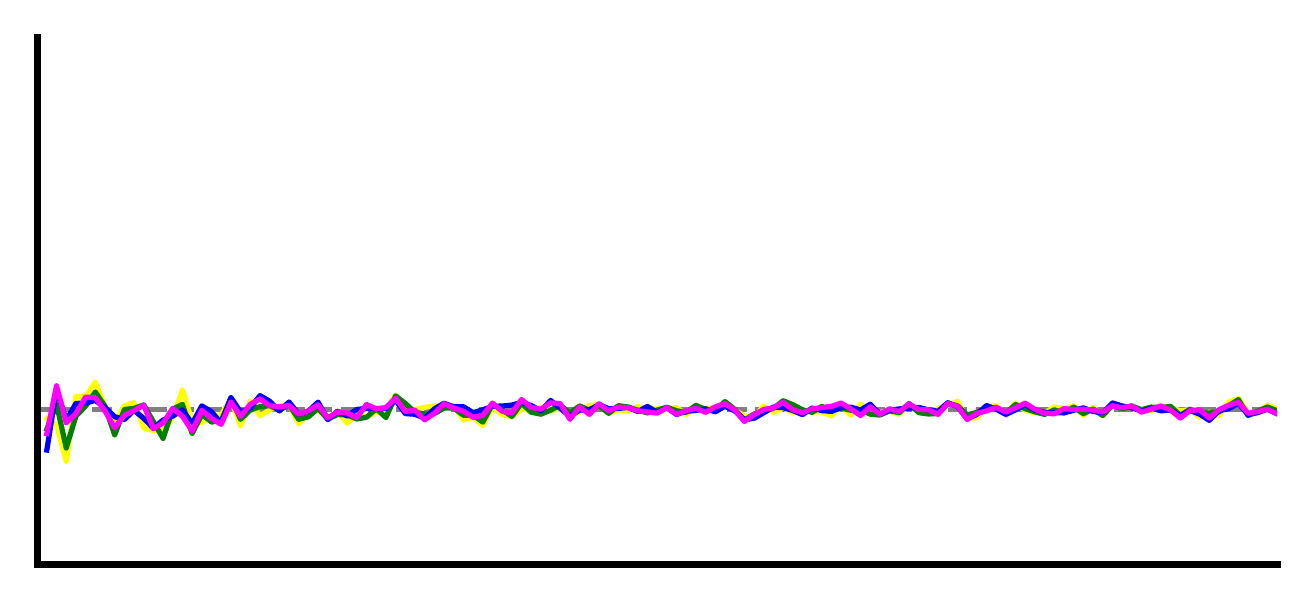}
  };
  \node[anchor=south west,inner sep=0] (A) at (0,0.95)
  {
    \pdfliteral{ 1 w}\includegraphics[width=0.75in,height=0.35in,page=1] {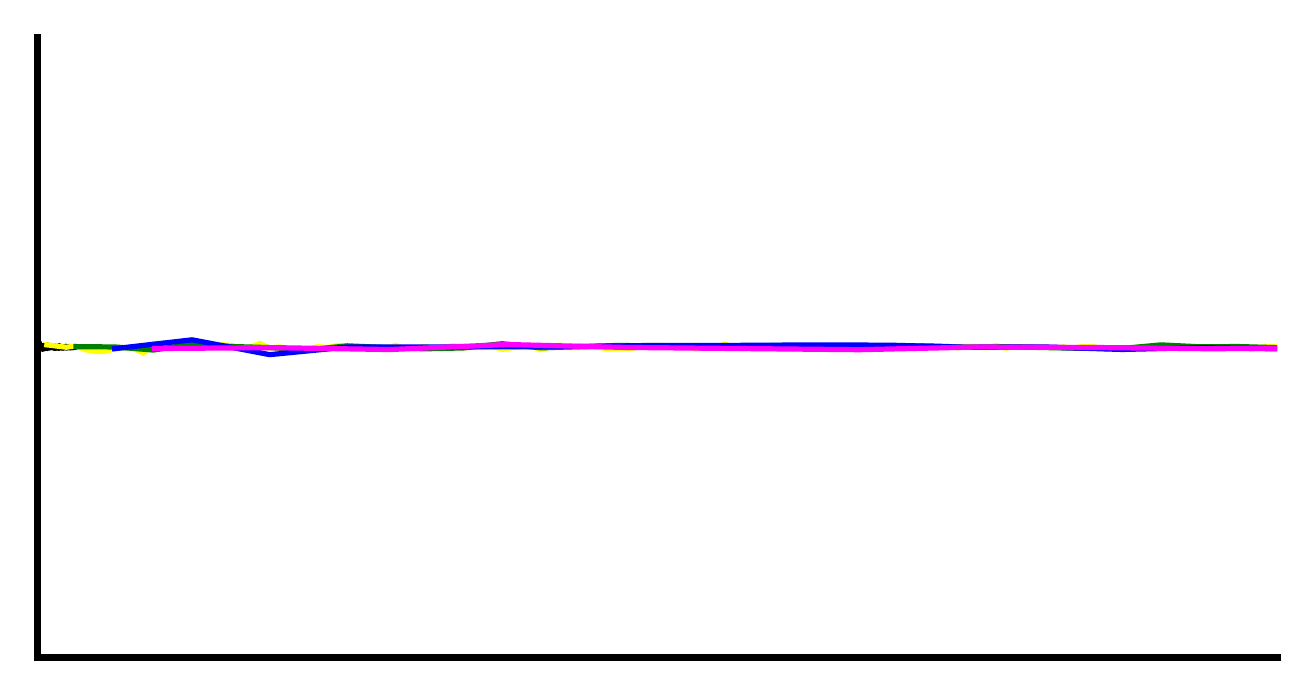}
  };
    \begin{scope}
       \filldraw[black] (0.05,1.7) circle (0pt) node[anchor=west] {\tiny power};
   \filldraw[black] (1.5,0.12) circle (0pt) node[anchor=west] {\tiny $\nu$};
     \filldraw[black] (1.5,1.1) circle (0pt) node[anchor=west] {\tiny $\nu$};
   \filldraw[black] (0.05,0.77) circle (0pt) node[anchor=west] {\tiny anisotropy}; 
     \filldraw[black] (0.75,1.5) circle (0pt) node[anchor=west] {\tiny radial mean};  
  \end{scope}
  \begin{scope}
  \filldraw[black] (0.01,1.5) circle (0pt) node[anchor=west] {\tiny $1$};
  \filldraw[black] (0.01,1.12) circle (0pt) node[anchor=west] {\tiny $0$};
  \filldraw[black] (0.01,0.2) circle (0pt) node[anchor=west] {\tiny $-10$dB};
  \end{scope}
\end{tikzpicture}
&
\begin{tikzpicture}
\node[anchor=south west,inner sep=0] (A) at (0,0)
  {
    \pdfliteral{ 1 w}\includegraphics[width=0.75in,height=0.35in,page=1] {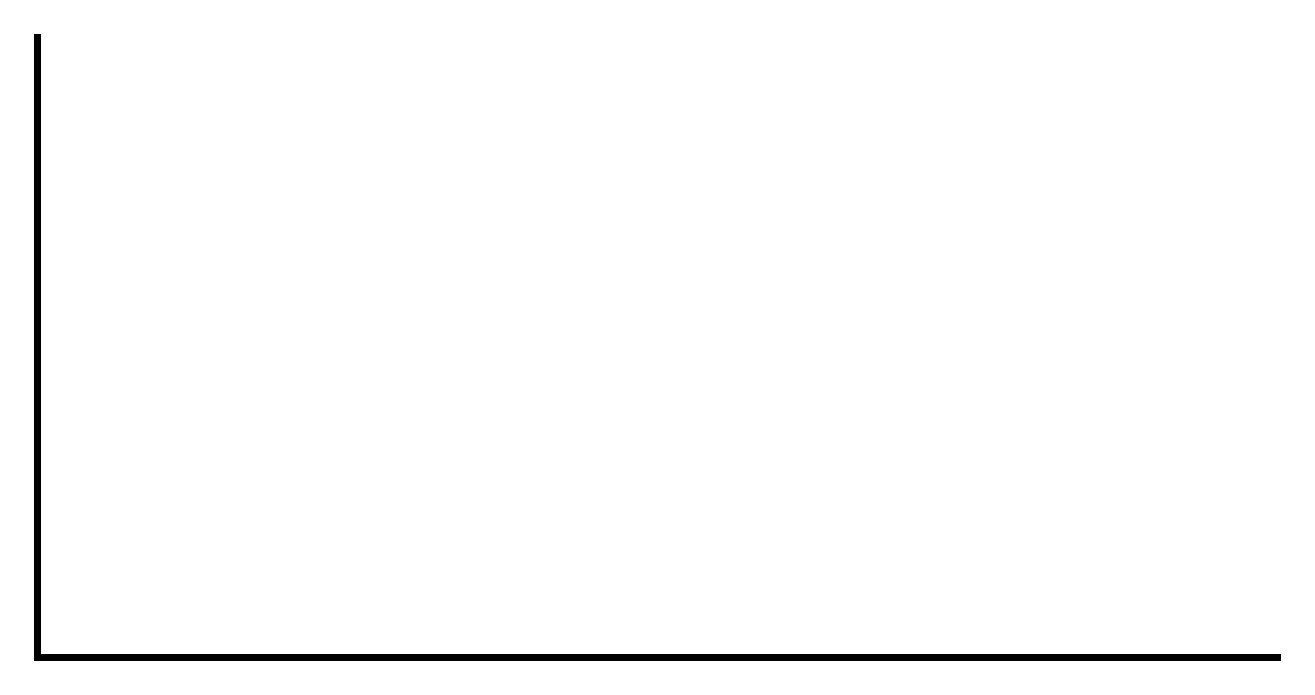}
  };
  \node[anchor=south west,inner sep=0] (A) at (0,0.95)
  {
    \pdfliteral{ 1 w}\includegraphics[width=0.75in,height=0.35in,page=1] {images/samplers/radial-mean-random.pdf}
  };
\end{tikzpicture}
&
\begin{tikzpicture}
  \node[anchor=south west,inner sep=0] (image) at (0,0)
  {
    \pdfliteral{ 1 w}\includegraphics[width=0.75in,page=1]{images/samplers/power-Pconstant-v1-random-proj0-1-n4096-001000.png}
  };
  \draw[mymagenta,thick] (0,0) -- (1.9,0) -- (1.9,1.9) -- (0,1.9) -- cycle;
\end{tikzpicture} 
&
\begin{tikzpicture}
  \node[anchor=south west,inner sep=0] (image) at (0,0)
  {
    \pdfliteral{ 1 w}\includegraphics[width=0.75in,page=1]{images/samplers/power-Pconstant-v1-random-proj0-1-n4096-001000.png}
  };
  \draw[green,thick] (0,0) -- (1.9,0) -- (1.9,1.9) -- (0,1.9) -- cycle;
\end{tikzpicture} 
&
\begin{tikzpicture}
  \node[anchor=south west,inner sep=0] (image) at (0,0)
  {
    \pdfliteral{ 1 w}\includegraphics[width=0.75in,page=1]{images/samplers/power-Pconstant-v1-random-proj0-1-n4096-001000.png}
  };
  \draw[skyblue,thick] (0,0) -- (1.9,0) -- (1.9,1.9) -- (0,1.9) -- cycle;
\end{tikzpicture} 
&
\begin{tikzpicture}
\node[anchor=south west,inner sep=0] (A) at (0,0)
  {
    \pdfliteral{ 1 w}\includegraphics[width=0.75in,height=0.35in,page=1] {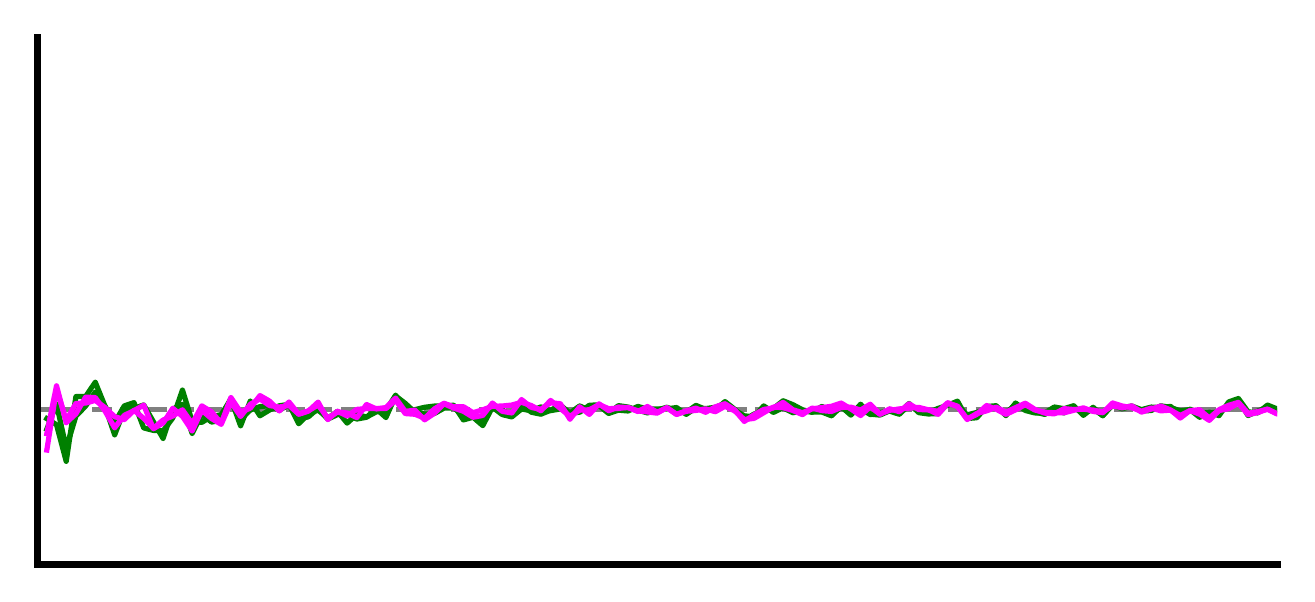}
  };
  \node[anchor=south west,inner sep=0] (A) at (0,0.95)
  {
    \pdfliteral{ 1 w}\includegraphics[width=0.75in,height=0.35in,page=1] {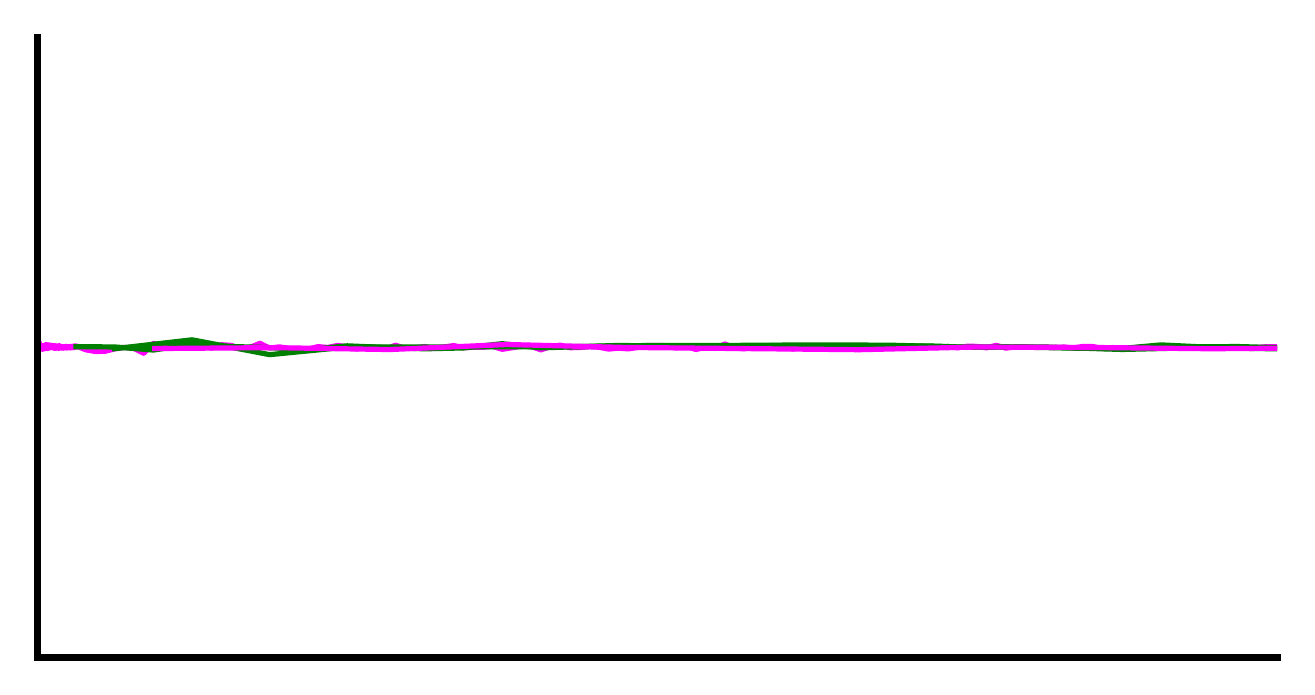}
  };
\end{tikzpicture}
&
\begin{tikzpicture}
\node[anchor=south west,inner sep=0] (A) at (0,0)
  {
    \pdfliteral{ 1 w}\includegraphics[width=0.75in,height=0.35in,page=1] {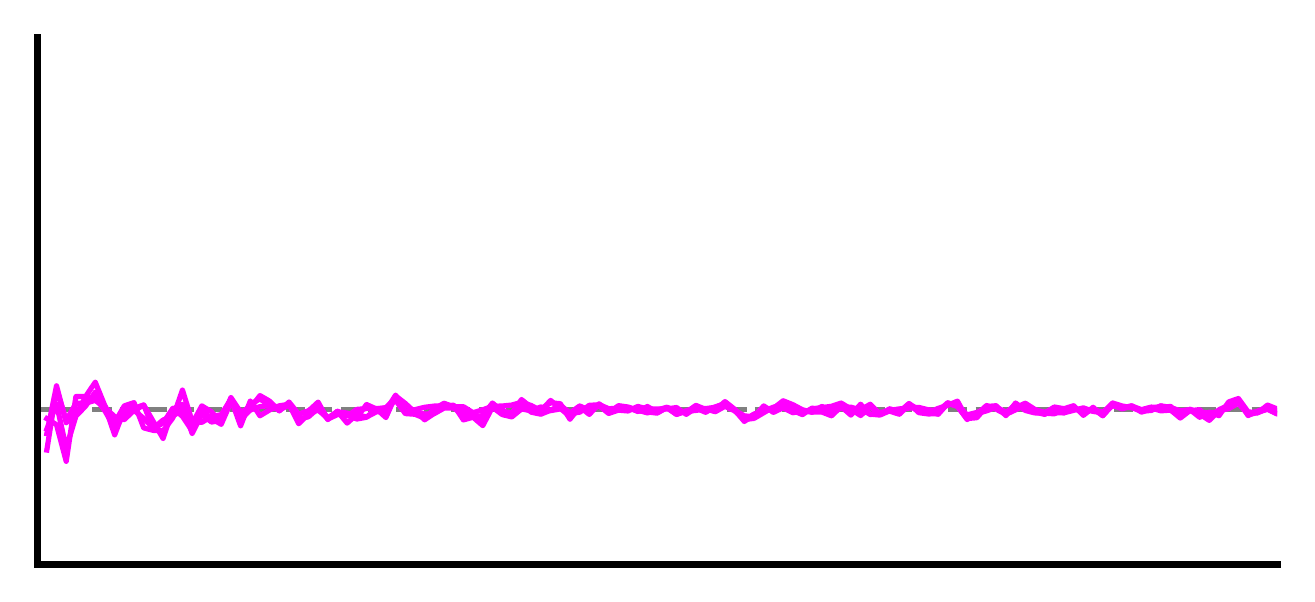}
  };
  \node[anchor=south west,inner sep=0] (A) at (0,0.95)
  {
    \pdfliteral{ 1 w}\includegraphics[width=0.75in,height=0.35in,page=1] {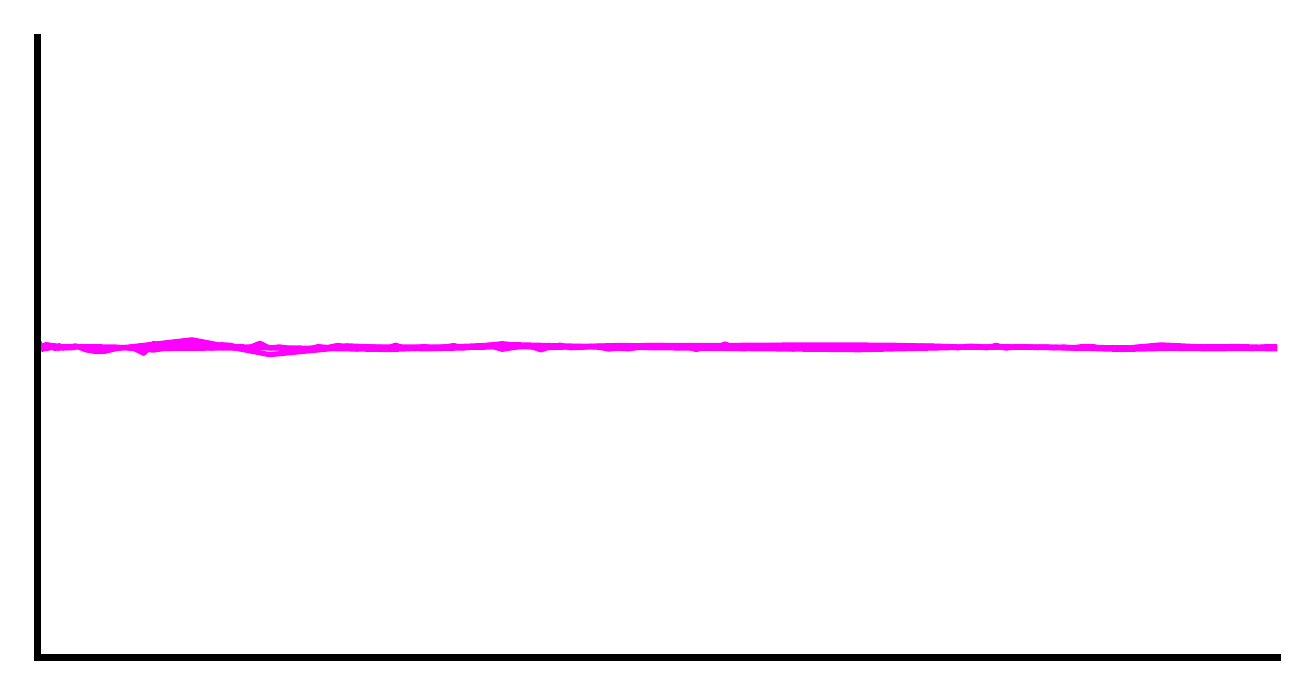}
  };
\end{tikzpicture}
\\
%%%%%%%%%%%%%%%%%%%%%%%%%%%%%%%%%
%%%%%%%%%%%%%%%%%%%%%%%%%%%%%%%%%
%%%%%%%%%%%%%%%%%%%%%%%%%%%%%%%%%
\rotatebox{90}{ Latinhypercube}
&
\begin{tikzpicture}
  \node[anchor=south west,inner sep=0] (image) at (0,0)
  {
    \pdfliteral{ 1 w}\includegraphics[width=0.75in,page=1]{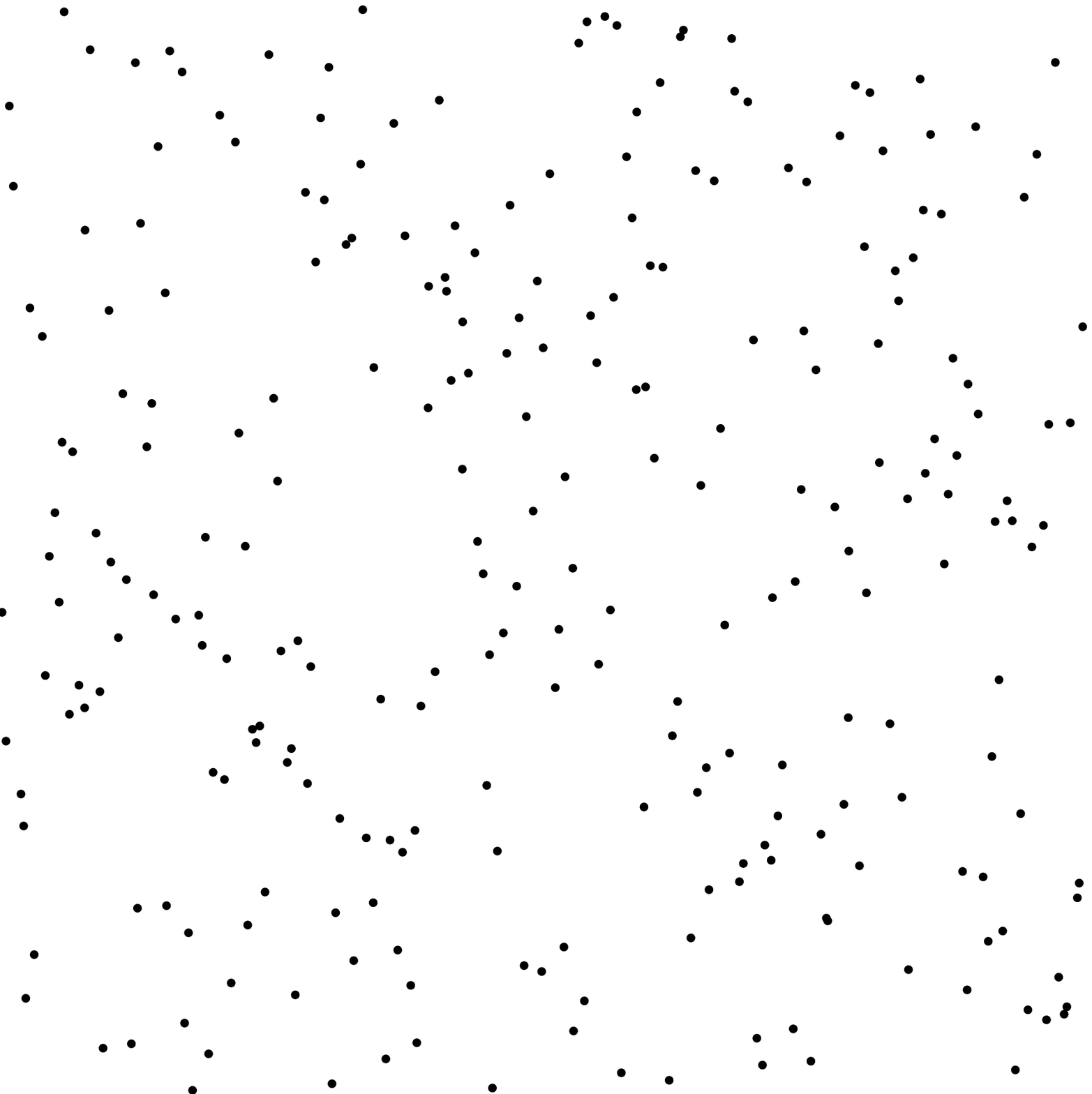}
  };
      \draw[black,thick] (0,0) -- (1.9,0) -- (1.9,1.9) -- (0,1.9) -- cycle;
\end{tikzpicture} 
&
\begin{tikzpicture}
  \node[anchor=south west,inner sep=0] (image) at (0,0)
  {
    \pdfliteral{ 1 w}\includegraphics[width=0.75in,page=1]{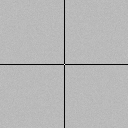}
  };
\end{tikzpicture} 
&
\begin{tikzpicture}
\node[anchor=south west,inner sep=0] (A) at (0,0)
  {
    \pdfliteral{ 1 w}\includegraphics[width=0.75in,height=0.35in,page=1] {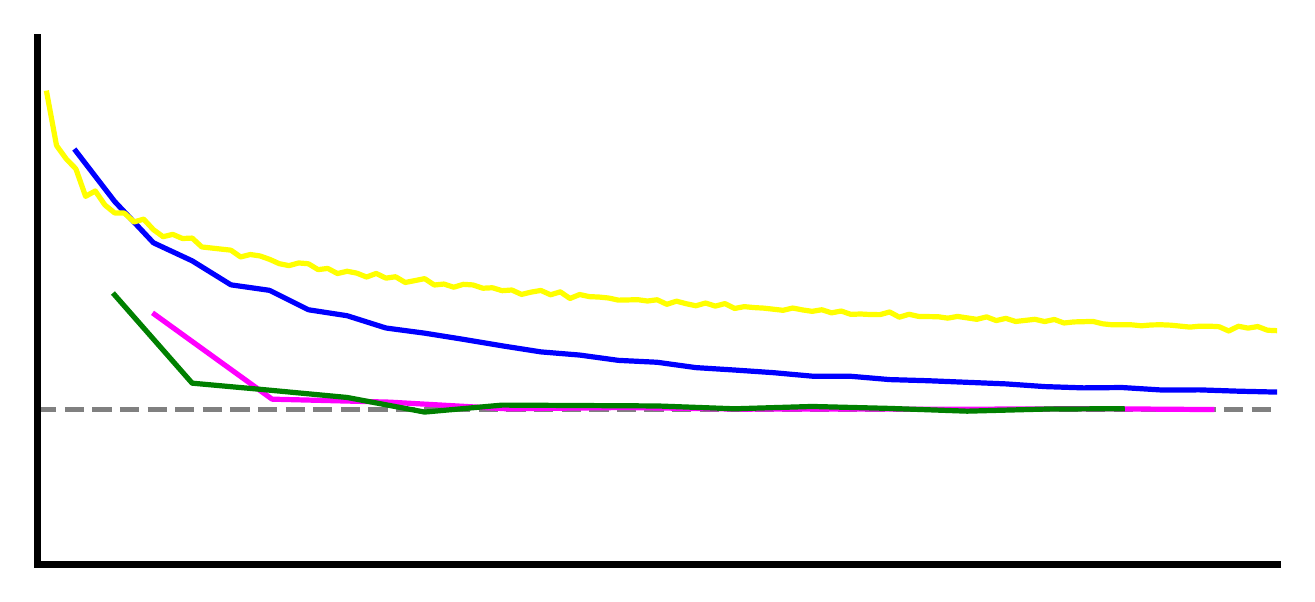}
  };
  \node[anchor=south west,inner sep=0] (A) at (0,0.95)
  {
    \pdfliteral{ 1 w}\includegraphics[width=0.75in,height=0.35in,page=1] {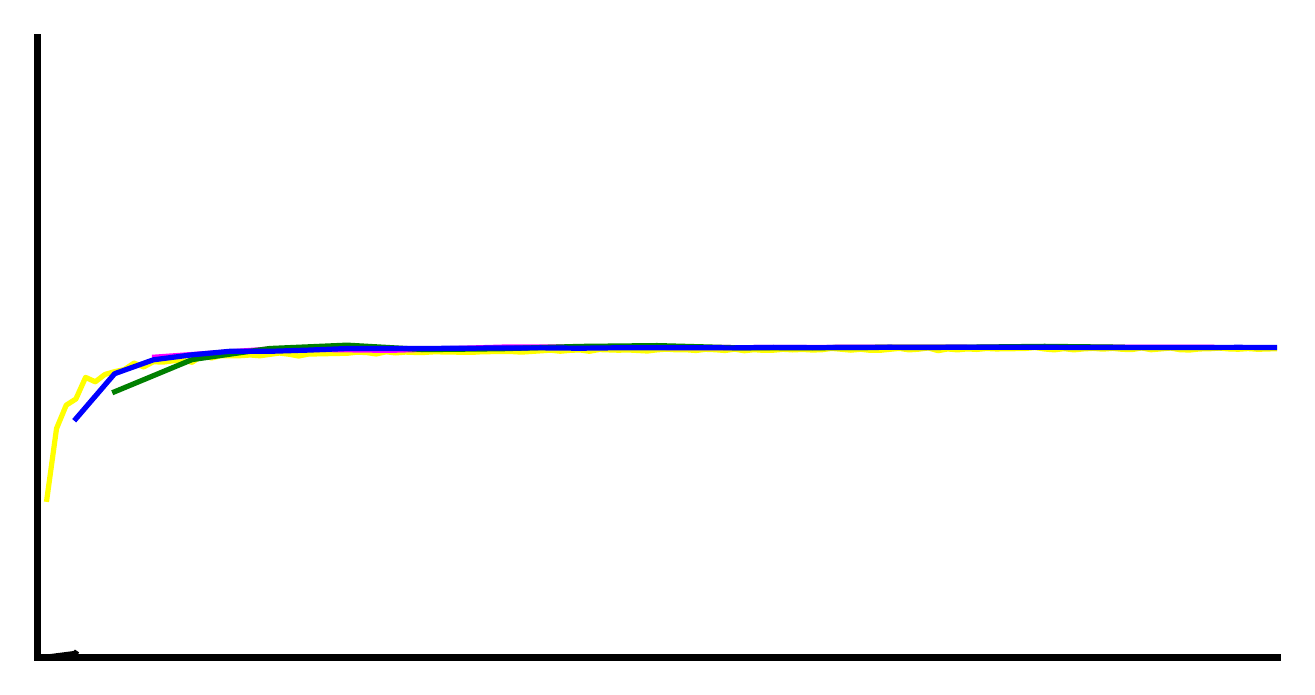}
  };
  \begin{scope}
   \filldraw[black] (0.05,0.77) circle (0pt) node[anchor=west] {\tiny anisotropy}; 
     \filldraw[black] (0.05,1.7) circle (0pt) node[anchor=west] {\tiny power};  
  \end{scope}
\end{tikzpicture}
&
\begin{tikzpicture}
\node[anchor=south west,inner sep=0] (A) at (0,0)
  {
    \pdfliteral{ 1 w}\includegraphics[width=0.75in,height=0.35in,page=1] {images/samplers/radial1d-blank.pdf}
  };
  \node[anchor=south west,inner sep=0] (A) at (0,0.95)
  {
    \pdfliteral{ 1 w}\includegraphics[width=0.75in,height=0.35in,page=1] {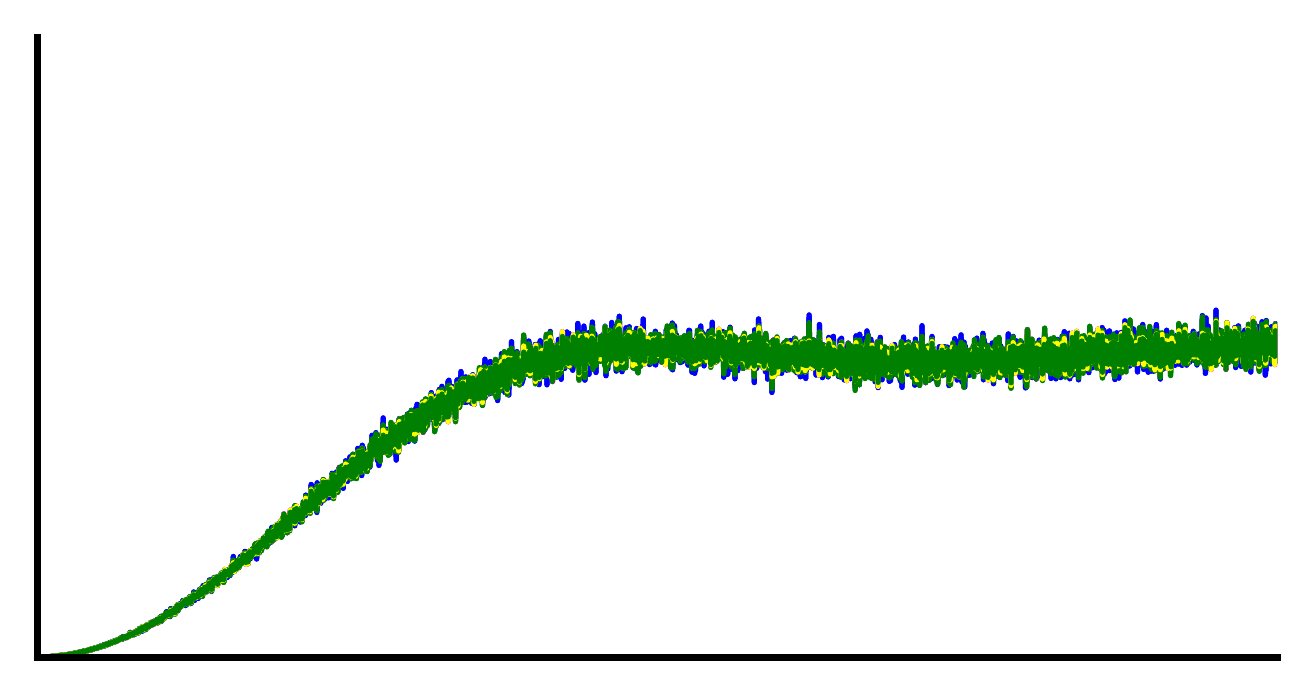}
  };
\end{tikzpicture}
&
\begin{tikzpicture}
  \node[anchor=south west,inner sep=0] (image) at (0,0)
  {
    \pdfliteral{ 1 w}\includegraphics[width=0.75in,page=1]{images/samplers/power-Pconstant-v1-nrooks-proj0-1-n4096-001000.png}
  };
    \draw[mymagenta,thick] (0,0) -- (1.9,0) -- (1.9,1.9) -- (0,1.9) -- cycle;
\end{tikzpicture} 
&
\begin{tikzpicture}
  \node[anchor=south west,inner sep=0] (image) at (0,0)
  {
    \pdfliteral{ 1 w}\includegraphics[width=0.75in,page=1]{images/samplers/power-Pconstant-v1-nrooks-proj0-1-n4096-001000.png}
  };
    \draw[green,thick] (0,0) -- (1.9,0) -- (1.9,1.9) -- (0,1.9) -- cycle;
\end{tikzpicture} 
&
\begin{tikzpicture}
  \node[anchor=south west,inner sep=0] (image) at (0,0)
  {
    \pdfliteral{ 1 w}\includegraphics[width=0.75in,page=1]{images/samplers/power-Pconstant-v1-nrooks-proj0-1-n4096-001000.png}
  };
   \draw[skyblue,thick] (0,0) -- (1.9,0) -- (1.9,1.9) -- (0,1.9) -- cycle;
\end{tikzpicture} 
&
\begin{tikzpicture}
\node[anchor=south west,inner sep=0] (A) at (0,0)
  {
    \pdfliteral{ 1 w}\includegraphics[width=0.75in,height=0.35in,page=1] {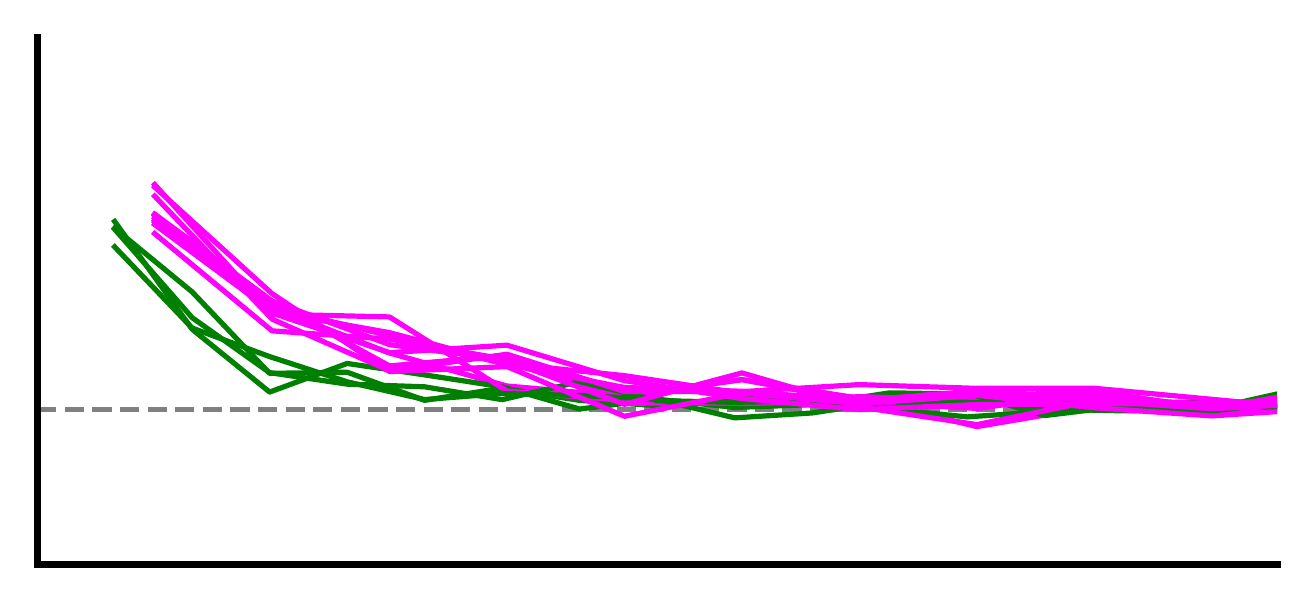}
  };
  \node[anchor=south west,inner sep=0] (A) at (0,0.95)
  {
    \pdfliteral{ 1 w}\includegraphics[width=0.75in,height=0.35in,page=1] {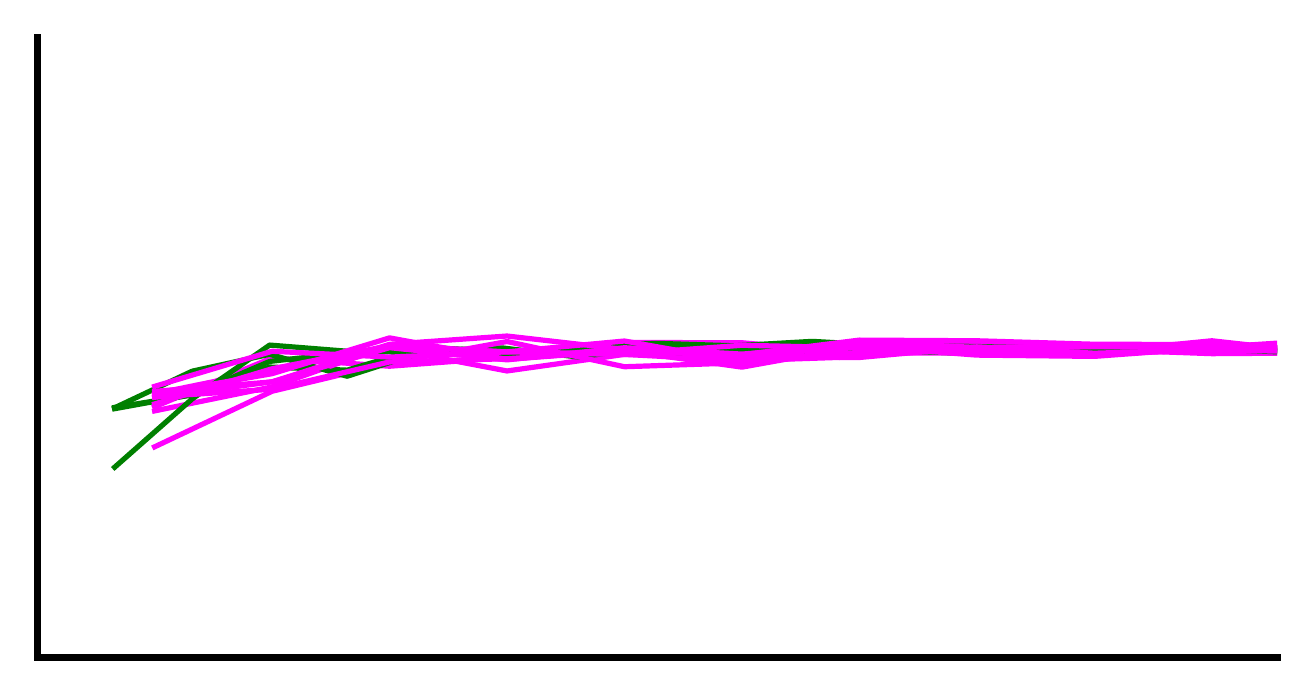}
  };
\end{tikzpicture}
&
\begin{tikzpicture}
\node[anchor=south west,inner sep=0] (A) at (0,0)
  {
    \pdfliteral{ 1 w}\includegraphics[width=0.75in,height=0.35in,page=1] {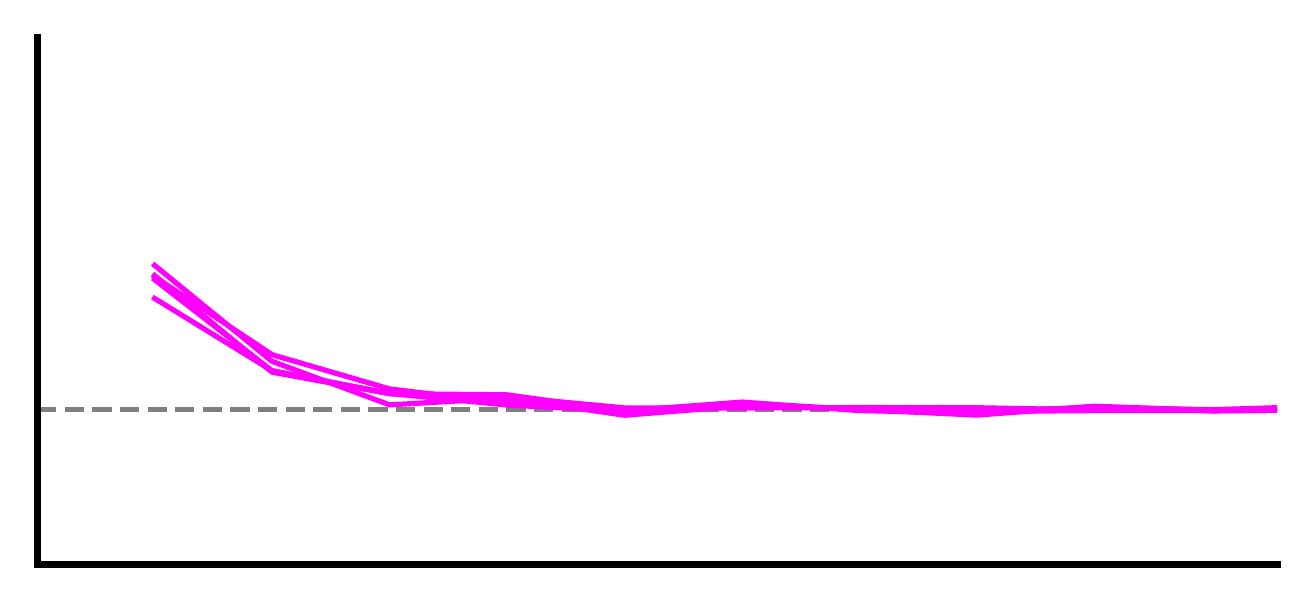}
  };
  \node[anchor=south west,inner sep=0] (A) at (0,0.95)
  {
    \pdfliteral{ 1 w}\includegraphics[width=0.75in,height=0.35in,page=1] {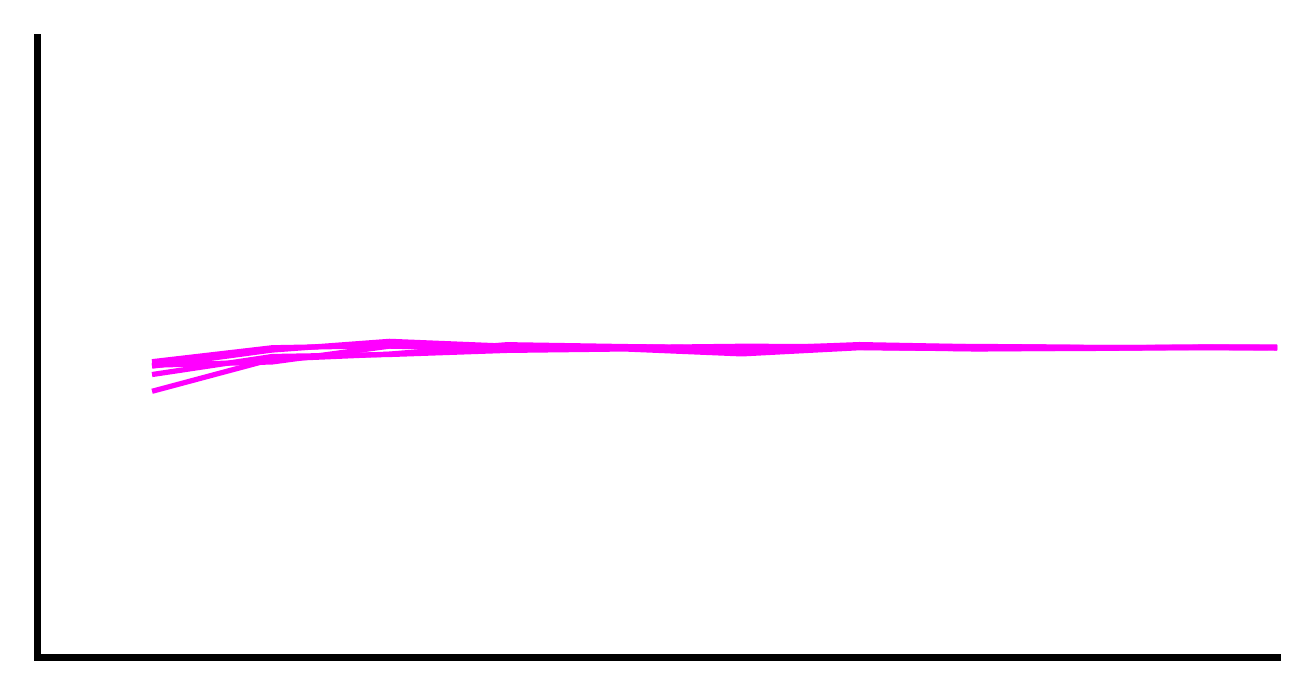}
  };
\end{tikzpicture}
\\
%%%%%%%%%%%%%%%%%%%%%%%%%%%%%%%%%
%%%%%%%%%%%%%%%%%%%%%%%%%%%%%%%%%
%%%%%%%%%%%%%%%%%%%%%%%%%%%%%%%%%
\rotatebox{90}{\qquad Jittered}
&
\begin{tikzpicture}
  \node[anchor=south west,inner sep=0] (image) at (0,0)
  {
    \pdfliteral{ 1 w}\includegraphics[width=0.75in,page=1]{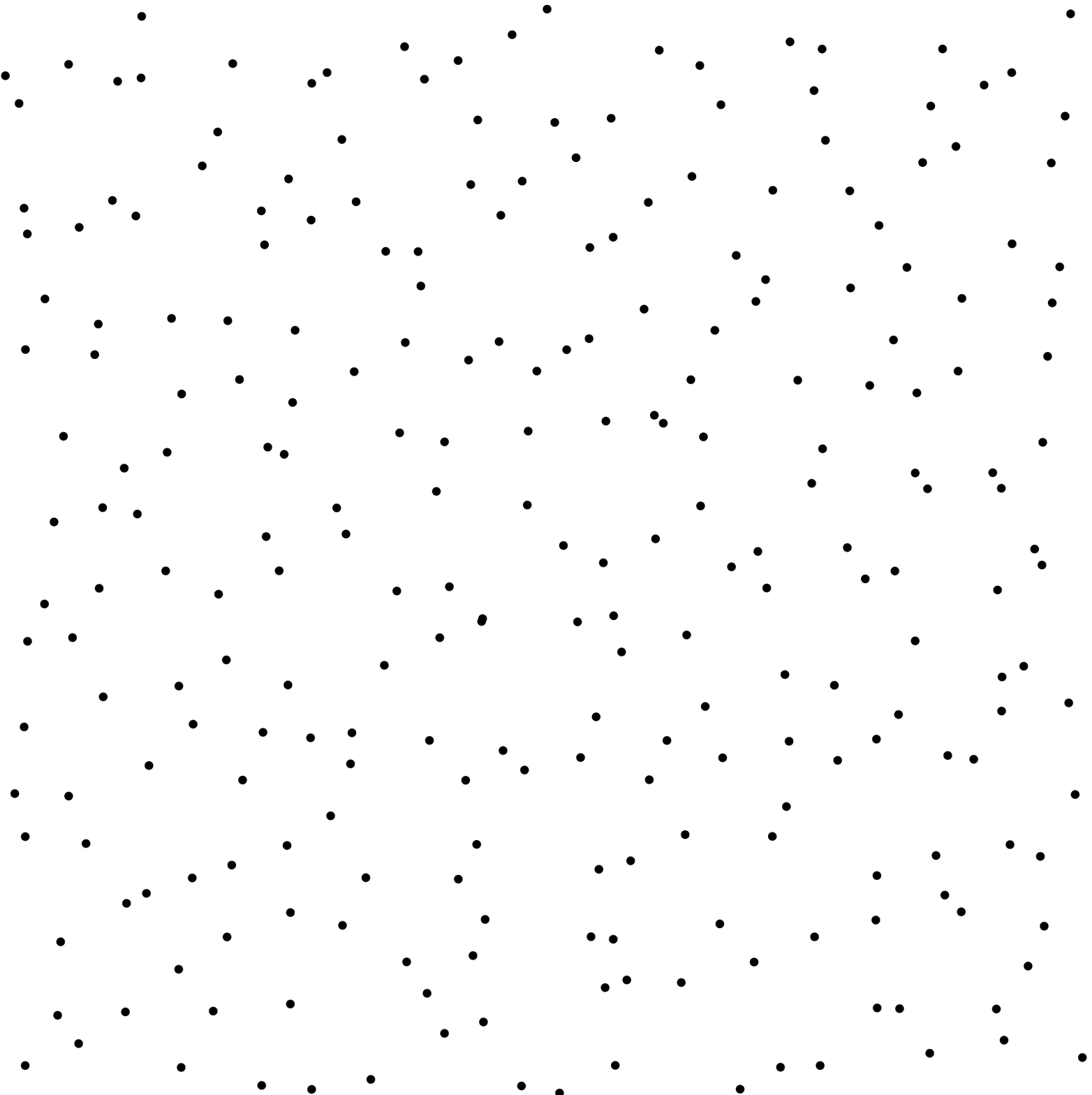}
  };
      \draw[black,thick] (0,0) -- (1.9,0) -- (1.9,1.9) -- (0,1.9) -- cycle;
\end{tikzpicture} 
&
\begin{tikzpicture}
  \node[anchor=south west,inner sep=0] (image) at (0,0)
  {
    \pdfliteral{ 1 w}\includegraphics[width=0.75in,page=1]{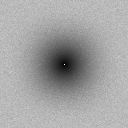}
  };
\end{tikzpicture} 
&
\begin{tikzpicture}
\node[anchor=south west,inner sep=0] (A) at (0,0)
  {
    \pdfliteral{ 1 w}\includegraphics[width=0.75in,height=0.35in,page=1] {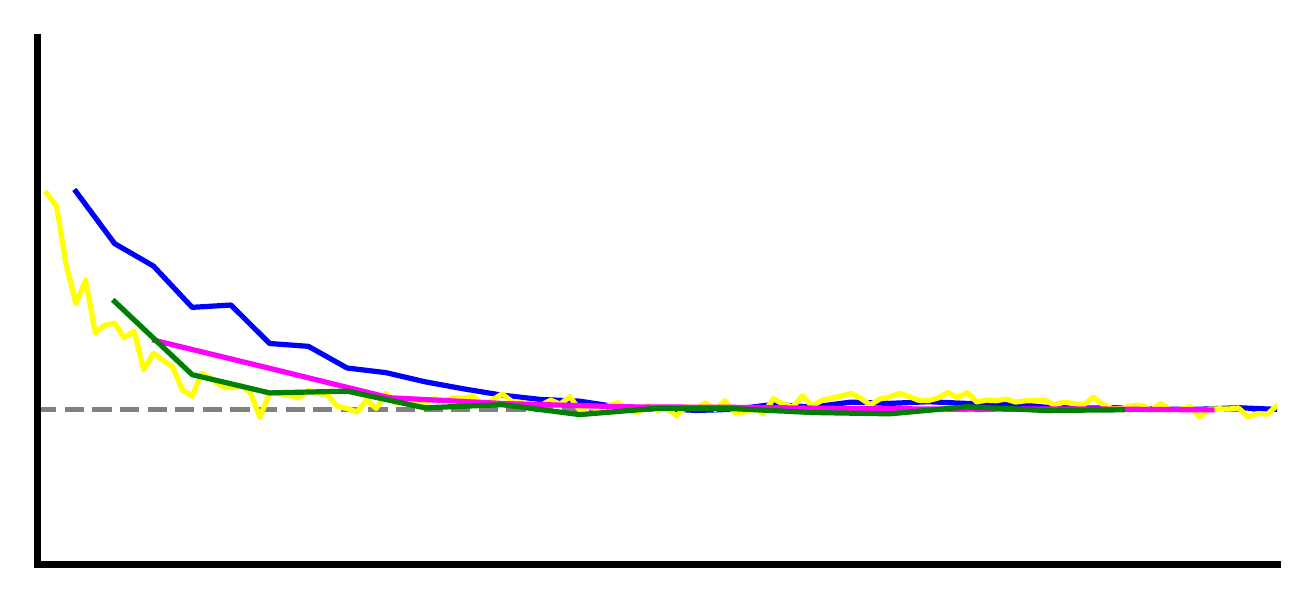}
  };
  \node[anchor=south west,inner sep=0] (A) at (0,0.95)
  {
    \pdfliteral{ 1 w}\includegraphics[width=0.75in,height=0.35in,page=1] {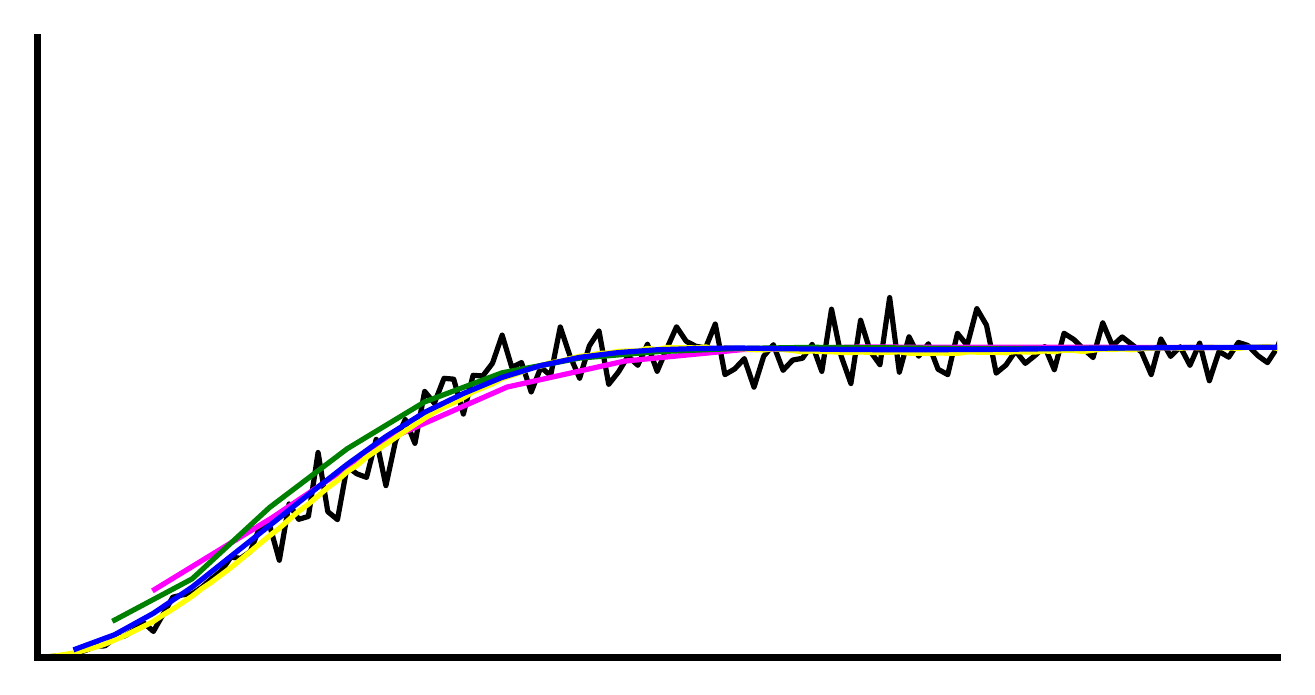}
  };
  \begin{scope}
   \filldraw[black] (0.05,0.77) circle (0pt) node[anchor=west] {\tiny anisotropy}; 
     \filldraw[black] (0.05,1.7) circle (0pt) node[anchor=west] {\tiny power};  
  \end{scope}
\end{tikzpicture}
&
\begin{tikzpicture}
\node[anchor=south west,inner sep=0] (A) at (0,0)
  {
    \pdfliteral{ 1 w}\includegraphics[width=0.75in,height=0.35in,page=1] {images/samplers/radial1d-blank.pdf}
  };
  \node[anchor=south west,inner sep=0] (A) at (0,0.95)
  {
    \pdfliteral{ 1 w}\includegraphics[width=0.75in,height=0.35in,page=1] {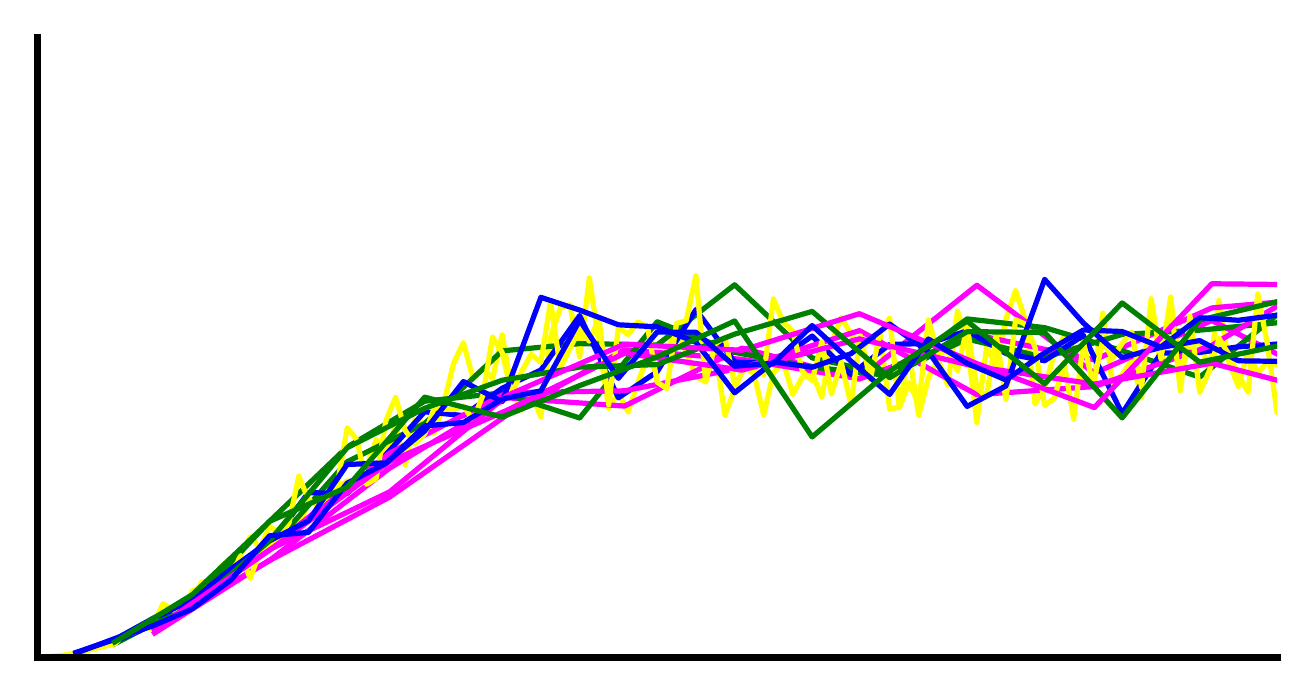}
  };
\end{tikzpicture}
&
\begin{tikzpicture}
  \node[anchor=south west,inner sep=0] (image) at (0,0)
  {
    \pdfliteral{ 1 w}\includegraphics[width=0.75in,page=1]{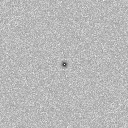}
  };
    \draw[mymagenta,thick] (0,0) -- (1.9,0) -- (1.9,1.9) -- (0,1.9) -- cycle;
\end{tikzpicture} 
&
\begin{tikzpicture}
  \node[anchor=south west,inner sep=0] (image) at (0,0)
  {
    \pdfliteral{ 1 w}\includegraphics[width=0.75in,page=1]{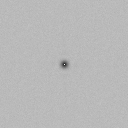}
  };
    \draw[green,thick] (0,0) -- (1.9,0) -- (1.9,1.9) -- (0,1.9) -- cycle;
\end{tikzpicture} 
&
\begin{tikzpicture}
  \node[anchor=south west,inner sep=0] (image) at (0,0)
  {
    \pdfliteral{ 1 w}\includegraphics[width=0.75in,page=1]{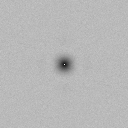}
  };
   \draw[skyblue,thick] (0,0) -- (1.9,0) -- (1.9,1.9) -- (0,1.9) -- cycle;
\end{tikzpicture} 
&
\begin{tikzpicture}
\node[anchor=south west,inner sep=0] (A) at (0,0)
  {
    \pdfliteral{ 1 w}\includegraphics[width=0.75in,height=0.35in,page=1] {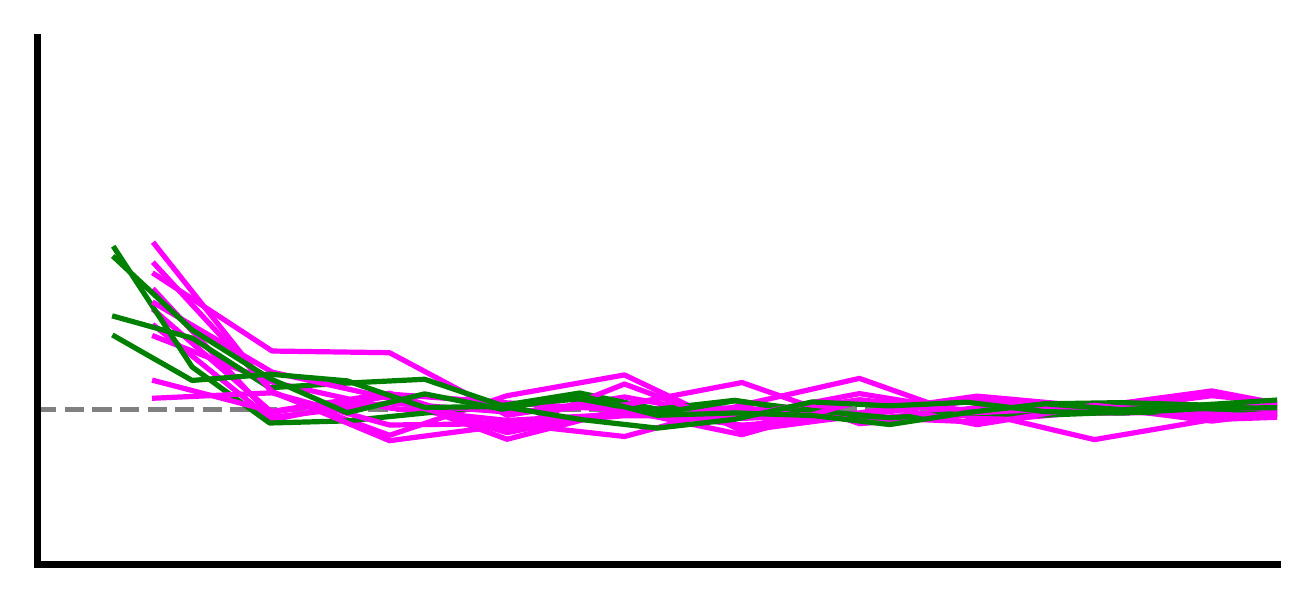}
  };
  \node[anchor=south west,inner sep=0] (A) at (0,0.95)
  {
    \pdfliteral{ 1 w}\includegraphics[width=0.75in,height=0.35in,page=1] {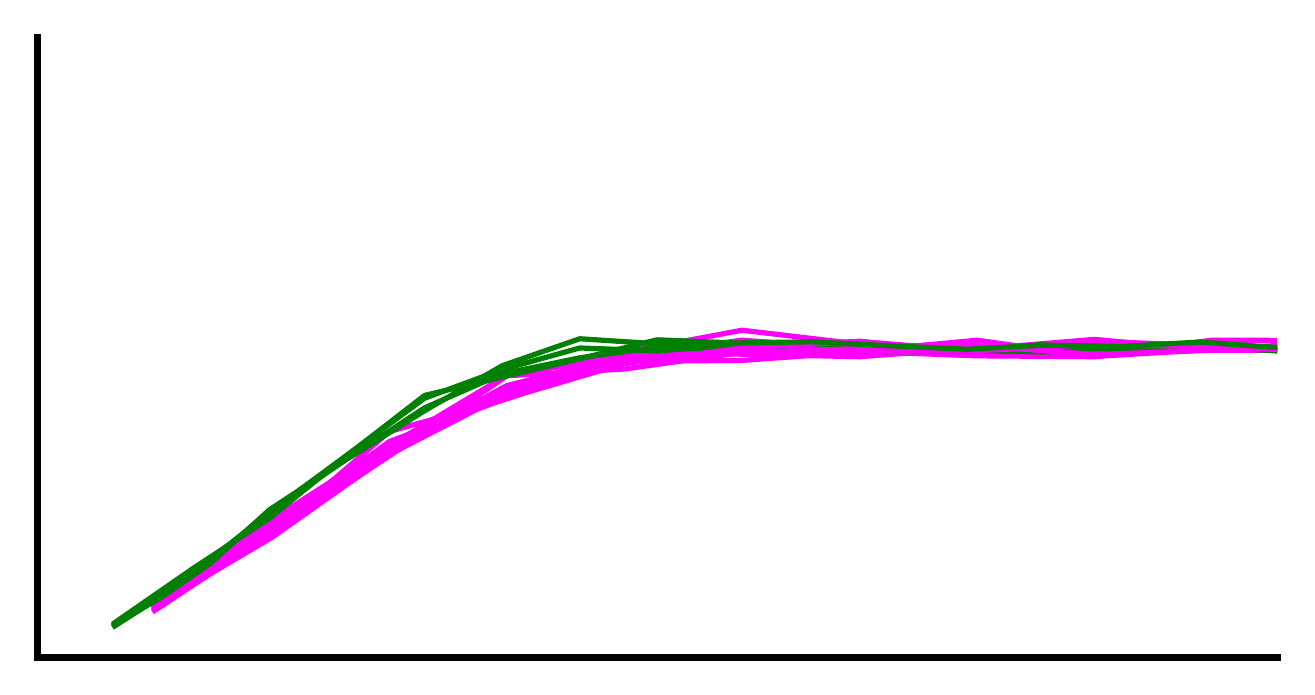}
  };
\end{tikzpicture}
&
\begin{tikzpicture}
\node[anchor=south west,inner sep=0] (A) at (0,0)
  {
    \pdfliteral{ 1 w}\includegraphics[width=0.75in,height=0.35in,page=1] {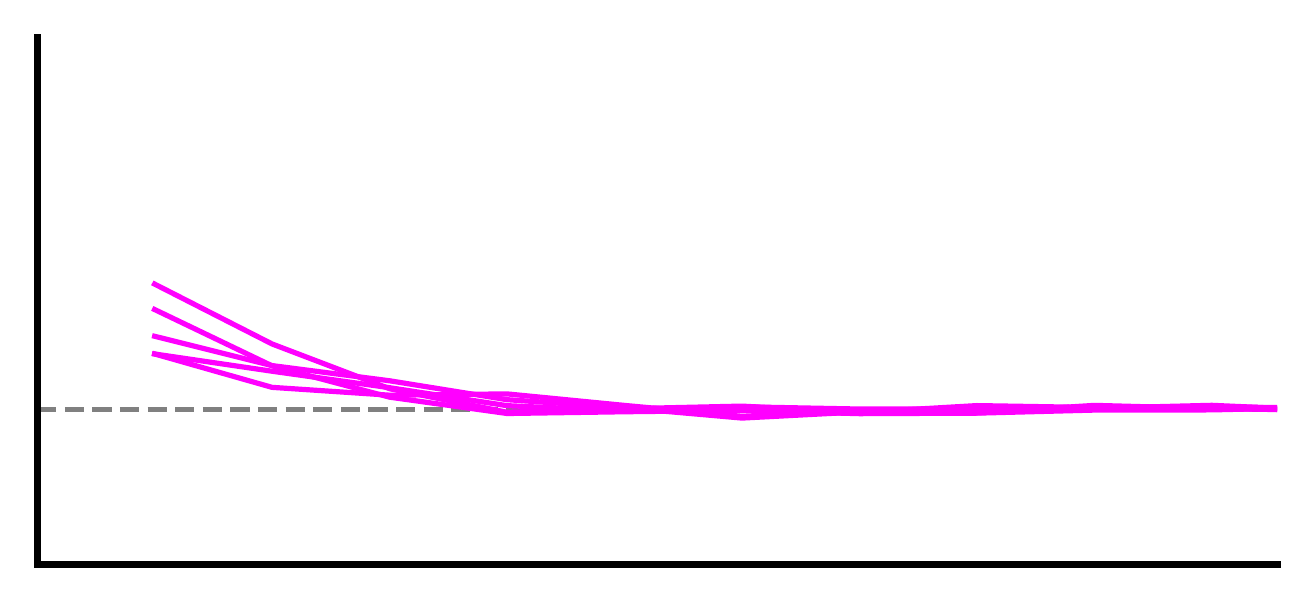}
  };
  \node[anchor=south west,inner sep=0] (A) at (0,0.95)
  {
    \pdfliteral{ 1 w}\includegraphics[width=0.75in,height=0.35in,page=1] {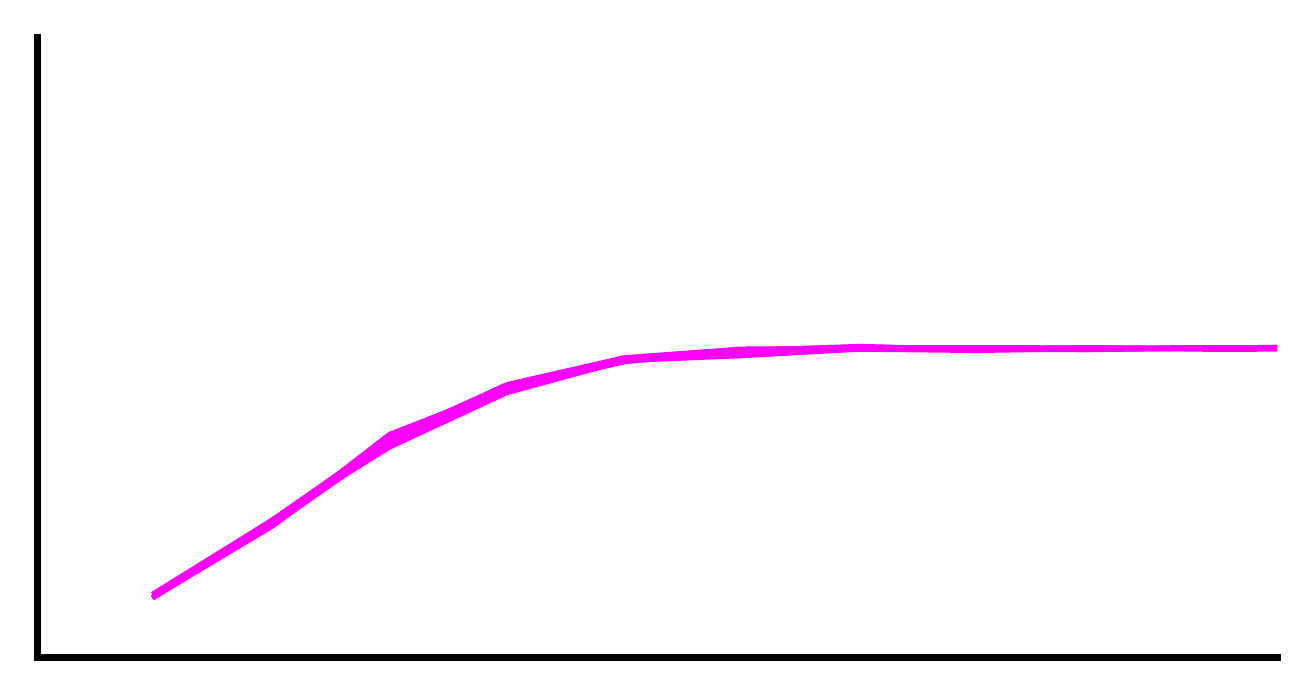}
  };
\end{tikzpicture}
\\
%%%%%%%%%%%%%%%%%%%%%%%%%%%%%%%%%
%%%%%%%%%%%%%%%%%%%%%%%%%%%%%%%%%
%%%%%%%%%%%%%%%%%%%%%%%%%%%%%%%%%
\rotatebox{90}{\quad Polyhex~\shortcite{wachtel2014fast}}
&
\begin{tikzpicture}
  \node[anchor=south west,inner sep=0] (image) at (0,0)
  {
    \pdfliteral{ 1 w}\includegraphics[width=0.75in,page=1]{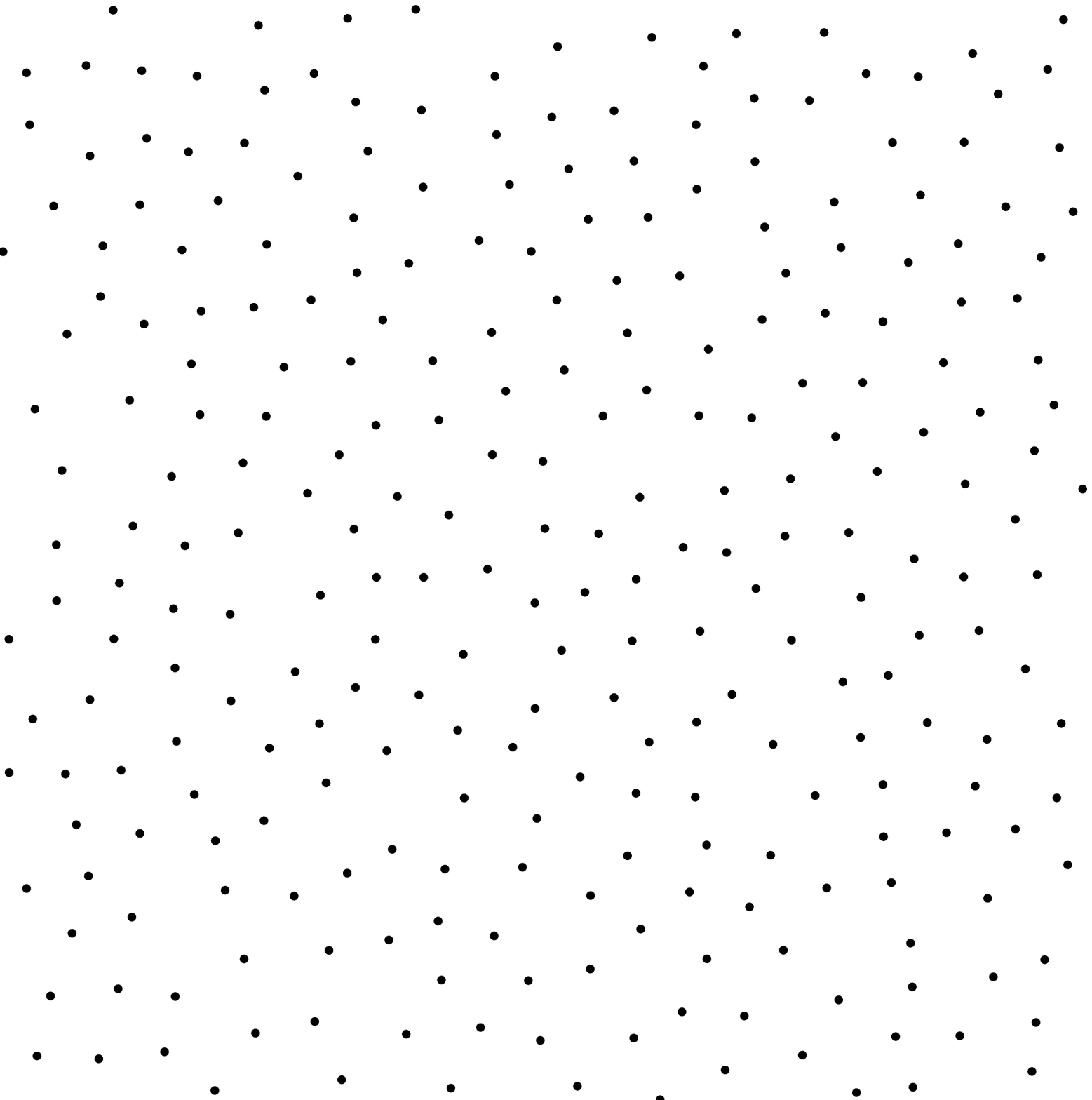}
  };
      \draw[black,thick] (0,0) -- (1.9,0) -- (1.9,1.9) -- (0,1.9) -- cycle;
\end{tikzpicture} 
&
\begin{tikzpicture}
  \node[anchor=south west,inner sep=0] (image) at (0,0)
  {
    \pdfliteral{ 1 w}\includegraphics[width=0.75in,page=1]{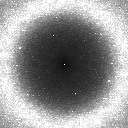}
  };
\end{tikzpicture} 
&
\begin{tikzpicture}
\node[anchor=south west,inner sep=0] (A) at (0,0)
  {
    \pdfliteral{ 1 w}\includegraphics[width=0.75in,height=0.35in,page=1] {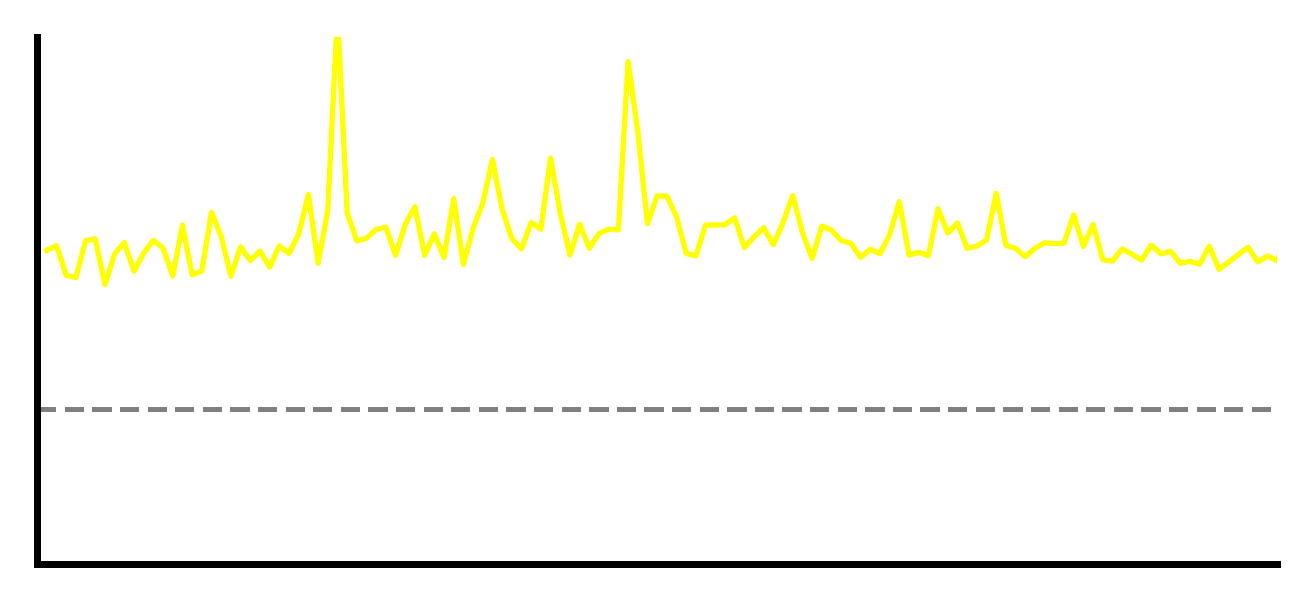}
  };
  \node[anchor=south west,inner sep=0] (A) at (0,0.95)
  {
    \pdfliteral{ 1 w}\includegraphics[width=0.75in,height=0.35in,page=1] {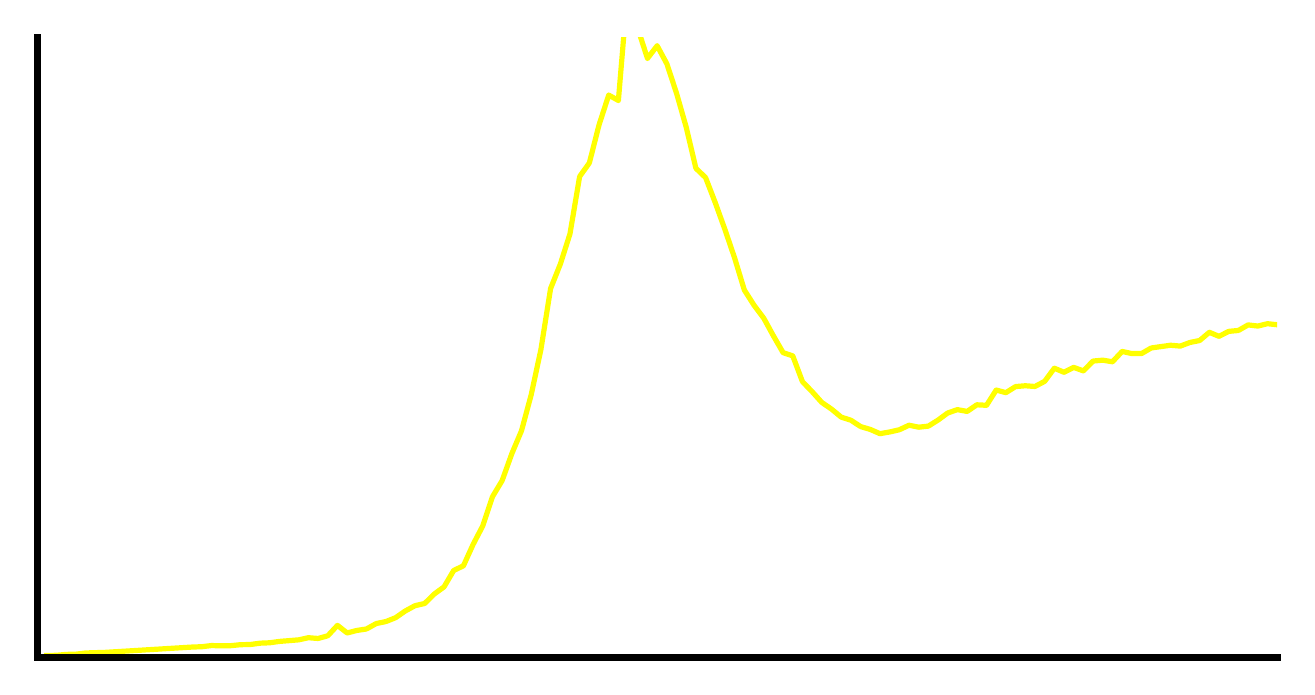}
  };
  \begin{scope}
   \filldraw[black] (0.05,0.77) circle (0pt) node[anchor=west] {\tiny anisotropy}; 
     \filldraw[black] (0.05,1.7) circle (0pt) node[anchor=west] {\tiny power};  
  \end{scope}
\end{tikzpicture}
&
\begin{tikzpicture}
\node[anchor=south west,inner sep=0] (A) at (0,0)
  {
    \pdfliteral{ 1 w}\includegraphics[width=0.75in,height=0.35in,page=1] {images/samplers/radial1d-blank.pdf}
  };
  \node[anchor=south west,inner sep=0] (A) at (0,0.95)
  {
    \pdfliteral{ 1 w}\includegraphics[width=0.75in,height=0.35in,page=1] {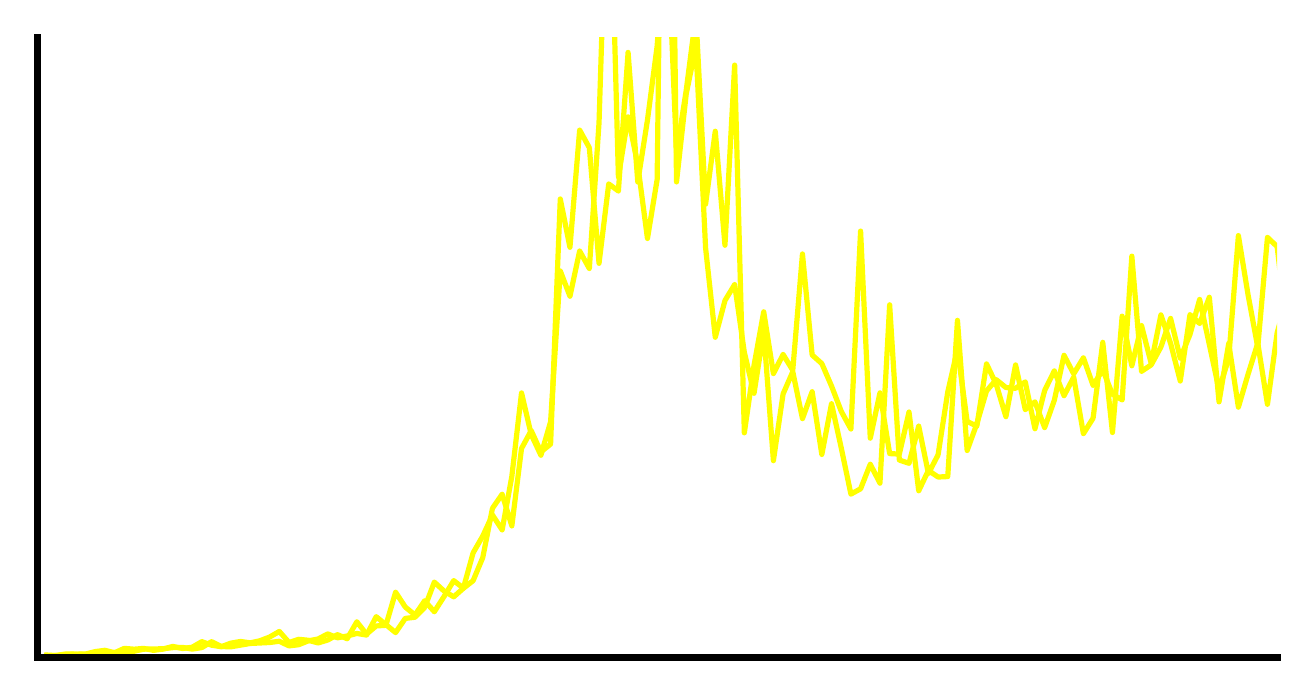}
  };
\end{tikzpicture}
&
\begin{tikzpicture}
  \node[anchor=south west,inner sep=0] (image) at (0,0)
  {
    \pdfliteral{ 1 w}\includegraphics[width=0.75in,page=1]{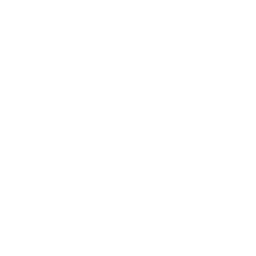}
  };
    \draw[mymagenta,thick] (0,0) -- (1.9,0) -- (1.9,1.9) -- (0,1.9) -- cycle;
\end{tikzpicture} 
&
\begin{tikzpicture}
  \node[anchor=south west,inner sep=0] (image) at (0,0)
  {
    \pdfliteral{ 1 w}\includegraphics[width=0.75in,page=1]{images/samplers/white.png}
  };
    \draw[green,thick] (0,0) -- (1.9,0) -- (1.9,1.9) -- (0,1.9) -- cycle;
\end{tikzpicture} 
&
\begin{tikzpicture}
  \node[anchor=south west,inner sep=0] (image) at (0,0)
  {
    \pdfliteral{ 1 w}\includegraphics[width=0.75in,page=1]{images/samplers/white.png}
  };
   \draw[skyblue,thick] (0,0) -- (1.9,0) -- (1.9,1.9) -- (0,1.9) -- cycle;
\end{tikzpicture} 
&
\begin{tikzpicture}
  \node[anchor=south west,inner sep=0] (image) at (0,0)
  {
    \pdfliteral{ 1 w}\includegraphics[width=0.75in,page=1]{images/samplers/white.png}
  };
 %  \draw[skyblue,thick] (0,0) -- (1.9,0) -- (1.9,1.9) -- (0,1.9) -- cycle;
\end{tikzpicture} 
&
\begin{tikzpicture}
  \node[anchor=south west,inner sep=0] (image) at (0,0)
  {
    \pdfliteral{ 1 w}\includegraphics[width=0.75in,page=1]{images/samplers/white.png}
  };
%   \draw[skyblue,thick] (0,0) -- (1.9,0) -- (1.9,1.9) -- (0,1.9) -- cycle;
\end{tikzpicture} 
% \begin{tikzpicture}
% \node[anchor=south west,inner sep=0] (A) at (0,0)
%   {
%     \pdfliteral{ 1 w}\includegraphics[width=0.75in,height=0.35in,page=1] {images/samplers/radial-mean-jitter2d-n4096.pdf}
%   };
%   \node[anchor=south west,inner sep=0] (A) at (0,0.95)
%   {
%     \pdfliteral{ 1 w}\includegraphics[width=0.75in,height=0.35in,page=1] {images/samplers/radial-mean-jitter2d-n4096.pdf}
%   };
% \end{tikzpicture}
% &
% \begin{tikzpicture}
% \node[anchor=south west,inner sep=0] (A) at (0,0)
%   {
%     \pdfliteral{ 1 w}\includegraphics[width=0.75in,height=0.35in,page=1] {images/samplers/radial-mean-jitter2d-n4096.pdf}
%   };
%   \node[anchor=south west,inner sep=0] (A) at (0,0.95)
%   {
%     \pdfliteral{ 1 w}\includegraphics[width=0.75in,height=0.35in,page=1] {images/samplers/radial-mean-jitter2d-n4096.pdf}
%   };
% \end{tikzpicture}
\\
%%%%%%%%%%%%%%%%%%%%%%%%%%%%%%%%%
%%%%%%%%%%%%%%%%%%%%%%%%%%%%%%%%%
%%%%%%%%%%%%%%%%%%%%%%%%%%%%%%%%%
\rotatebox{90}{\qquad Halton}
&
\begin{tikzpicture}
  \node[anchor=south west,inner sep=0] (image) at (0,0)
  {
    \pdfliteral{ 1 w}\includegraphics[width=0.75in,page=1]{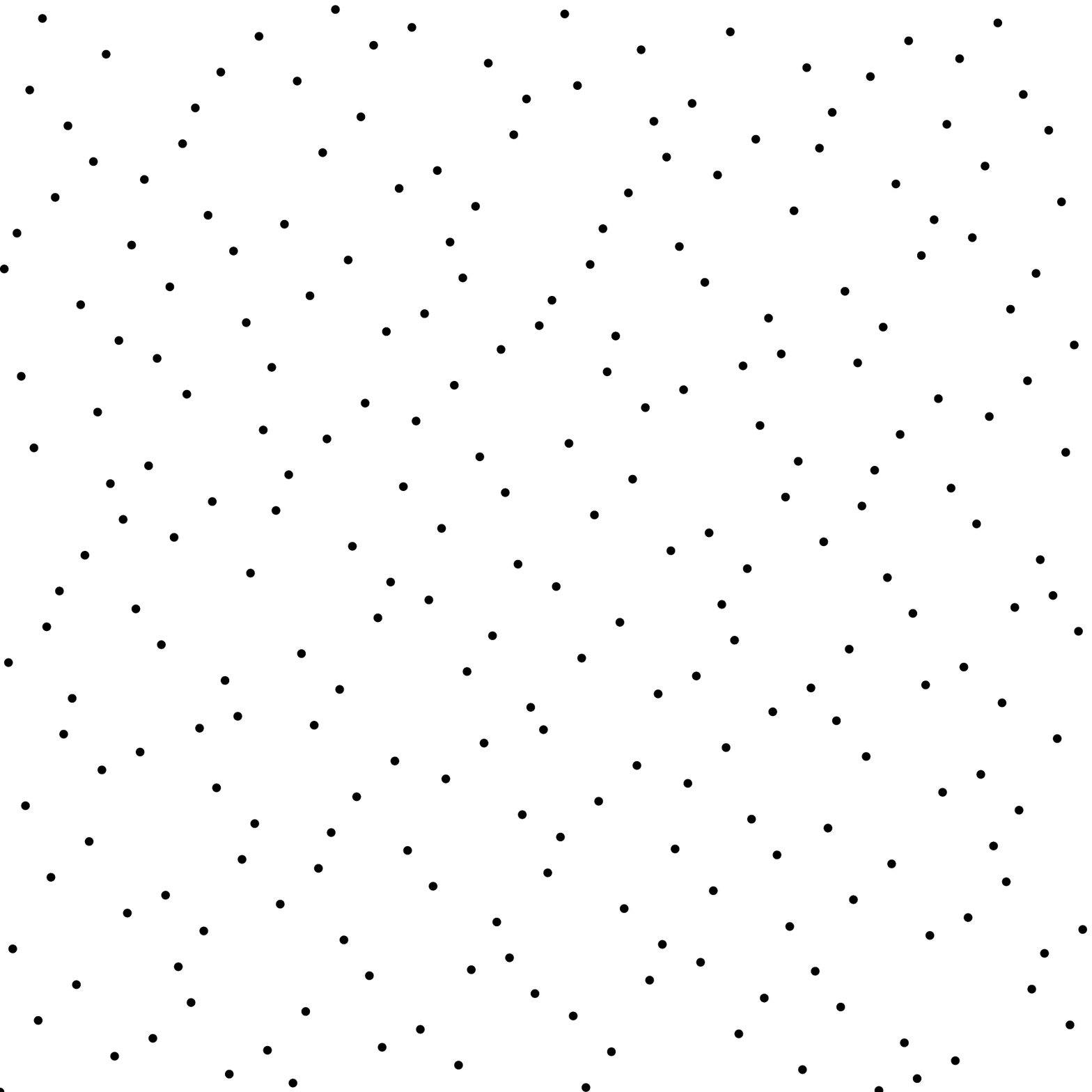}
  };
      \draw[black,thick] (0,0) -- (1.9,0) -- (1.9,1.9) -- (0,1.9) -- cycle;
\end{tikzpicture} 
&
\begin{tikzpicture}
  \node[anchor=south west,inner sep=0] (image) at (0,0)
  {
    \pdfliteral{ 1 w}\includegraphics[width=0.75in,page=1]{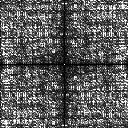}
  };
\end{tikzpicture} 
&
\begin{tikzpicture}
\node[anchor=south west,inner sep=0] (A) at (0,0)
  {
    \pdfliteral{ 1 w}\includegraphics[width=0.75in,height=0.35in,page=1] {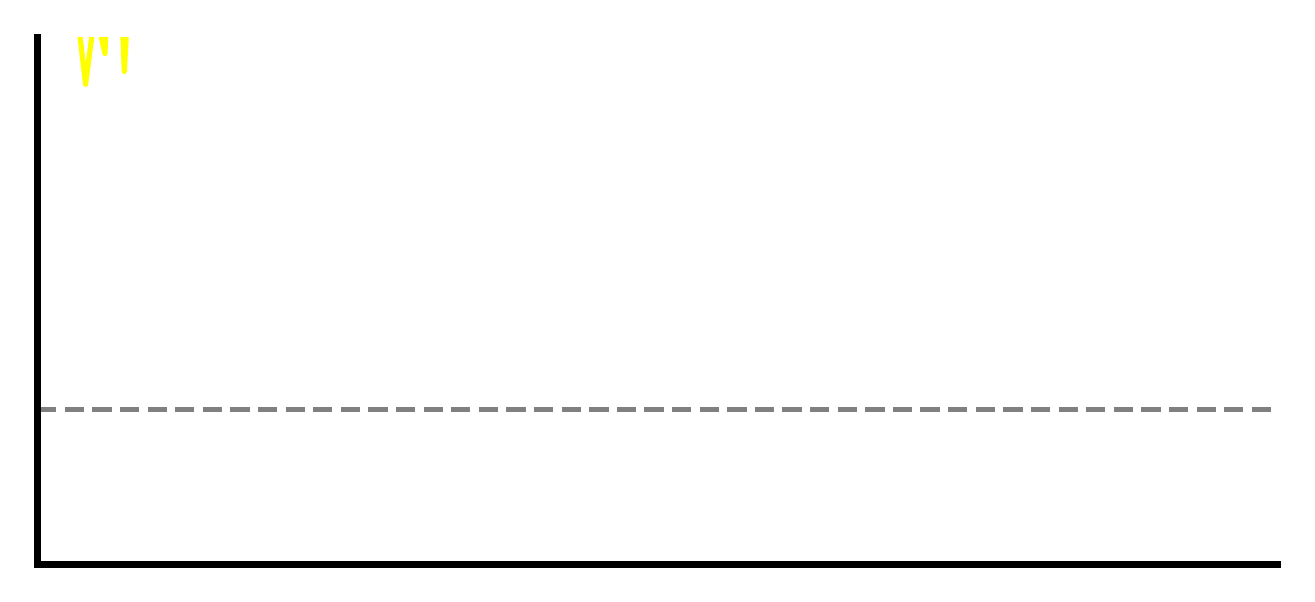}
  };
  \node[anchor=south west,inner sep=0] (A) at (0,0.95)
  {
    \pdfliteral{ 1 w}\includegraphics[width=0.75in,height=0.35in,page=1] {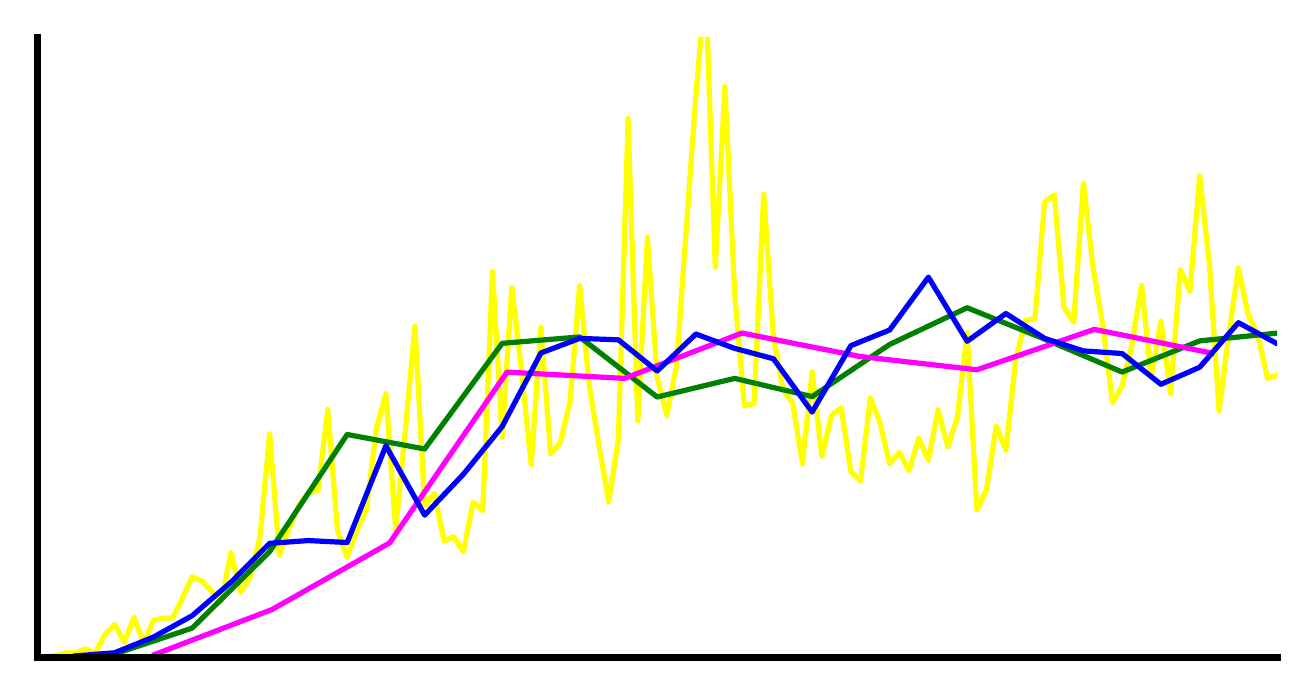}
  };
  \begin{scope}
   \filldraw[black] (0.05,0.77) circle (0pt) node[anchor=west] {\tiny anisotropy}; 
     \filldraw[black] (0.05,1.7) circle (0pt) node[anchor=west] {\tiny power};  
  \end{scope}
\end{tikzpicture}
&
\begin{tikzpicture}
\node[anchor=south west,inner sep=0] (A) at (0,0)
  {
    \pdfliteral{ 1 w}\includegraphics[width=0.75in,height=0.35in,page=1] {images/samplers/radial1d-blank.pdf}
  };
  \node[anchor=south west,inner sep=0] (A) at (0,0.95)
  {
    \pdfliteral{ 1 w}\includegraphics[width=0.75in,height=0.35in,page=1] {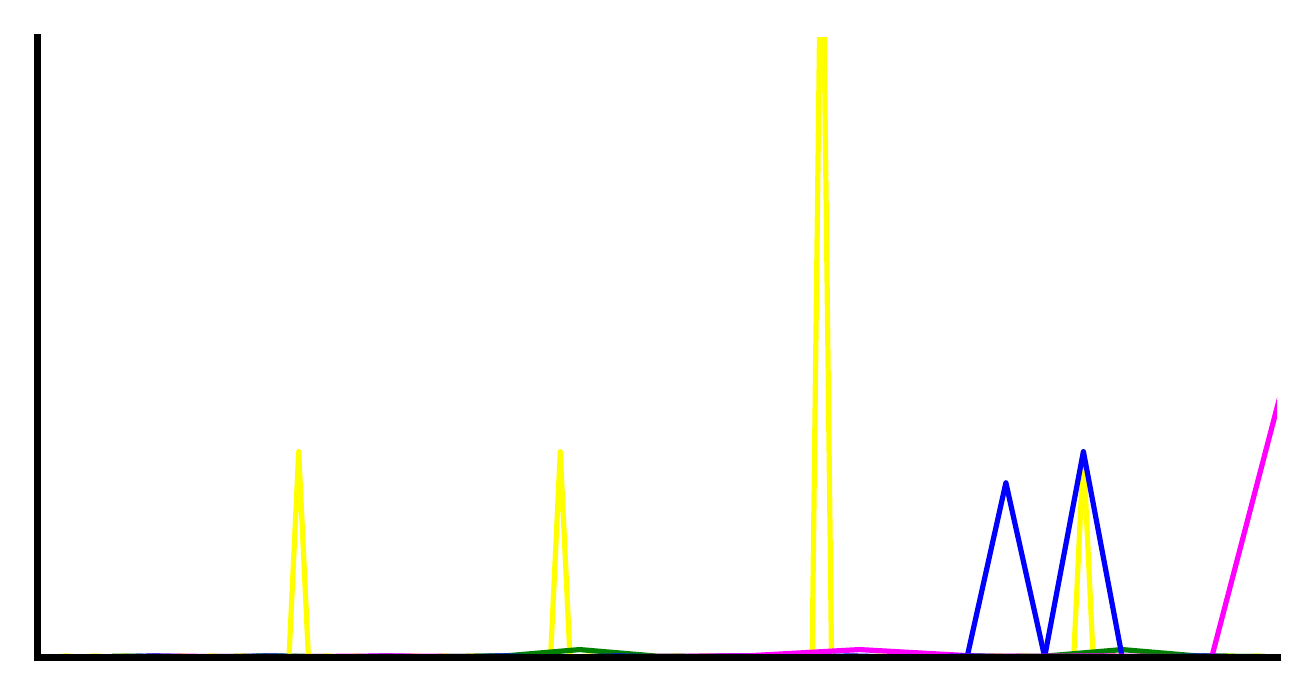}
  };
\end{tikzpicture}
&
\begin{tikzpicture}
  \node[anchor=south west,inner sep=0] (image) at (0,0)
  {
    \pdfliteral{ 1 w}\includegraphics[width=0.75in,page=1]{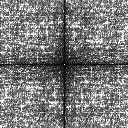}
  };
    \draw[mymagenta,thick] (0,0) -- (1.9,0) -- (1.9,1.9) -- (0,1.9) -- cycle;
\end{tikzpicture} 
&
\begin{tikzpicture}
  \node[anchor=south west,inner sep=0] (image) at (0,0)
  {
    \pdfliteral{ 1 w}\includegraphics[width=0.75in,page=1]{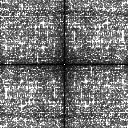}
  };
    \draw[green,thick] (0,0) -- (1.9,0) -- (1.9,1.9) -- (0,1.9) -- cycle;
\end{tikzpicture} 
&
\begin{tikzpicture}
  \node[anchor=south west,inner sep=0] (image) at (0,0)
  {
    \pdfliteral{ 1 w}\includegraphics[width=0.75in,page=1]{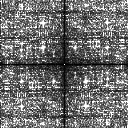}
  };
   \draw[skyblue,thick] (0,0) -- (1.9,0) -- (1.9,1.9) -- (0,1.9) -- cycle;
\end{tikzpicture} 
&
\begin{tikzpicture}
\node[anchor=south west,inner sep=0] (A) at (0,0)
  {
    \pdfliteral{ 1 w}\includegraphics[width=0.75in,height=0.35in,page=1] {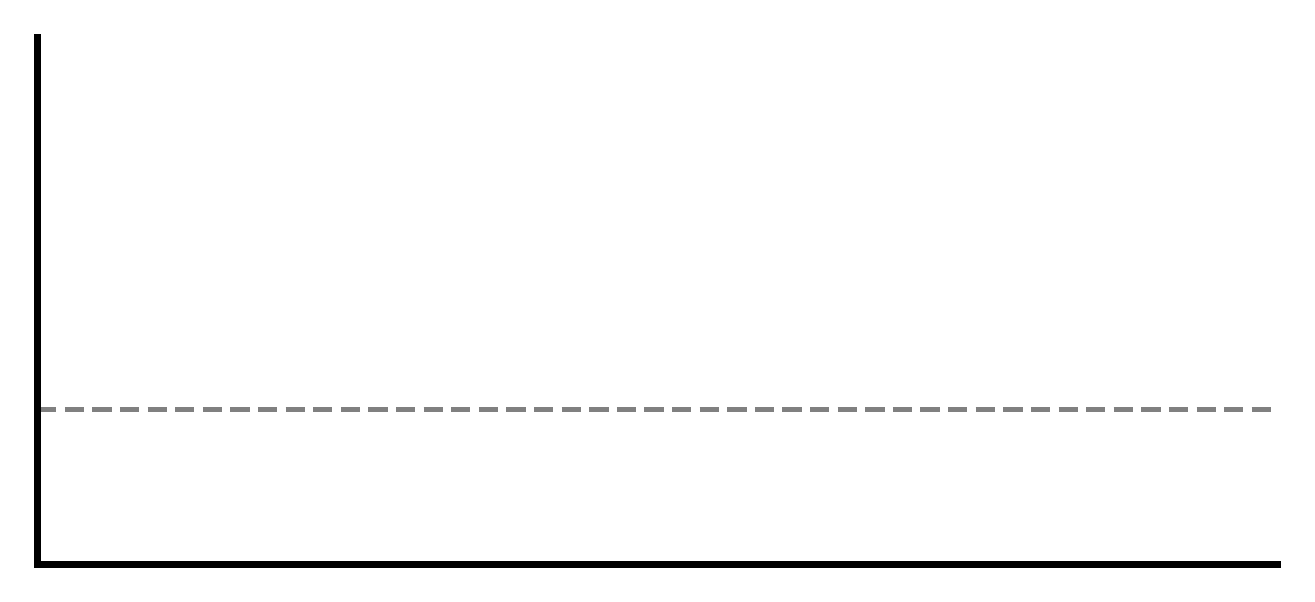}
  };
  \node[anchor=south west,inner sep=0] (A) at (0,0.95)
  {
    \pdfliteral{ 1 w}\includegraphics[width=0.75in,height=0.35in,page=1] {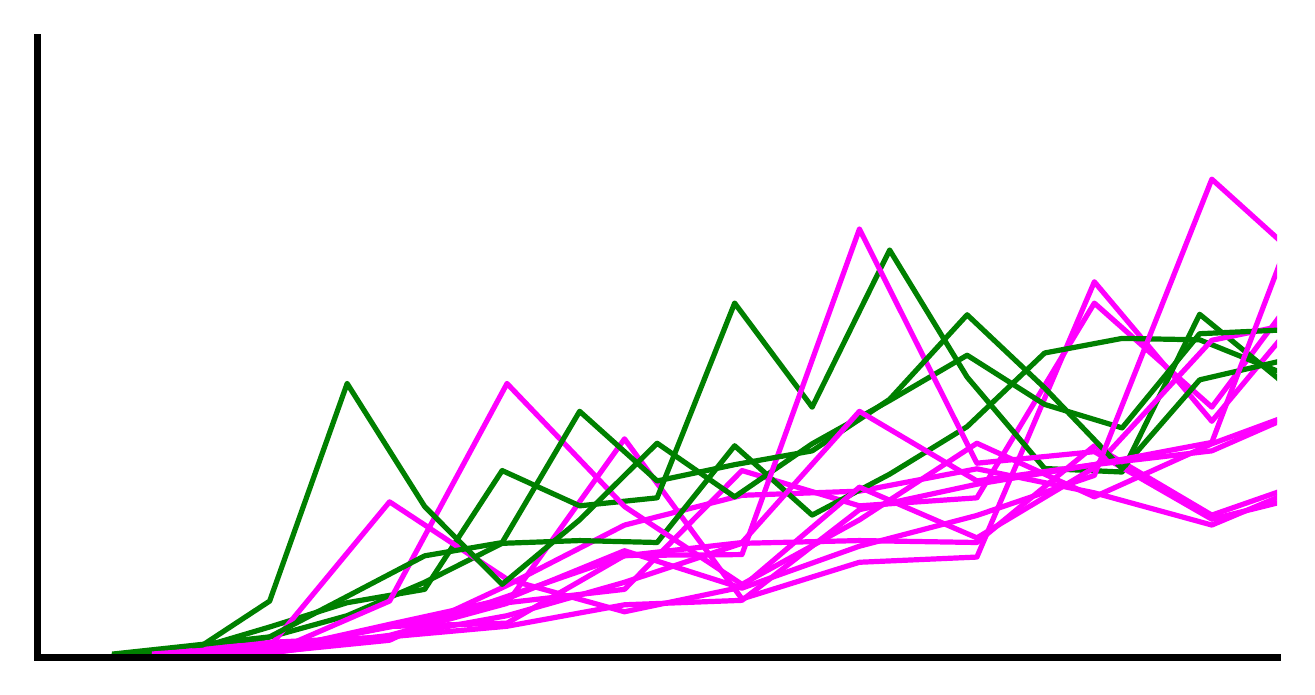}
  };
\end{tikzpicture}
&
\begin{tikzpicture}
\node[anchor=south west,inner sep=0] (A) at (0,0)
  {
    \pdfliteral{ 1 w}\includegraphics[width=0.75in,height=0.35in,page=1] {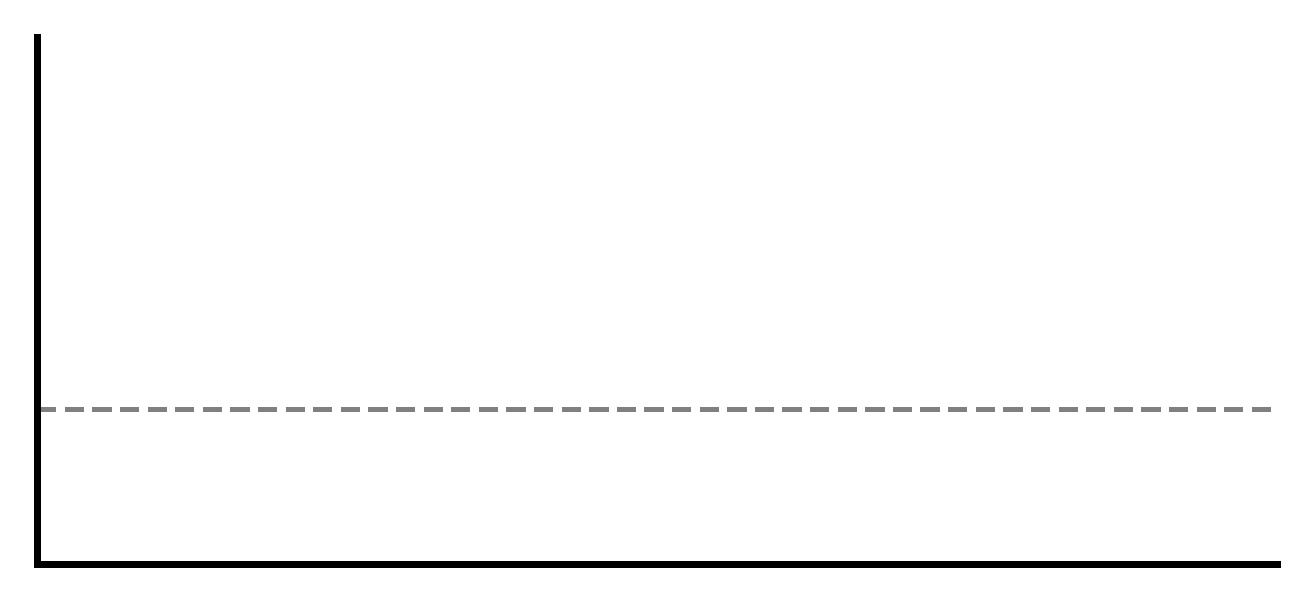}
  };
  \node[anchor=south west,inner sep=0] (A) at (0,0.95)
  {
    \pdfliteral{ 1 w}\includegraphics[width=0.75in,height=0.35in,page=1] {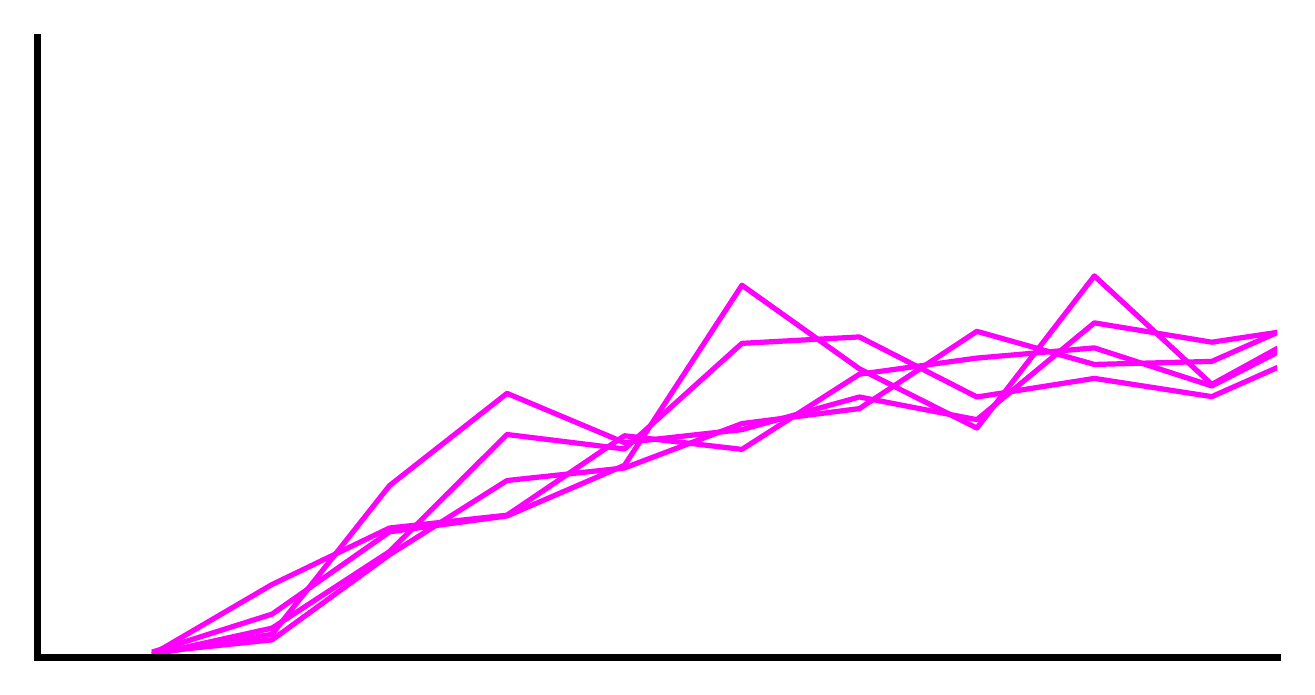}
  };
\end{tikzpicture}
\\
%%%%%%%%%%%%%%%%%%%%%%%%%%%%%%%%%
%%%%%%%%%%%%%%%%%%%%%%%%%%%%%%%%%
%%%%%%%%%%%%%%%%%%%%%%%%%%%%%%%%%
\rotatebox{90}{\quad BNLD~\shortcite{perrier18eg}}
&
\begin{tikzpicture}
  \node[anchor=south west,inner sep=0] (image) at (0,0)
  {
    \pdfliteral{ 1 w}\includegraphics[width=0.75in,page=1]{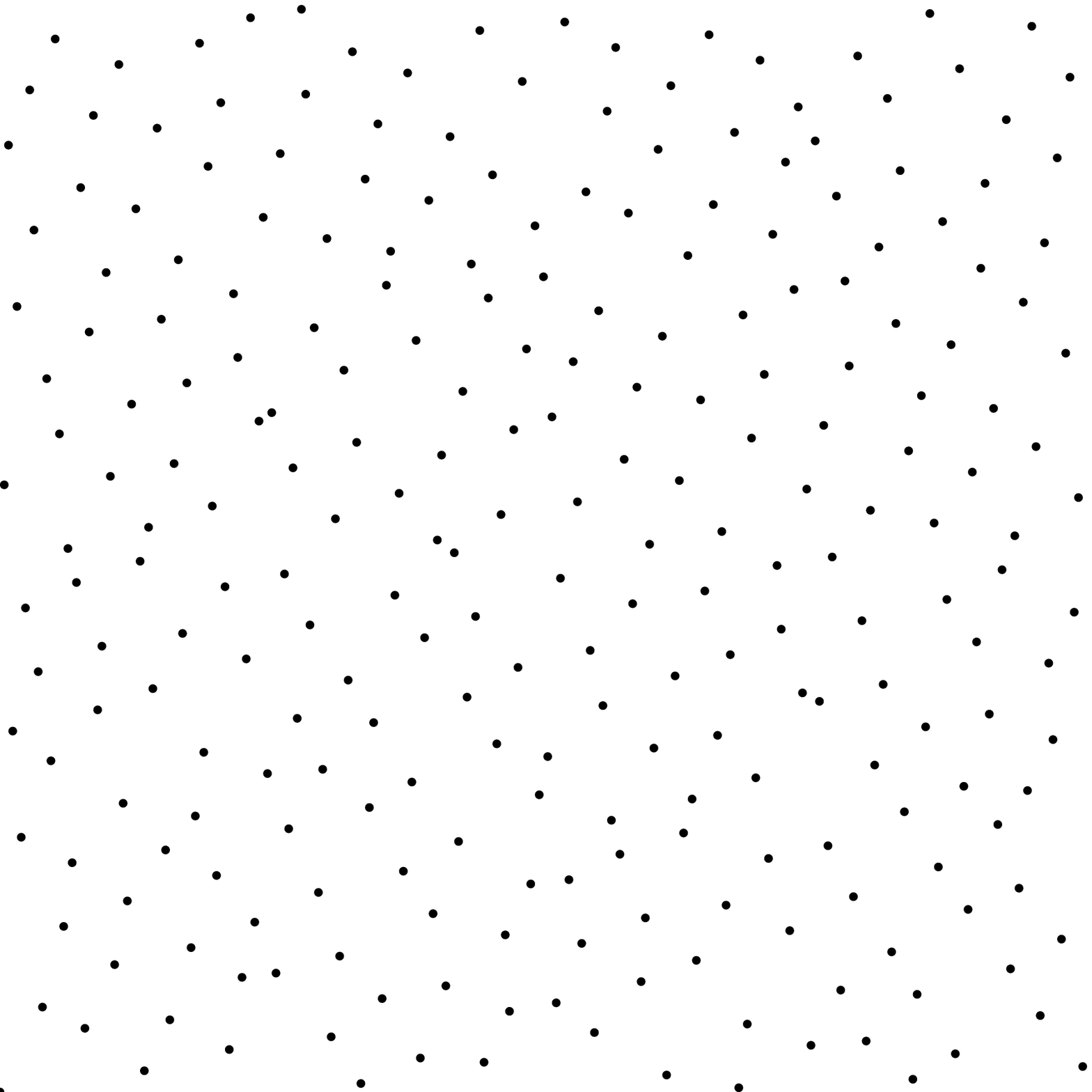}
  };
      \draw[black,thick] (0,0) -- (1.9,0) -- (1.9,1.9) -- (0,1.9) -- cycle;
\end{tikzpicture} 
&
\begin{tikzpicture}
  \node[anchor=south west,inner sep=0] (image) at (0,0)
  {
    \pdfliteral{ 1 w}\includegraphics[width=0.75in,page=1]{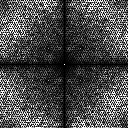}
  };
\end{tikzpicture} 
&
\begin{tikzpicture}
\node[anchor=south west,inner sep=0] (A) at (0,0)
  {
    \pdfliteral{ 1 w}\includegraphics[width=0.75in,height=0.35in,page=1] {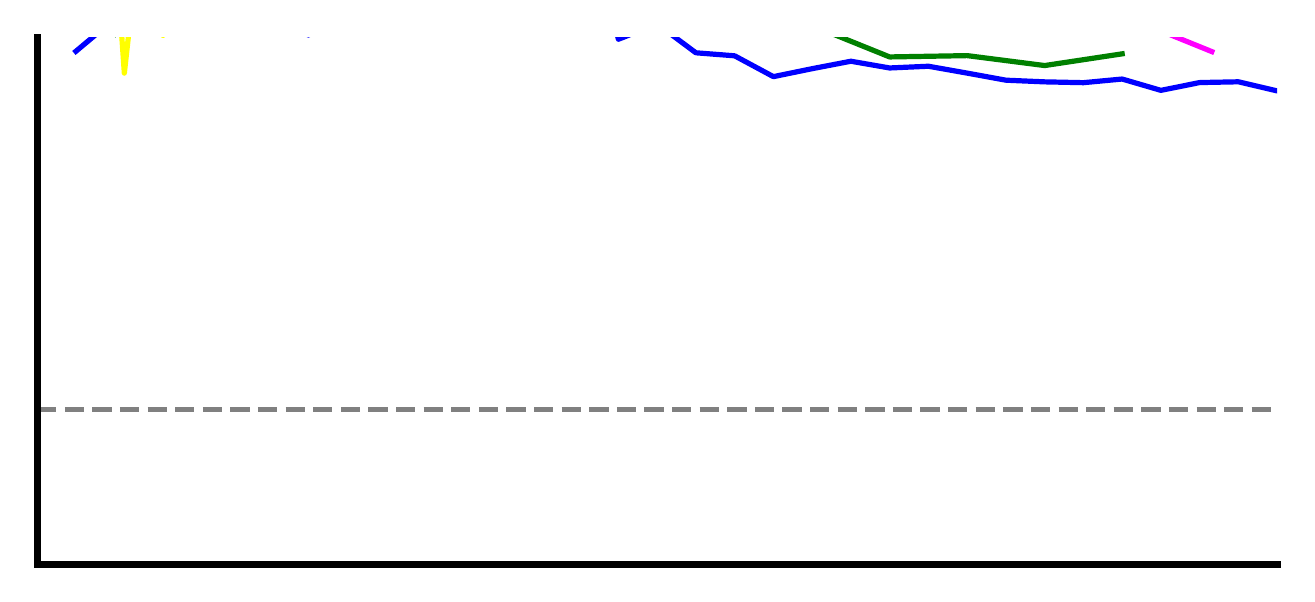}
  };
  \node[anchor=south west,inner sep=0] (A) at (0,0.95)
  {
    \pdfliteral{ 1 w}\includegraphics[width=0.75in,height=0.35in,page=1] {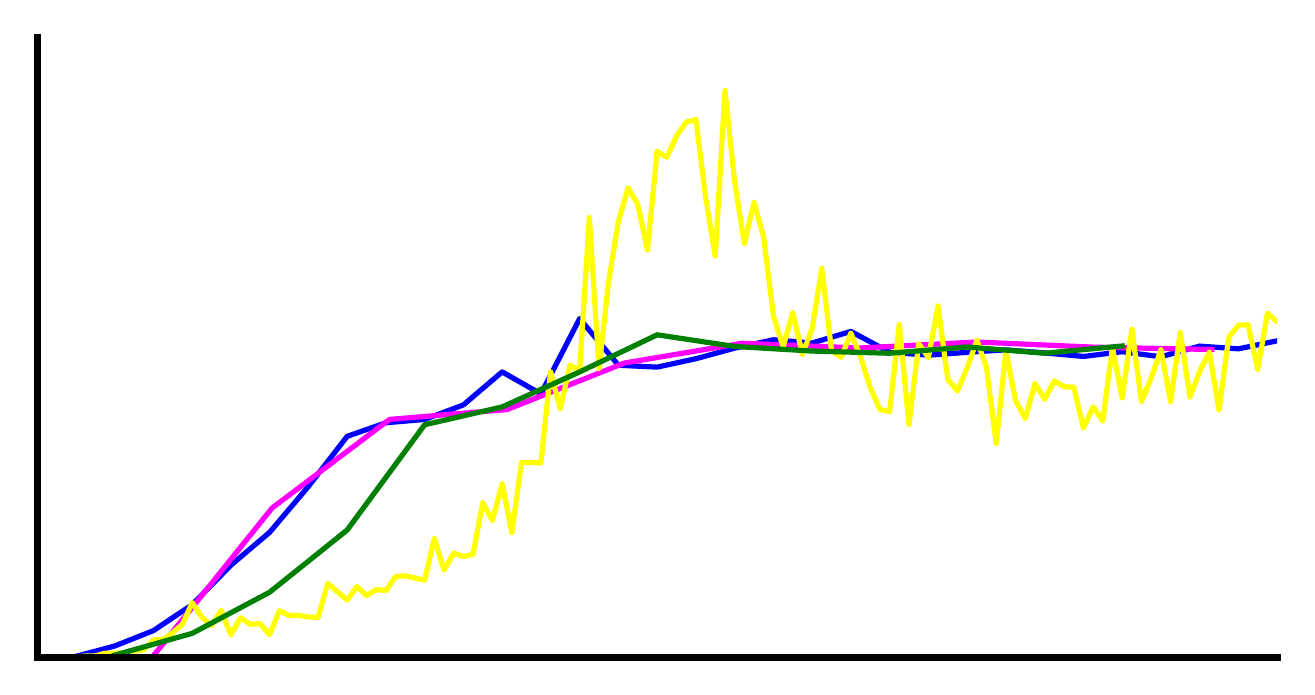}
  };
  \begin{scope}
   \filldraw[black] (0.05,0.77) circle (0pt) node[anchor=west] {\tiny anisotropy}; 
     \filldraw[black] (0.05,1.7) circle (0pt) node[anchor=west] {\tiny power};  
  \end{scope}
\end{tikzpicture}
&
\begin{tikzpicture}
\node[anchor=south west,inner sep=0] (A) at (0,0)
  {
    \pdfliteral{ 1 w}\includegraphics[width=0.75in,height=0.35in,page=1] {images/samplers/radial1d-blank.pdf}
  };
  \node[anchor=south west,inner sep=0] (A) at (0,0.95)
  {
    \pdfliteral{ 1 w}\includegraphics[width=0.75in,height=0.35in,page=1] {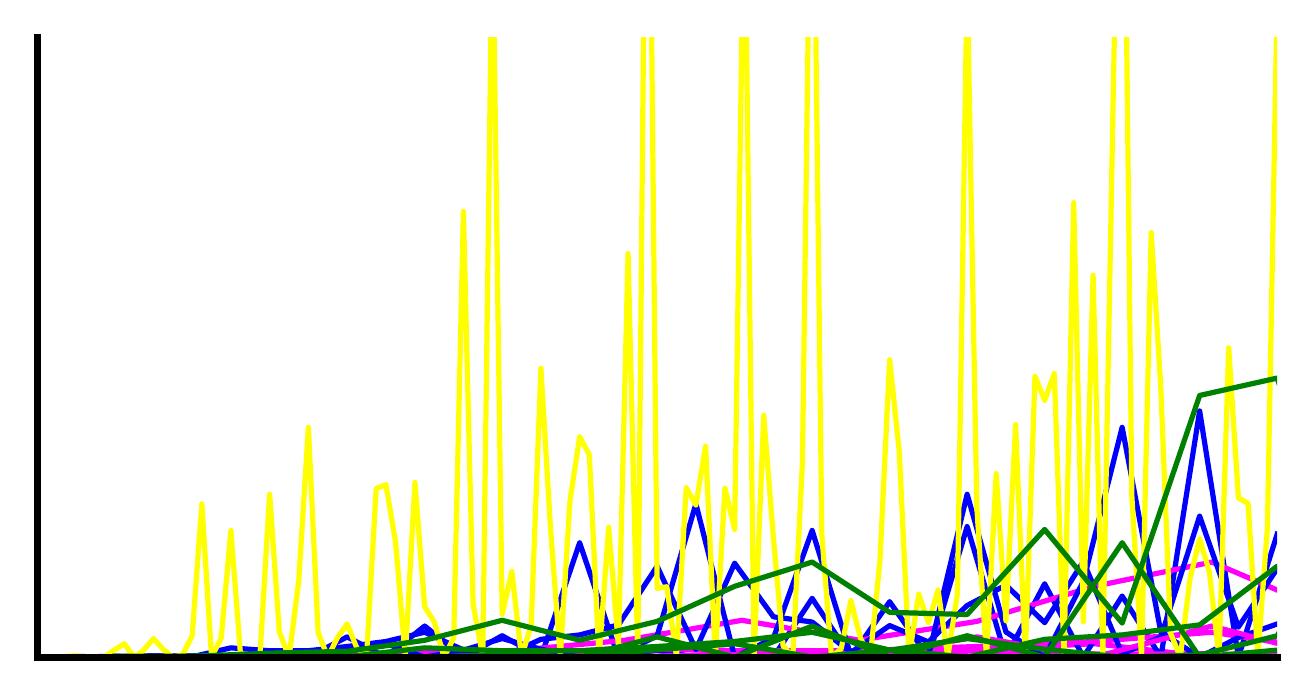}
  };
\end{tikzpicture}
&
\begin{tikzpicture}
  \node[anchor=south west,inner sep=0] (image) at (0,0)
  {
    \pdfliteral{ 1 w}\includegraphics[width=0.75in,page=1]{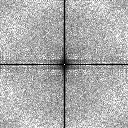}
  };
    \draw[mymagenta,thick] (0,0) -- (1.9,0) -- (1.9,1.9) -- (0,1.9) -- cycle;
\end{tikzpicture} 
&
\begin{tikzpicture}
  \node[anchor=south west,inner sep=0] (image) at (0,0)
  {
    \pdfliteral{ 1 w}\includegraphics[width=0.75in,page=1]{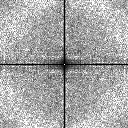}
  };
    \draw[green,thick] (0,0) -- (1.9,0) -- (1.9,1.9) -- (0,1.9) -- cycle;
\end{tikzpicture} 
&
\begin{tikzpicture}
  \node[anchor=south west,inner sep=0] (image) at (0,0)
  {
    \pdfliteral{ 1 w}\includegraphics[width=0.75in,page=1]{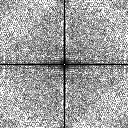}
  };
   \draw[skyblue,thick] (0,0) -- (1.9,0) -- (1.9,1.9) -- (0,1.9) -- cycle;
\end{tikzpicture} 
&
\begin{tikzpicture}
\node[anchor=south west,inner sep=0] (A) at (0,0)
  {
    \pdfliteral{ 1 w}\includegraphics[width=0.75in,height=0.35in,page=1] {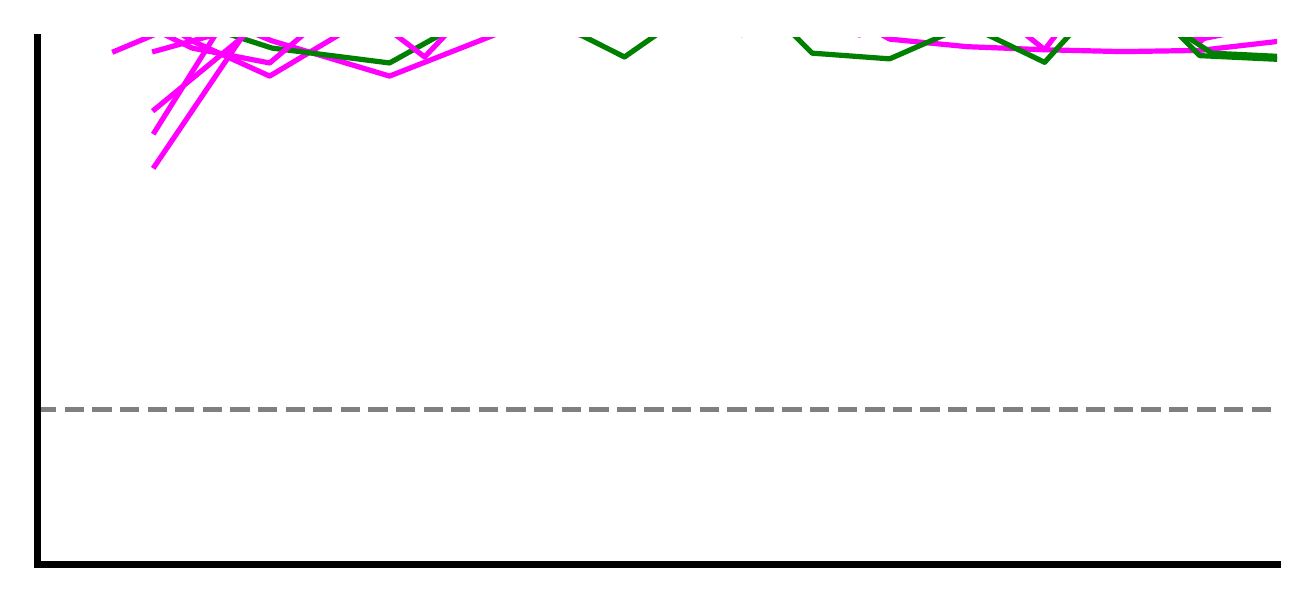}
  };
  \node[anchor=south west,inner sep=0] (A) at (0,0.95)
  {
    \pdfliteral{ 1 w}\includegraphics[width=0.75in,height=0.35in,page=1] {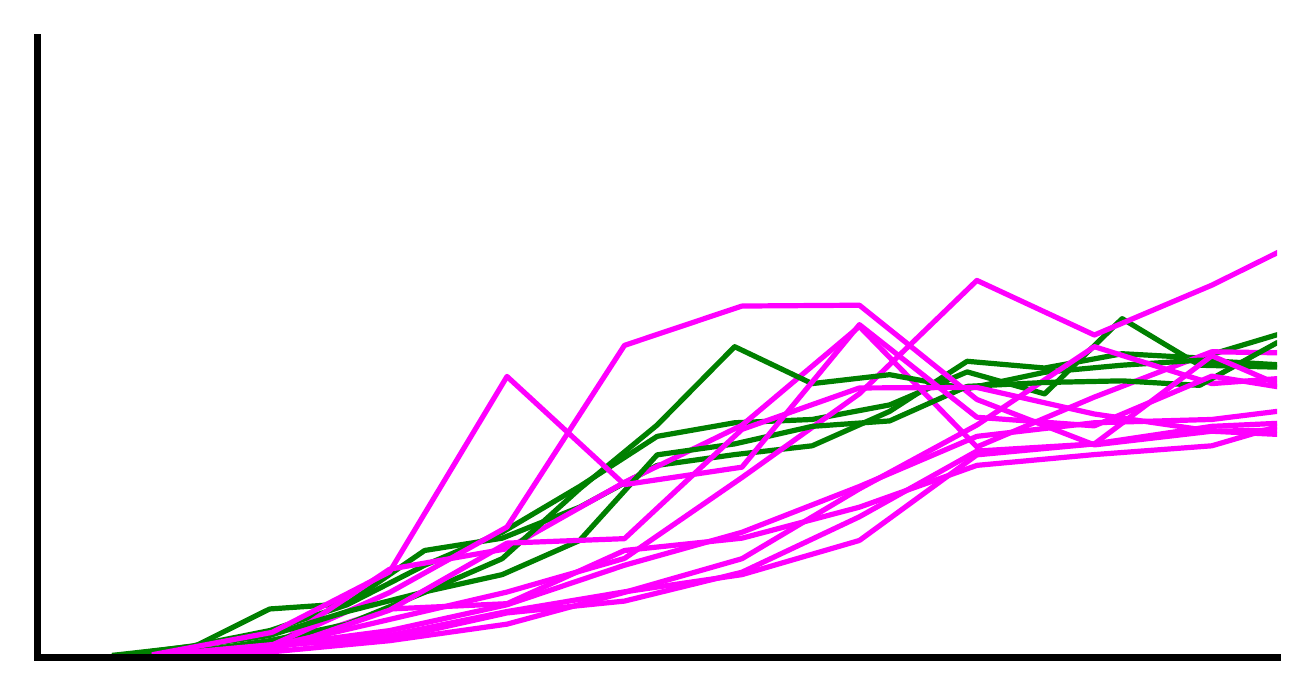}
  };
\end{tikzpicture}
&
\begin{tikzpicture}
\node[anchor=south west,inner sep=0] (A) at (0,0)
  {
    \pdfliteral{ 1 w}\includegraphics[width=0.75in,height=0.35in,page=1] {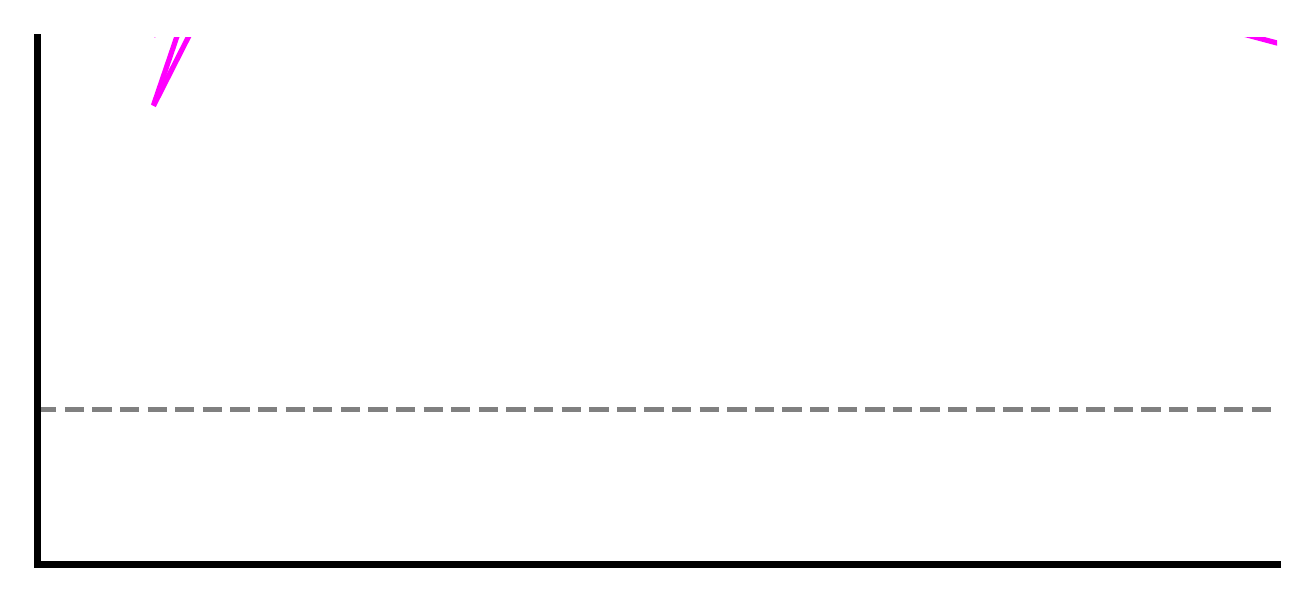}
  };
  \node[anchor=south west,inner sep=0] (A) at (0,0.95)
  {
    \pdfliteral{ 1 w}\includegraphics[width=0.75in,height=0.35in,page=1] {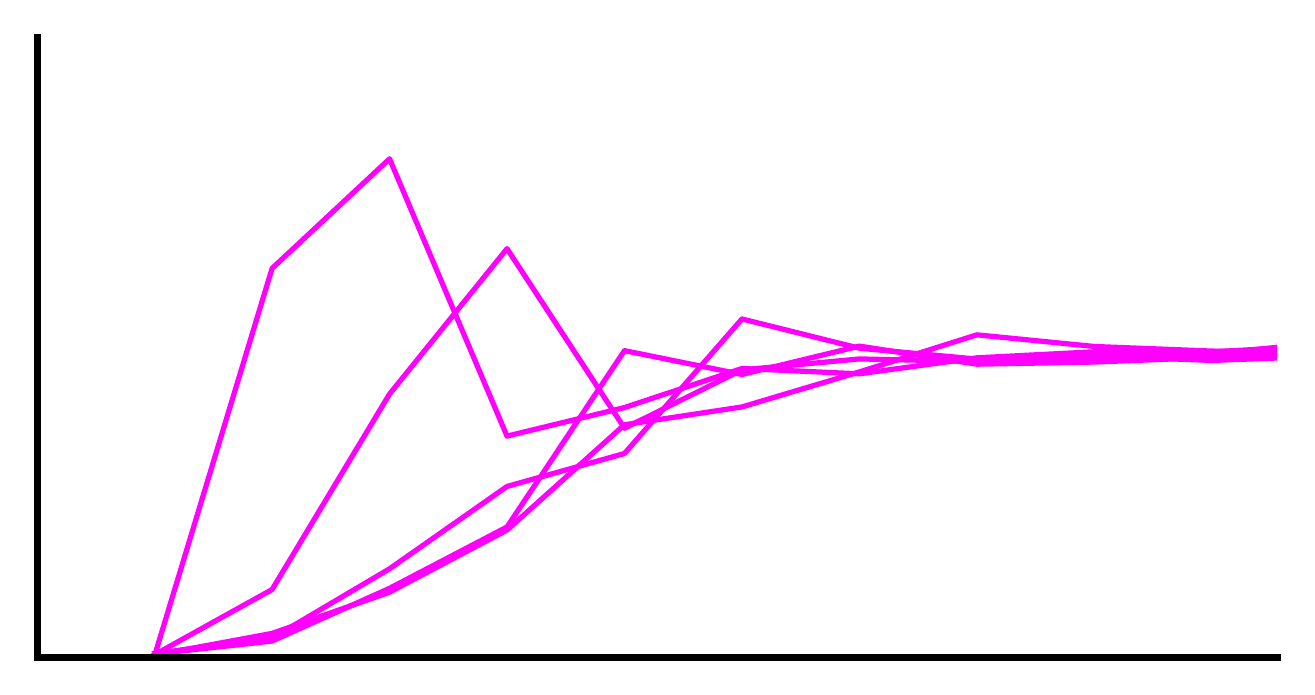}
  };
\end{tikzpicture}
\\
%%%%%%%%%%%%%%%%%%%%%%%%%%%%%%%%%
%%%%%%%%%%%%%%%%%%%%%%%%%%%%%%%%%
%%%%%%%%%%%%%%%%%%%%%%%%%%%%%%%%%
\rotatebox{90}{\quad Our (jittered)}
&
\begin{tikzpicture}
  \node[anchor=south west,inner sep=0] (image) at (0,0)
  {
    \pdfliteral{ 1 w}\includegraphics[width=0.75in,page=1]{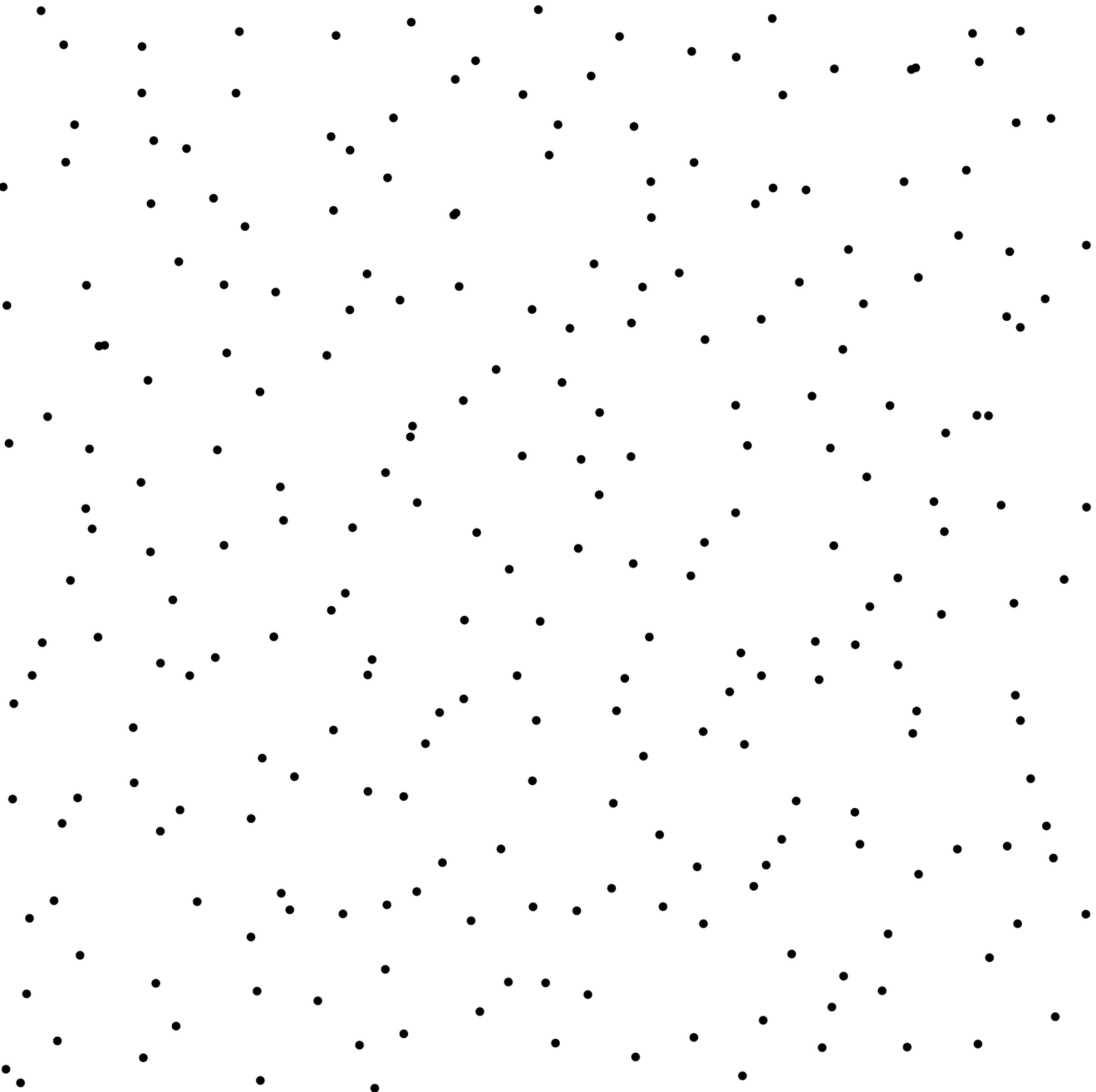}
  };
      \draw[black,thick] (0,0) -- (1.9,0) -- (1.9,1.9) -- (0,1.9) -- cycle;
\end{tikzpicture} 
&
\begin{tikzpicture}
  \node[anchor=south west,inner sep=0] (image) at (0,0)
  {
    \pdfliteral{ 1 w}\includegraphics[width=0.75in,page=1]{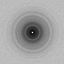}
  };
\end{tikzpicture} 
&
\begin{tikzpicture}
\node[anchor=south west,inner sep=0] (A) at (0,0)
  {
    \pdfliteral{ 1 w}\includegraphics[width=0.75in,height=0.35in,page=1] {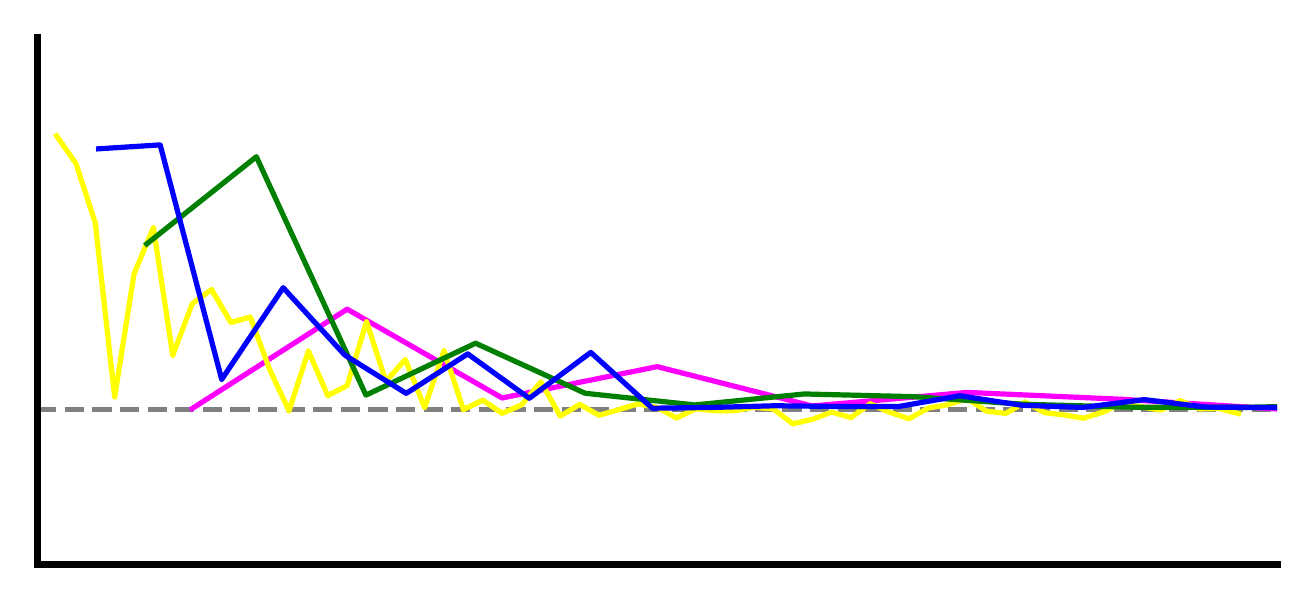}
  };
  \node[anchor=south west,inner sep=0] (A) at (0,0.95)
  {
    \pdfliteral{ 1 w}\includegraphics[width=0.75in,height=0.35in,page=1] {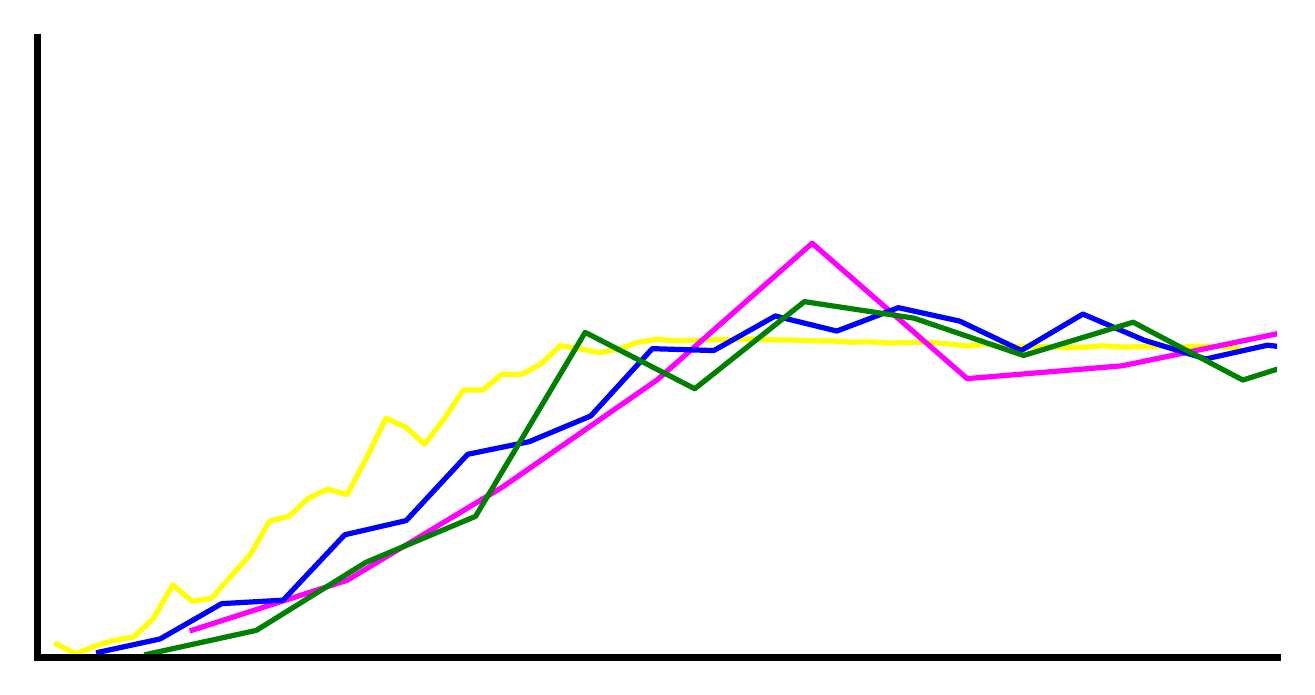}
  };
  \begin{scope}
   \filldraw[black] (0.05,0.77) circle (0pt) node[anchor=west] {\tiny anisotropy}; 
     \filldraw[black] (0.05,1.7) circle (0pt) node[anchor=west] {\tiny power};  
  \end{scope}
\end{tikzpicture}
&
\begin{tikzpicture}
\node[anchor=south west,inner sep=0] (A) at (0,0)
  {
    \pdfliteral{ 1 w}\includegraphics[width=0.75in,height=0.35in,page=1] {images/samplers/radial1d-blank.pdf}
  };
  \node[anchor=south west,inner sep=0] (A) at (0,0.95)
  {
    \pdfliteral{ 1 w}\includegraphics[width=0.75in,height=0.35in,page=1] {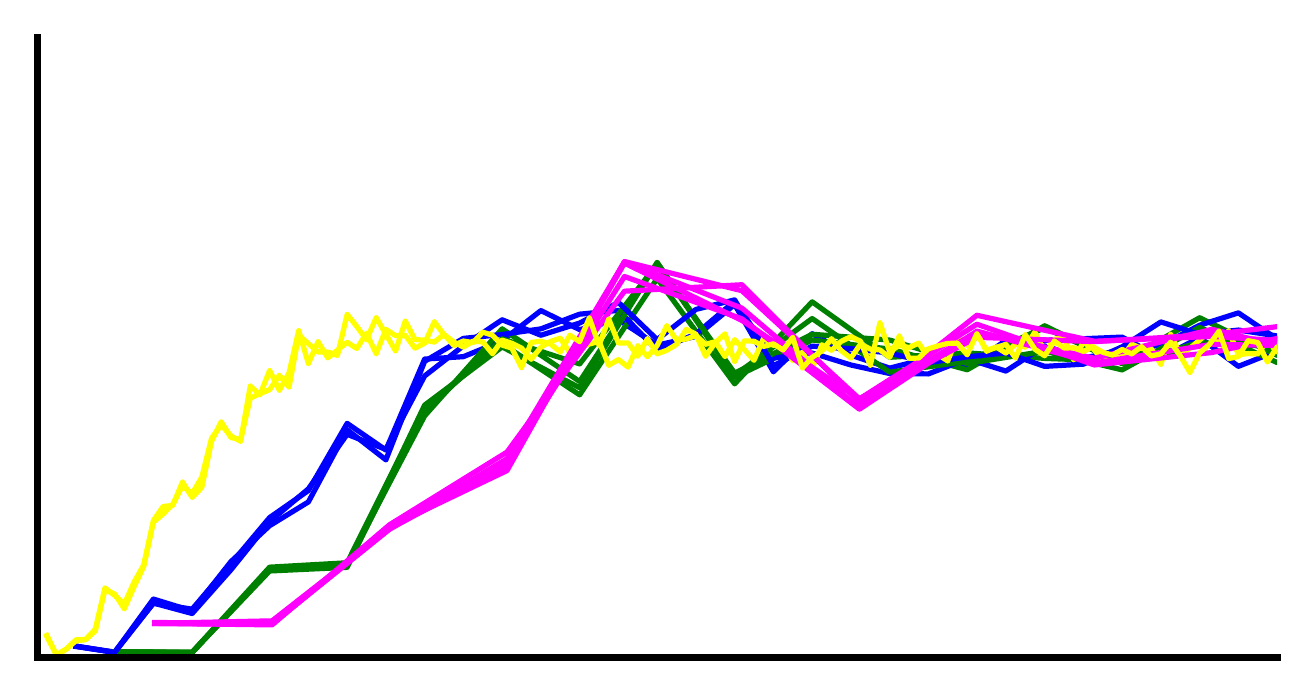}
  };
\end{tikzpicture}
&
\begin{tikzpicture}
  \node[anchor=south west,inner sep=0] (image) at (0,0)
  {
    \pdfliteral{ 1 w}\includegraphics[width=0.75in,page=1]{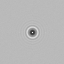}
  };
    \draw[mymagenta,thick] (0,0) -- (1.9,0) -- (1.9,1.9) -- (0,1.9) -- cycle;
\end{tikzpicture} 
&
\begin{tikzpicture}
  \node[anchor=south west,inner sep=0] (image) at (0,0)
  {
    \pdfliteral{ 1 w}\includegraphics[width=0.75in,page=1]{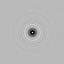}
  };
    \draw[green,thick] (0,0) -- (1.9,0) -- (1.9,1.9) -- (0,1.9) -- cycle;
\end{tikzpicture} 
&
\begin{tikzpicture}
  \node[anchor=south west,inner sep=0] (image) at (0,0)
  {
    \pdfliteral{ 1 w}\includegraphics[width=0.75in,page=1]{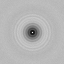}
  };
   \draw[skyblue,thick] (0,0) -- (1.9,0) -- (1.9,1.9) -- (0,1.9) -- cycle;
\end{tikzpicture} 
&
\begin{tikzpicture}
\node[anchor=south west,inner sep=0] (A) at (0,0)
  {
    \pdfliteral{ 1 w}\includegraphics[width=0.75in,height=0.35in,page=1] {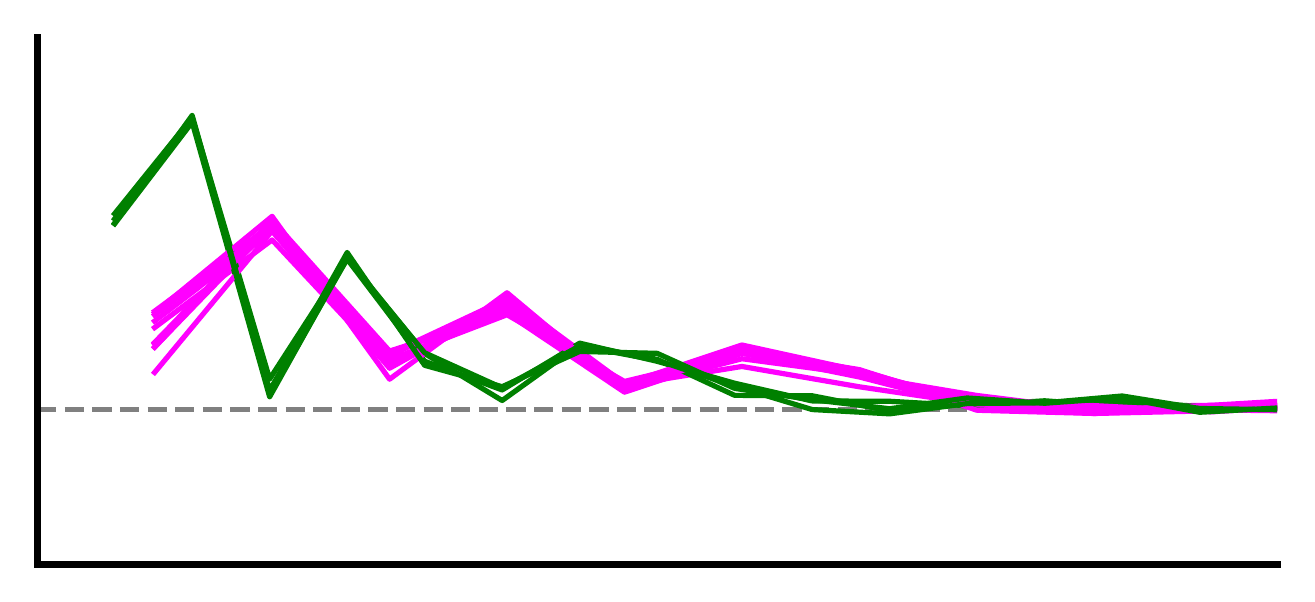}
  };
  \node[anchor=south west,inner sep=0] (A) at (0,0.95)
  {
    \pdfliteral{ 1 w}\includegraphics[width=0.75in,height=0.35in,page=1] {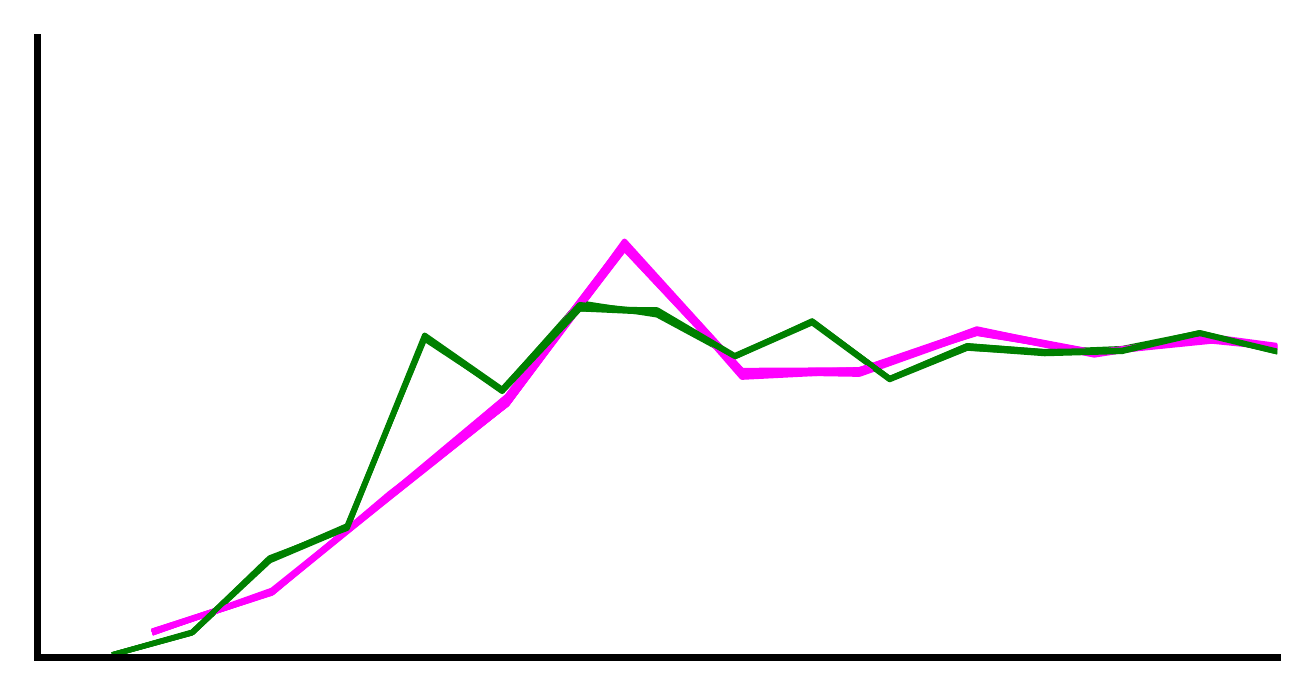}
  };
\end{tikzpicture}
&
\begin{tikzpicture}
\node[anchor=south west,inner sep=0] (A) at (0,0)
  {
    \pdfliteral{ 1 w}\includegraphics[width=0.75in,height=0.35in,page=1] {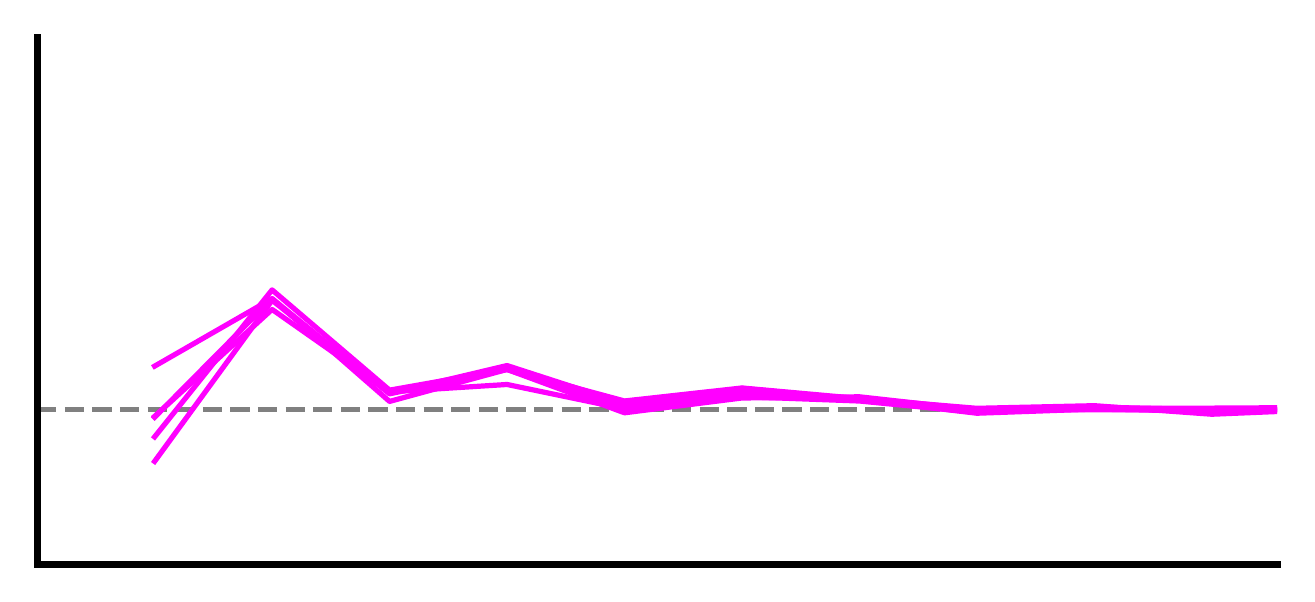}
  };
  \node[anchor=south west,inner sep=0] (A) at (0,0.95)
  {
    \pdfliteral{ 1 w}\includegraphics[width=0.75in,height=0.35in,page=1] {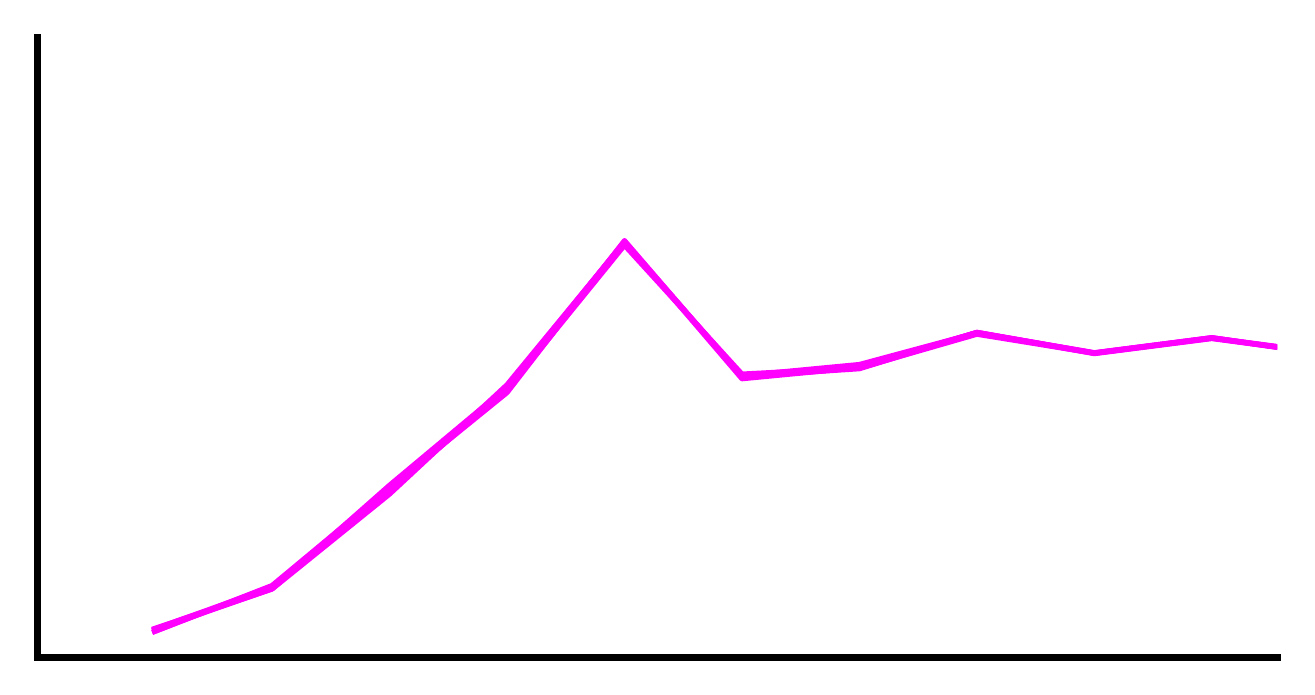}
  };
\end{tikzpicture}
\\
%%%%%%%%%%%%%%%%%%%%%%%%%%%%%%%%%
%%%%%%%%%%%%%%%%%%%%%%%%%%%%%%%%%
%%%%%%%%%%%%%%%%%%%%%%%%%%%%%%%%%
\rotatebox{90}{\quad Our (BNOT)}
&
\begin{tikzpicture}
  \node[anchor=south west,inner sep=0] (image) at (0,0)
  {
    \pdfliteral{ 1 w}\includegraphics[width=0.75in,page=1]{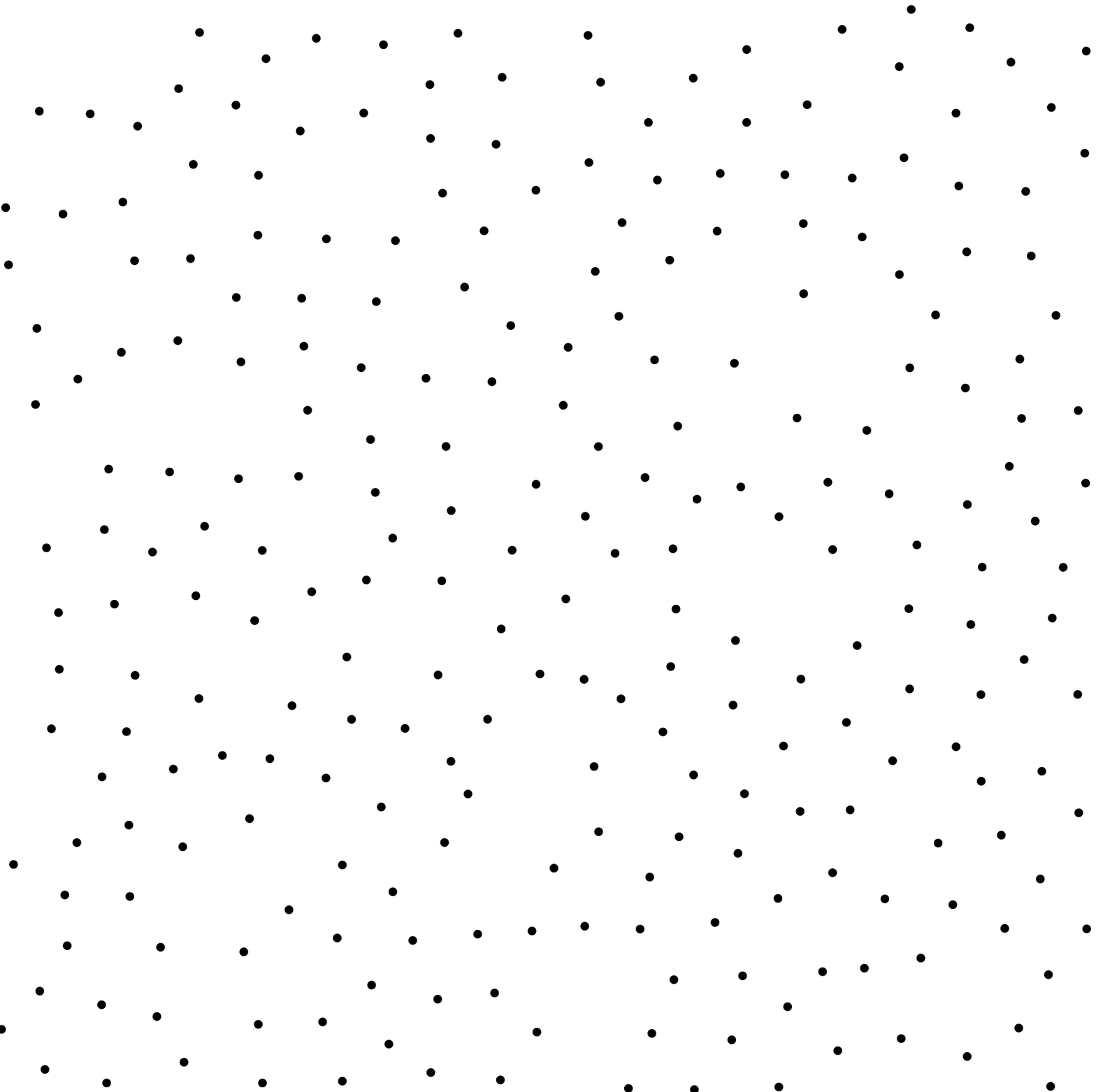}
  };
      \draw[black,thick] (0,0) -- (1.9,0) -- (1.9,1.9) -- (0,1.9) -- cycle;
\end{tikzpicture} 
&
\begin{tikzpicture}
  \node[anchor=south west,inner sep=0] (image) at (0,0)
  {
    \pdfliteral{ 1 w}\includegraphics[width=0.75in,page=1]{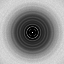}
  };
\end{tikzpicture} 
&
\begin{tikzpicture}
\node[anchor=south west,inner sep=0] (A) at (0,0)
  {
    \pdfliteral{ 1 w}\includegraphics[width=0.75in,height=0.35in,page=1] {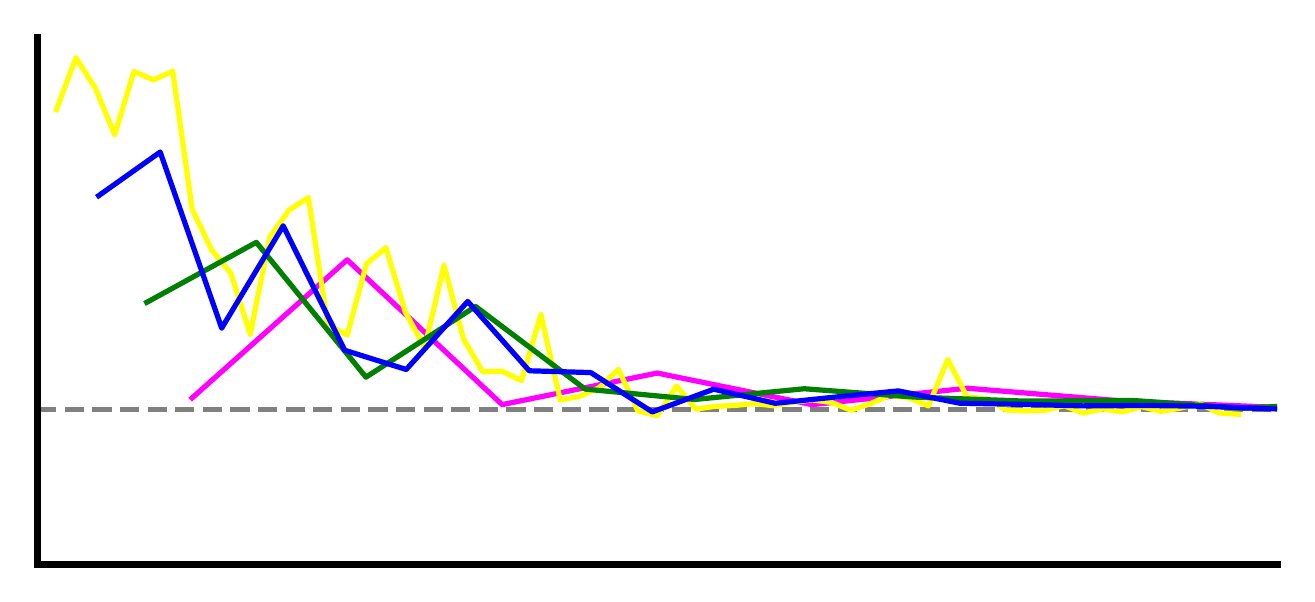}
  };
  \node[anchor=south west,inner sep=0] (A) at (0,0.95)
  {
    \pdfliteral{ 1 w}\includegraphics[width=0.75in,height=0.35in,page=1] {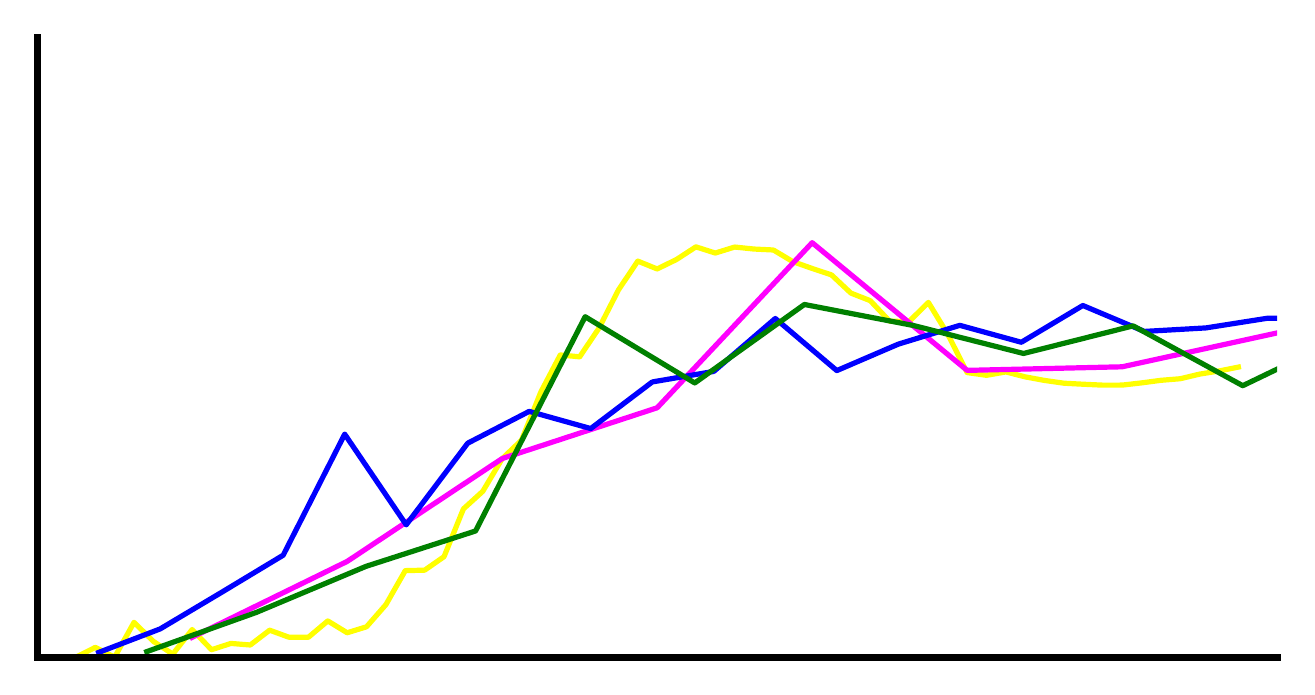}
  };
  \begin{scope}
   \filldraw[black] (0.05,0.77) circle (0pt) node[anchor=west] {\tiny anisotropy}; 
     \filldraw[black] (0.05,1.7) circle (0pt) node[anchor=west] {\tiny power};  
  \end{scope}
\end{tikzpicture}
&
\begin{tikzpicture}
\node[anchor=south west,inner sep=0] (A) at (0,0)
  {
    \pdfliteral{ 1 w}\includegraphics[width=0.75in,height=0.35in,page=1] {images/samplers/radial1d-blank.pdf}
  };
  \node[anchor=south west,inner sep=0] (A) at (0,0.95)
  {
    \pdfliteral{ 1 w}\includegraphics[width=0.75in,height=0.35in,page=1] {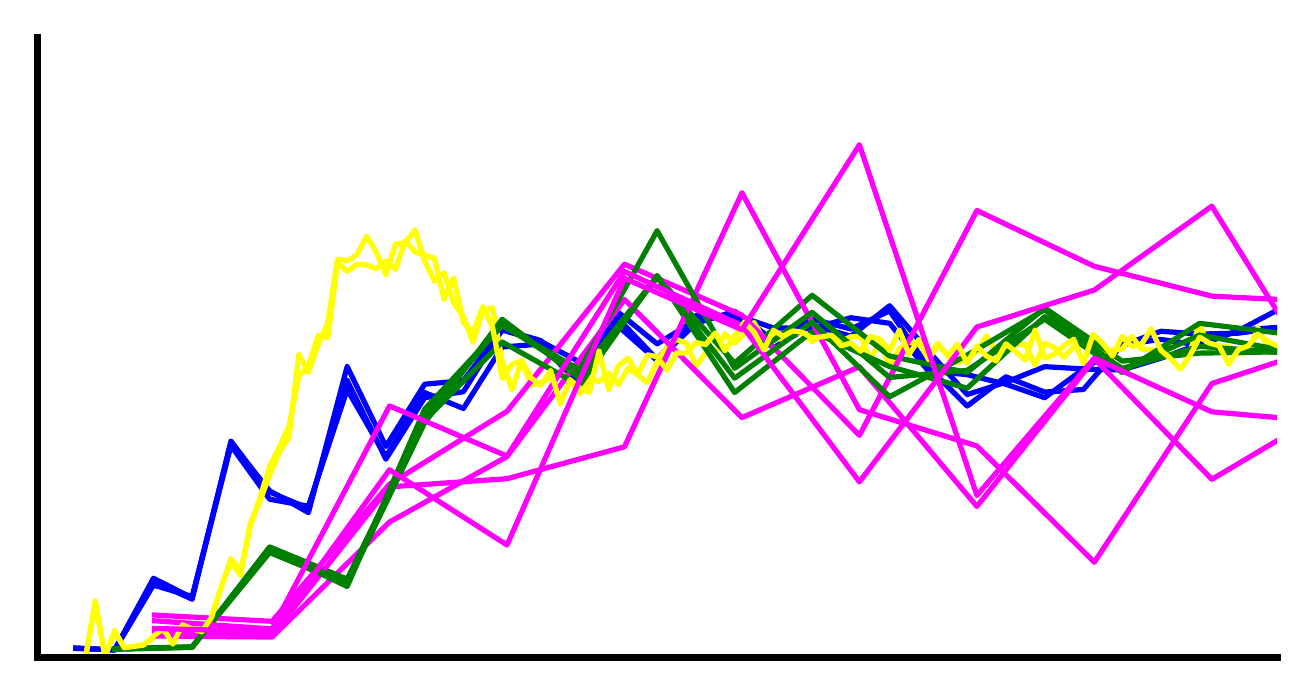}
  };
\end{tikzpicture}
&
\begin{tikzpicture}
  \node[anchor=south west,inner sep=0] (image) at (0,0)
  {
    \pdfliteral{ 1 w}\includegraphics[width=0.75in,page=1]{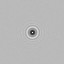}
  };
    \draw[mymagenta,thick] (0,0) -- (1.9,0) -- (1.9,1.9) -- (0,1.9) -- cycle;
\end{tikzpicture} 
&
\begin{tikzpicture}
  \node[anchor=south west,inner sep=0] (image) at (0,0)
  {
    \pdfliteral{ 1 w}\includegraphics[width=0.75in,page=1]{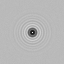}
  };
    \draw[green,thick] (0,0) -- (1.9,0) -- (1.9,1.9) -- (0,1.9) -- cycle;
\end{tikzpicture} 
&
\begin{tikzpicture}
  \node[anchor=south west,inner sep=0] (image) at (0,0)
  {
    \pdfliteral{ 1 w}\includegraphics[width=0.75in,page=1]{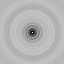}
  };
   \draw[skyblue,thick] (0,0) -- (1.9,0) -- (1.9,1.9) -- (0,1.9) -- cycle;
\end{tikzpicture} 
&
\begin{tikzpicture}
\node[anchor=south west,inner sep=0] (A) at (0,0)
  {
    \pdfliteral{ 1 w}\includegraphics[width=0.75in,height=0.35in,page=1] {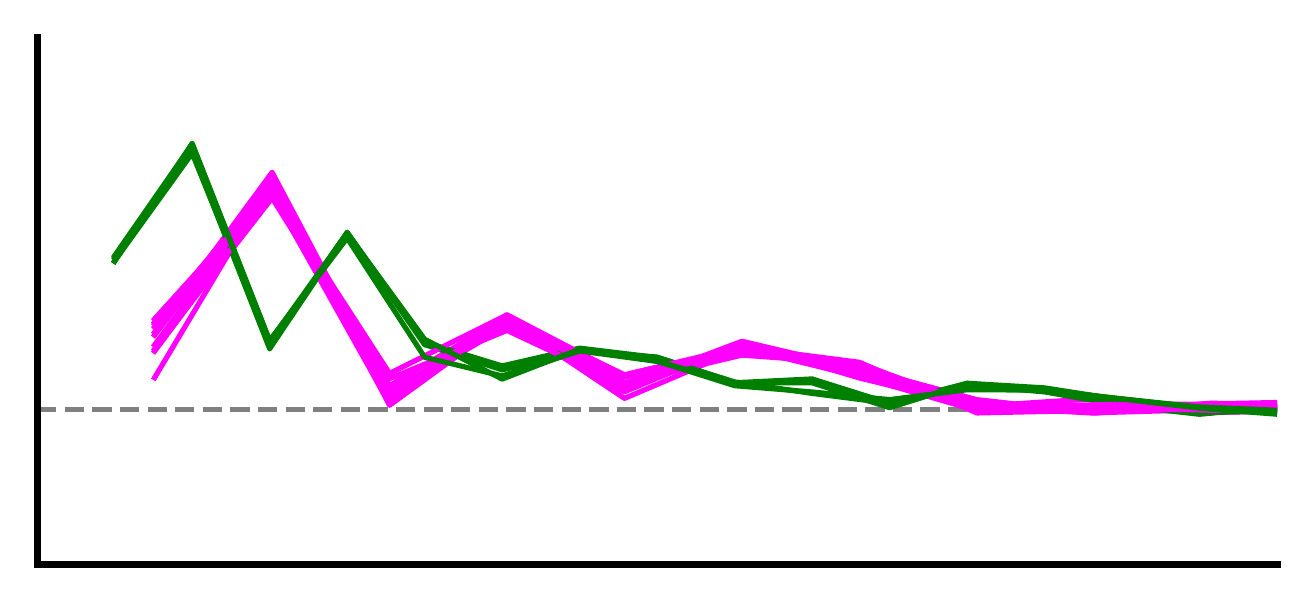}
  };
  \node[anchor=south west,inner sep=0] (A) at (0,0.95)
  {
    \pdfliteral{ 1 w}\includegraphics[width=0.75in,height=0.35in,page=1] {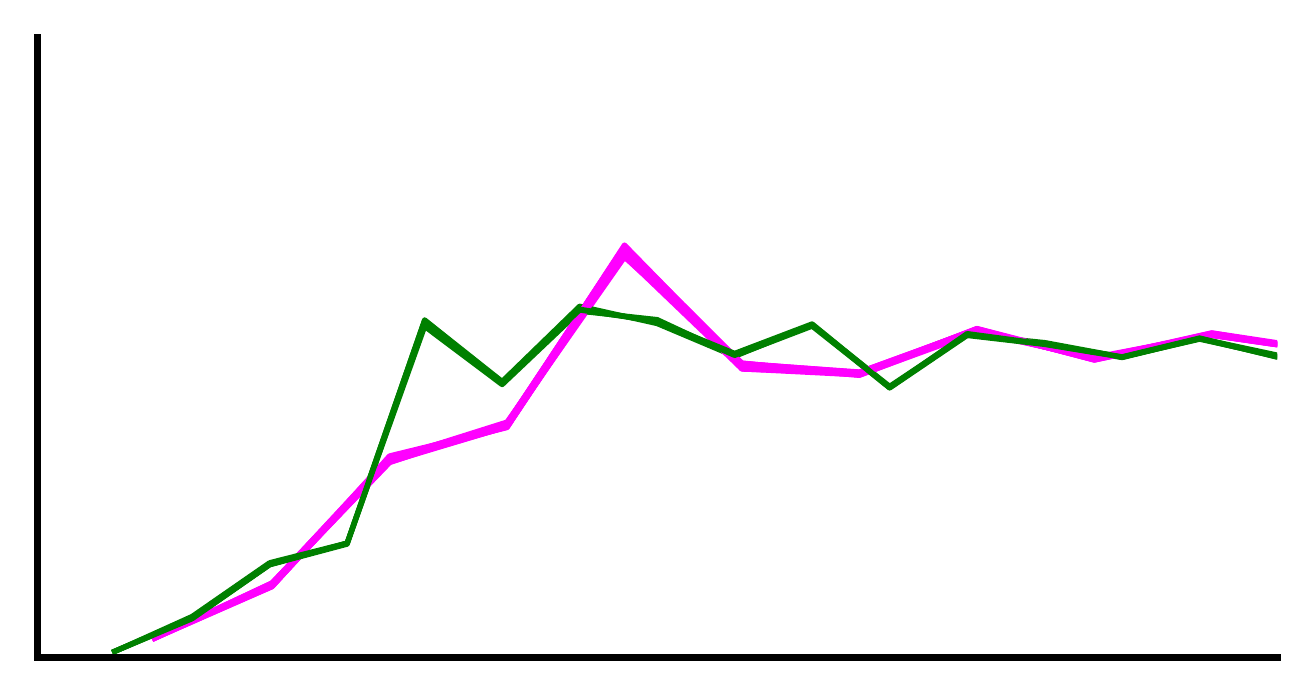}
  };
\end{tikzpicture}
&
\begin{tikzpicture}
\node[anchor=south west,inner sep=0] (A) at (0,0)
  {
    \pdfliteral{ 1 w}\includegraphics[width=0.75in,height=0.35in,page=1] {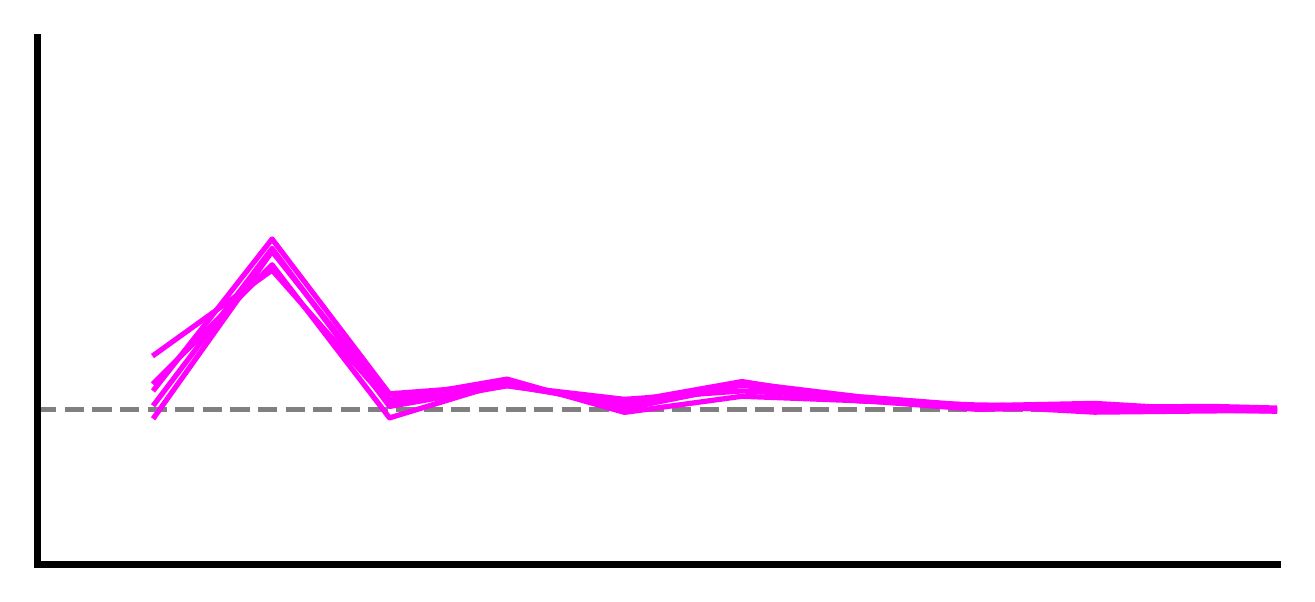}
  };
  \node[anchor=south west,inner sep=0] (A) at (0,0.95)
  {
    \pdfliteral{ 1 w}\includegraphics[width=0.75in,height=0.35in,page=1] {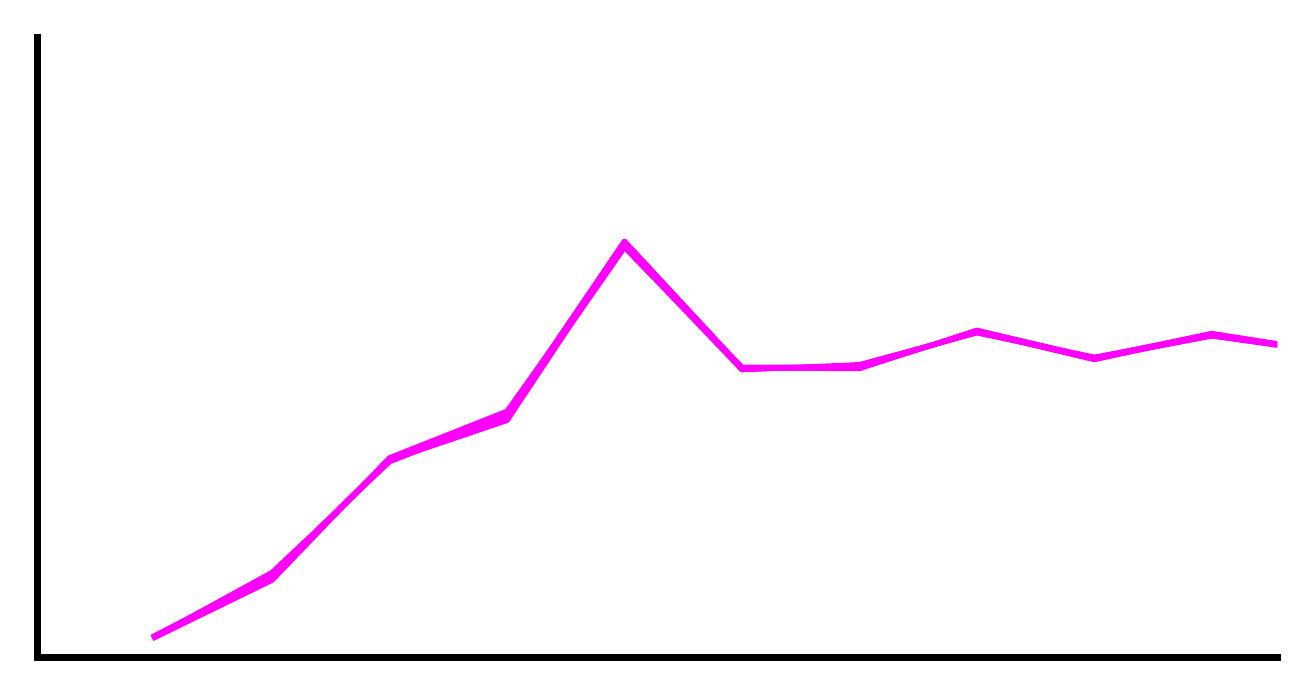}
  };
\end{tikzpicture}
\\
\end{tabular}%
\caption{
\label{fig:analysis-samplers}
Our multi-dimensional analysis (for $N=1024$ samples).
Every row shows a sample pattern, where patterns produced using our approach come last.
The first column shows a 2D realization.
The second column shows the power spectrum of the 2D pattern.
The third column shows the radial power mean and the anisotropy.
All radial profiles are appropriately scaled for a given sample count and the dimension to facilitate comparison.
In the \emph{anisotropy} radial plots, the dashed gray horizontal line shows the reference $-10$dB value, any sampler that deviates from this reference has some anisotropic structures present in it's spectrum. 
Colors encode different projections as explained in the legend.
The yellow line is for example the radial anisotropy of the 2D pattern.
All 1D projections are shown in the fourth column.
Columns five to seven show the average power spectra of all possible 2D projections, from 3D, 4D and 5D respectively.
For 3D this is the average of the $(x,y)$, $(y,z)$ and $(x,z)$ spectrum.
The last two columns show the radial averages of all 3D resp.\ 4D subspaces.
For 3D, this is the average of all 3D projections from 4D (green) and 5D (magenta).
Finally, the last column shows all 4D projections from 5D (magenta).
Please see \refSec{Analysis} for a discussion of the results for different metods (rows). 
%Our samplers are generated with a kernel receptive field of $\sigma=0.4$.
} 
\end{figure*}

%% file: images/fig-histogram-loss-poissondisk-all.tex
\begin{figure}[t!]

%\centering
\footnotesize
\hspace*{-1.15em}
\begin{tabular}{c@{}}
%
%\\
% & $\sigma = 0.2$ & $\sigma = 0.4$
\\
\begin{tikzpicture}
  \node[anchor=south west,inner sep=0] (image) at (0,0)
  {
    \pdfliteral{ 1 w}\includegraphics[width=2.75in,page=1]{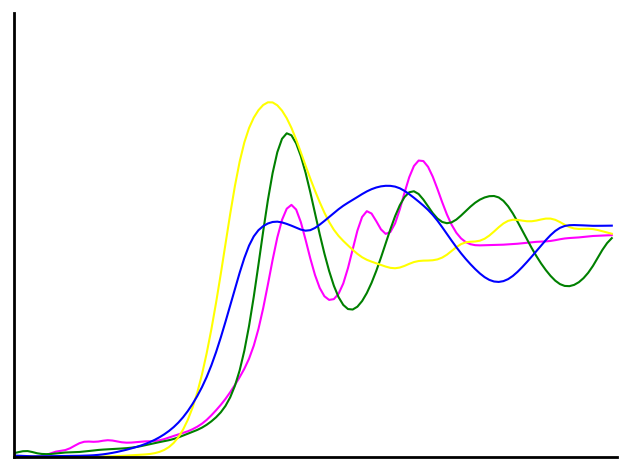}
  };
  %pointset-rdf-poissondisk-n625-d2-000000
  \begin{scope}
  \node[anchor=south west,inner sep=0] (lines) at (0.25,2.85)
 {
 \pdfliteral{ 1 w}\includegraphics[width=0.115\textwidth,page=1]{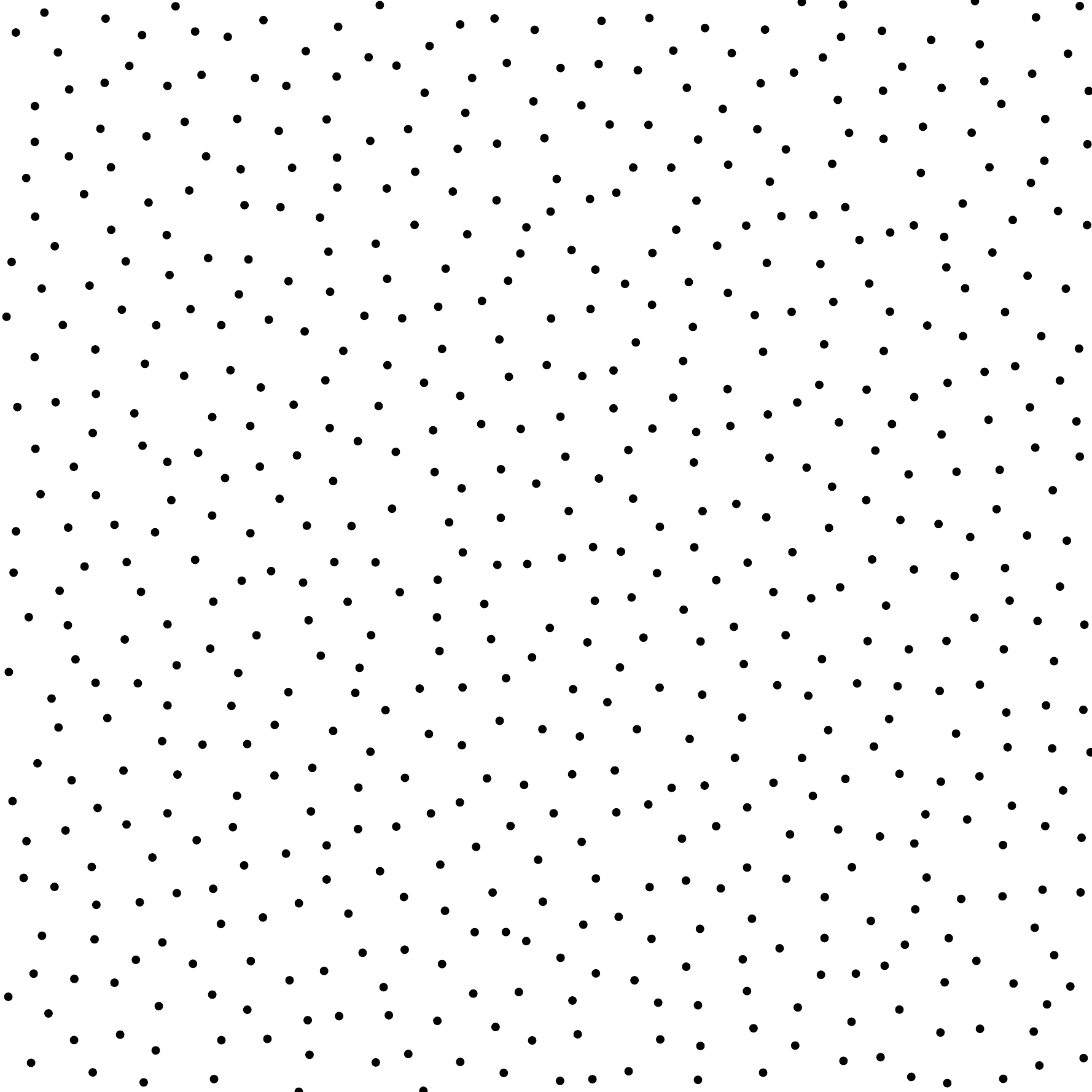}
  };
  \draw[gray,thick] (0.25,2.825) -- (2.35,2.825) -- (2.35,4.925) -- (0.25,4.925) -- cycle;
  \end{scope}
  
  %legends
 \begin{scope}
 \draw[mymagenta,thick] (5.75,5) -- (6.1,5);
\filldraw[black] (6.025,5) circle (0pt) node[anchor=west] {\tiny 5D};
\draw[green,thick] (5.75,4.5) -- (6.1,4.5);
\filldraw[black] (6.025,4.5) circle (0pt) node[anchor=west] {\tiny 4D};
\draw[blue,thick] (5.75,4) -- (6.1,4);
\filldraw[black] (6.025,4) circle (0pt) node[anchor=west] {\tiny 3D};
\draw[yellow,thick] (5.75,3.5) -- (6.1,3.5);
\filldraw[black] (6.025,3.5) circle (0pt) node[anchor=west] {\tiny 2D};
\draw[black,thick] (5.5,3.25) -- (6.5,3.25) -- (6.5,5.25) -- (5.5,5.25) -- cycle;
 \end{scope}
 
 %xlimit, ylimit
 \begin{scope}
  \filldraw[black] (0.05,-0.1) circle (0pt) node[anchor=west] { 0.0};
 \filldraw[black] (6.5,-0.1) circle (0pt) node[anchor=west] { 120.0};
 \filldraw[black] (-0.25,5) circle (0pt) node[anchor=west] { 2.0};
  \filldraw[black] (-0.25,0.1) circle (0pt) node[anchor=west] { 0.0};
 \end{scope}
 
 %axes labels
 \begin{scope}
  \filldraw[black] (3,-0.1) circle (0pt) node[anchor=west] { Distances};
 \filldraw[black] (-0.05,1.75) circle (0pt) node[anchor=west,rotate=90] {Normalized count};
 \end{scope}
 
\end{tikzpicture} 
\\
\end{tabular}%
\caption{
\label{fig:histogram-loss-poissondisk}
Differential analysis when using our approach with a \texttt{histogram} loss for dimensions upto $5D$. Here we are plotting (in green) the pair correlation function (PCF) for points generated using our approach, given target 2D Poisson disk PCF. Loss is defined as a sum over all 2D projections. 
%The blue curve denotes the target, the green curve the histogram our approach produces.
%We also obtained another blue noise target(Poisson Disk) samples for multiple dimensions using pair correlation function (radial histograms) as a loss function. 
} 
\end{figure}

%% file: images/fig-halftoning.tex
\begin{figure}[t!]

%\centering
\footnotesize
\hspace*{-1.15em}
\begin{tabular}{c@{\;}c@{\;}c@{\;}c@{}}
\\
& Mask & Spectrum & Halftoning
\\
\rotatebox{90}{\qquad\quad Random}
&
\begin{tikzpicture}
  \node[anchor=south west,inner sep=0] (image) at (0,0)
  {
    \pdfliteral{ 1 w}\includegraphics[width=1in,page=1]{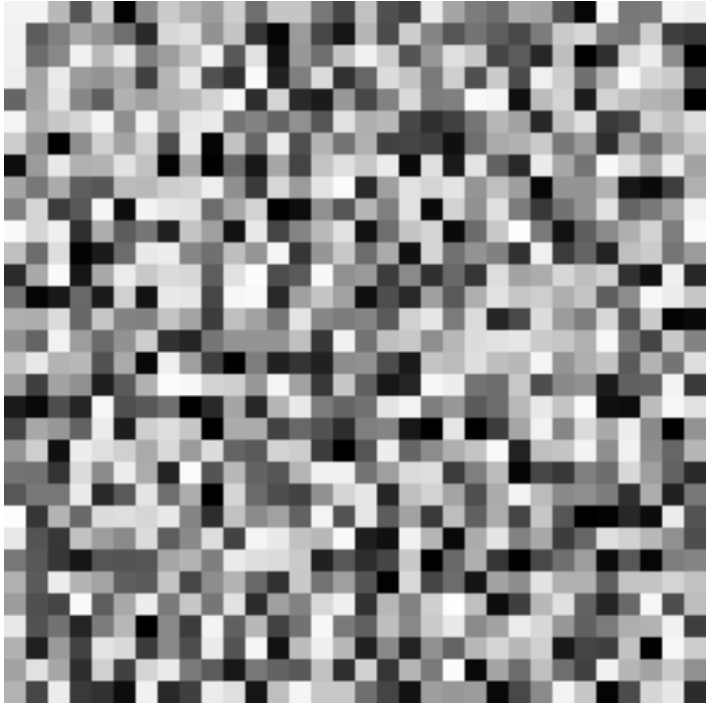}
  };
\end{tikzpicture} 
&
\begin{tikzpicture}
  \node[anchor=south west,inner sep=0] (image) at (0,0)
  {
    \pdfliteral{ 1 w}\includegraphics[width=1in,page=1]{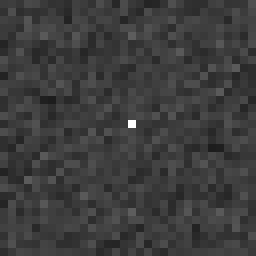}
  };
\end{tikzpicture} 
&
\begin{tikzpicture}
  \node[anchor=south west,inner sep=0] (image) at (0,0)
  {
    \pdfliteral{ 1 w}\includegraphics[width=1.3in,page=1]{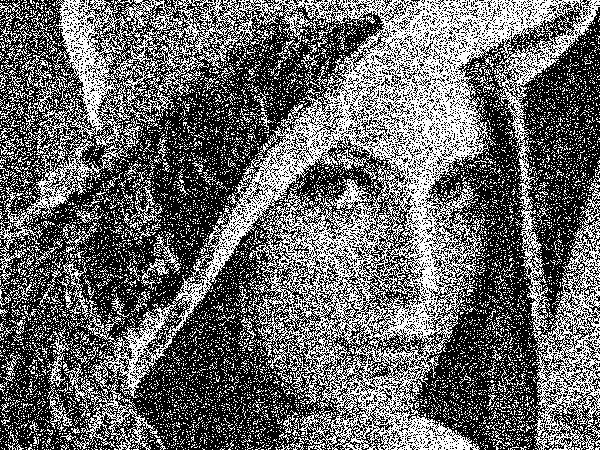}
  };
\end{tikzpicture} 
\\
\rotatebox{90}{(\texttt{spectrum(grid(s,x))}}
&
\begin{tikzpicture}
  \node[anchor=south west,inner sep=0] (image) at (0,0)
  {
    \pdfliteral{ 1 w}\includegraphics[width=1in,page=1]{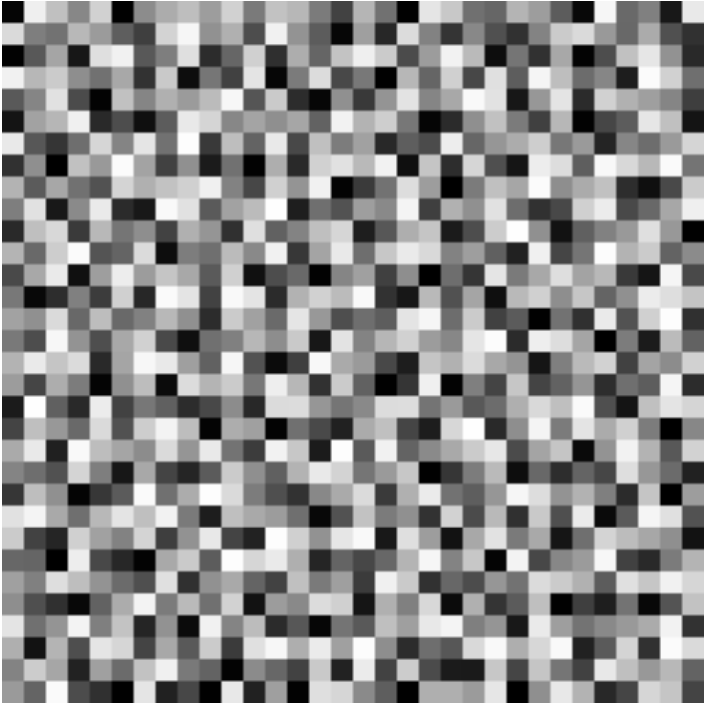}
  };
\end{tikzpicture} 
&
\begin{tikzpicture}
  \node[anchor=south west,inner sep=0] (image) at (0,0)
  {
    \pdfliteral{ 1 w}\includegraphics[width=1in,page=1]{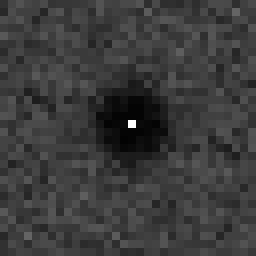}
  };
\end{tikzpicture} 
&
\begin{tikzpicture}
  \node[anchor=south west,inner sep=0] (image) at (0,0)
  {
    \pdfliteral{ 1 w}\includegraphics[width=1.3in,page=1]{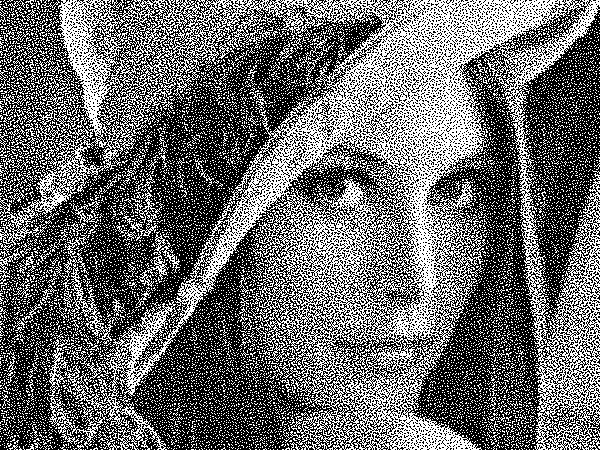}
  };
\end{tikzpicture} 
\end{tabular}%
\caption{
\label{fig:halftoning}
Comparison of a random and our learned masks \textbf{(bottom)}.
In a dithering mask \cite{georgiev2016dithered}, the 2D layout is fixed to a regular grid (\texttt{gridded}).
Our network has learned to filter the values so that the top spectrum \textbf{(second column)} turns into the bottom spectrum with a pronounced blue noise \ie no spatially nearby elements in the mask have similar values. 
Consequently, the artifacts using our sampling pattern in halftoning appears visually less suspicious \textbf{(third column)}.
} 
\end{figure}

%% file: images/fig-color-noise-spectralloss.tex
\begin{figure}[t!]

%\centering
\footnotesize
\hspace*{-1.15em}
\begin{tabular}{c@{\;}c@{\;}c@{}}
\\
Blue (noise) spectrum & Green (noise) spectrum & Pink (noise) spectrum
\\
%\rotatebox{90}{\qquad Random}
%&
\begin{tikzpicture}
  \node[anchor=south west,inner sep=0] (image) at (0,0)
  {
    \pdfliteral{ 1 w}\includegraphics[width=1.1in,page=1]{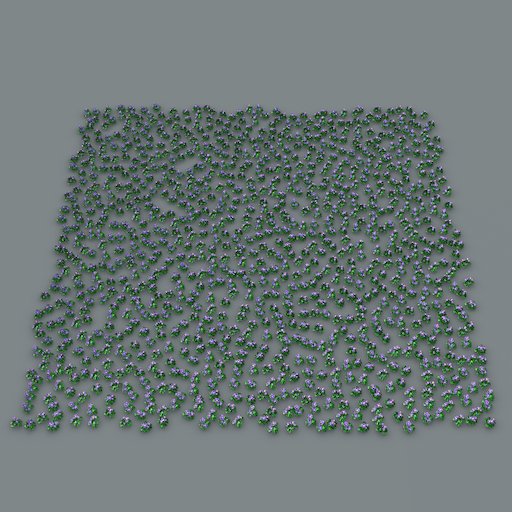}
  };
 \begin{scope}
  \node[anchor=south west,inner sep=0] (lines) at (0.025,1.7)
 {
 \pdfliteral{ 1 w}\includegraphics[width=0.06\textwidth,page=1]{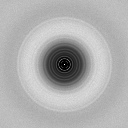}
  };
  \end{scope}
\end{tikzpicture} 
&
\begin{tikzpicture}
  \node[anchor=south west,inner sep=0] (image) at (0,0)
  {
    \pdfliteral{ 1 w}\includegraphics[width=1.1in,page=1]{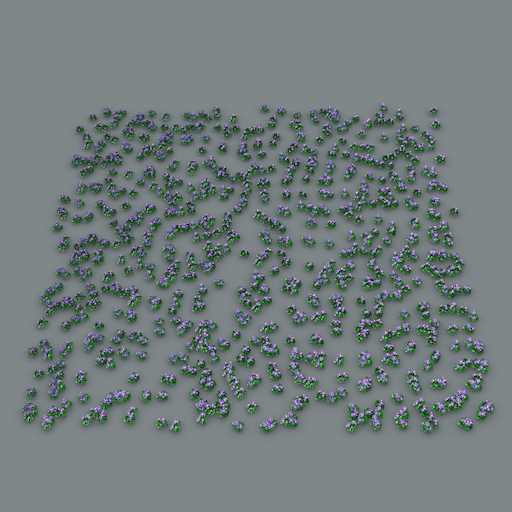}
  };
  \begin{scope}
  \node[anchor=south west,inner sep=0] (lines) at (0.025,1.7)
 {
 \pdfliteral{ 1 w}\includegraphics[width=0.06\textwidth,page=1]{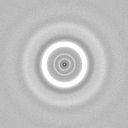}
  };
  \end{scope}
\end{tikzpicture} 
&
\begin{tikzpicture}
  \node[anchor=south west,inner sep=0] (image) at (0,0)
  {
    \pdfliteral{ 1 w}\includegraphics[width=1.1in,page=1]{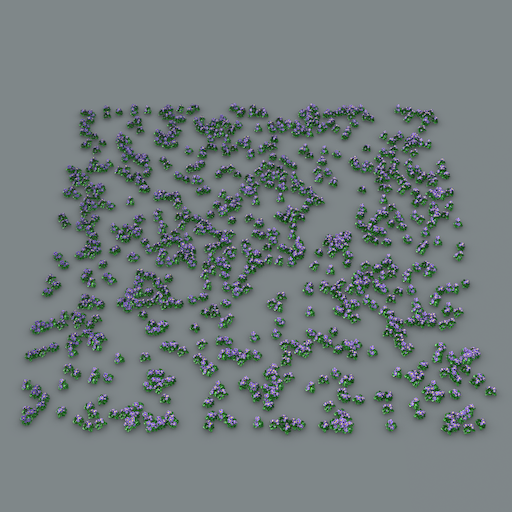}
  };
  \begin{scope}
  \node[anchor=south west,inner sep=0] (lines) at (0.025,1.7)
 {
 \pdfliteral{ 1 w}\includegraphics[width=0.06\textwidth,page=1]{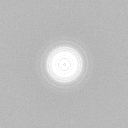}
  };
  \end{scope}
  \begin{scope}
%  \draw[black,thick] (0,1.7) -- (1.2,1.7) -- (1.2,2.8) -- (0,2.8) -- cycle;
  \end{scope}
\end{tikzpicture} 
\\
(a) & (b) & (c)
\\
\end{tabular}%
\caption{
\label{fig:color-noise-spectralloss}
Our network is capable of generating different colored noises. We demonstrate this by placing flowers at point locations generated from (a) blue (b) green and (c) pink noise (corresponding spectra in the insets) for $N=1024$ point samples.
%due to radially-averaged L2 loss.
} 
\end{figure}

%% file: images/fig-renderings.tex
%!TEX root = ../test-template.tex
%
\begin{figure*}[t!]

%\centering
\footnotesize
\hspace*{-1em}

\begin{tabular}{c@{\;}c@{\;}c@{\;}c@{\;}c@{\;}c@{}}
\\
& Reference & Our BNOT & Our Jittered & Jittered & Halton
\\
\rotatebox{90}{\qquad\qquad 3D ($N=64$)}
&
\begin{tikzpicture}
  \node[anchor=south west,inner sep=0] (image) at (0,0)
  {
    \pdfliteral{ 1 w}\includegraphics[width=1.75in,page=1]{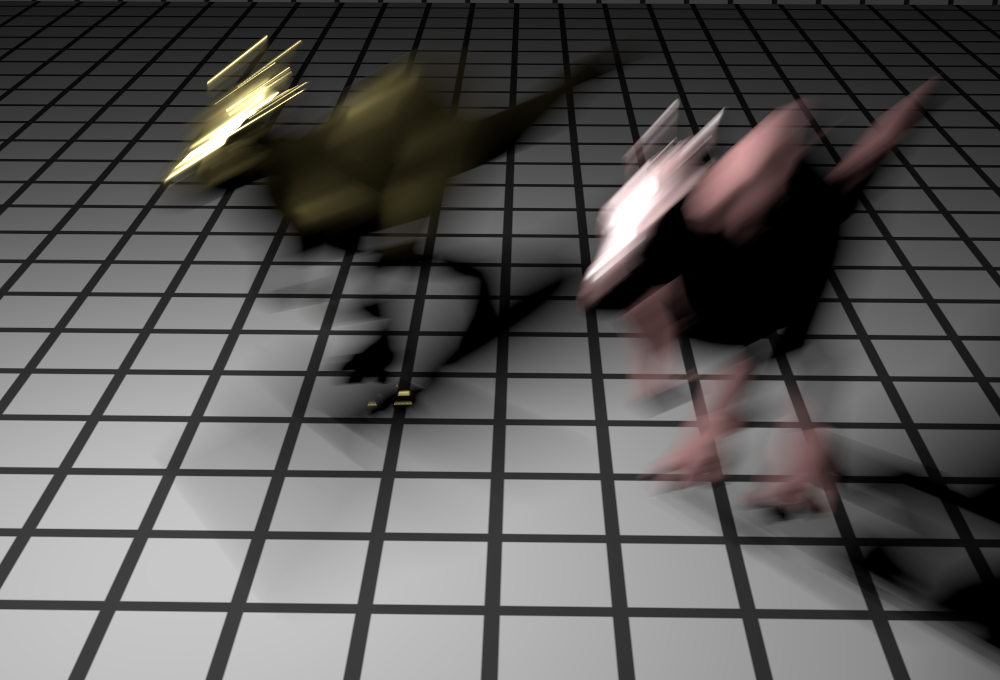}
  };
\end{tikzpicture} 
&
\begin{tikzpicture}
  \node[anchor=south west,inner sep=0] (image) at (0,0)
  {
    \pdfliteral{ 1 w}\includegraphics[width=1.2in,page=1]{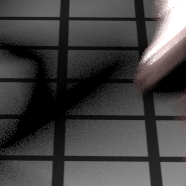}
  };
\end{tikzpicture} 
&
\begin{tikzpicture}
  \node[anchor=south west,inner sep=0] (image) at (0,0)
  {
    \pdfliteral{ 1 w}\includegraphics[width=1.2in,page=1]{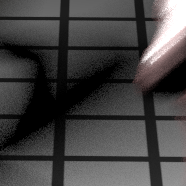}
    };
\end{tikzpicture} 
&
\begin{tikzpicture}
  \node[anchor=south west,inner sep=0] (image) at (0,0)
  {
    \pdfliteral{ 1 w}\includegraphics[width=1.2in,page=1]{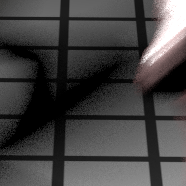}
    };
\end{tikzpicture} 
&
\begin{tikzpicture}
  \node[anchor=south west,inner sep=0] (image) at (0,0)
  {
    \pdfliteral{ 1 w}\includegraphics[width=1.2in,page=1]{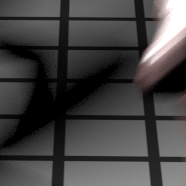}
    };
\end{tikzpicture} 
\\
& & MSE: $3.606\times 10^{-3}$ & MSE: $3.071\times 10^{-3}$ & MSE: $3.741\times 10^{-3}$ & MSE: $1.756\times 10^{-3}$
\\
%%%%%%%%%%%%%%%%%%%%%%%%%%%%
%%%%%%%%%%%%%%%%%%%%%%%%%%%%
%%%%%%%%%%%%%%%%%%%%%%%%%%%%
\rotatebox{90}{\qquad\qquad 4D ($N=256$)}
&
\begin{tikzpicture}
  \node[anchor=south west,inner sep=0] (image) at (0,0)
  {
    \pdfliteral{ 1 w}\includegraphics[width=1.75in,page=1]{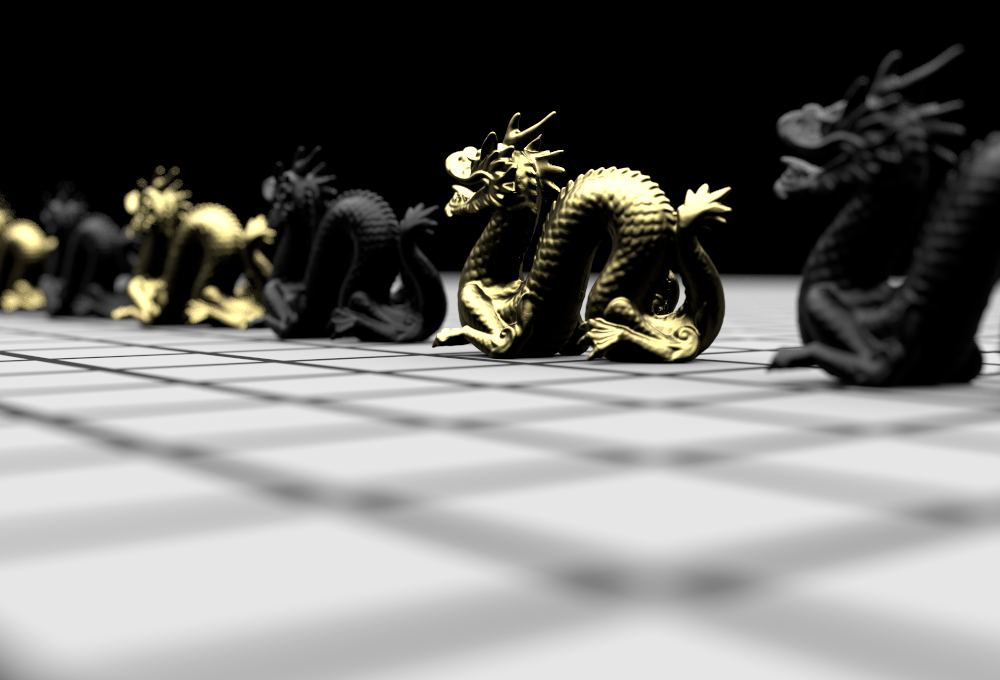}
  };
\end{tikzpicture} 
&
\begin{tikzpicture}
  \node[anchor=south west,inner sep=0] (image) at (0,0)
  {
    \pdfliteral{ 1 w}\includegraphics[width=1.2in,page=1]{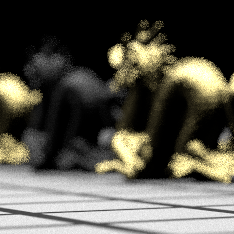}
  };
\end{tikzpicture} 
&
\begin{tikzpicture}
  \node[anchor=south west,inner sep=0] (image) at (0,0)
  {
    \pdfliteral{ 1 w}\includegraphics[width=1.2in,page=1]{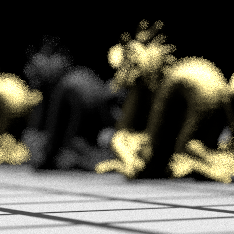}
    };
\end{tikzpicture} 
&
\begin{tikzpicture}
  \node[anchor=south west,inner sep=0] (image) at (0,0)
  {
    \pdfliteral{ 1 w}\includegraphics[width=1.2in,page=1]{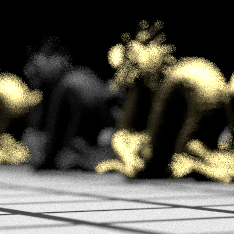}
    };
\end{tikzpicture} 
&
\begin{tikzpicture}
  \node[anchor=south west,inner sep=0] (image) at (0,0)
  {
    \pdfliteral{ 1 w}\includegraphics[width=1.2in,page=1]{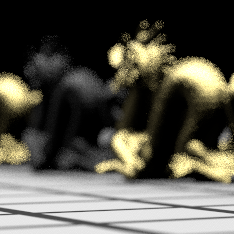}
    };
\end{tikzpicture} 
\\
& & MSE: $4.307\times 10^{-3}$ & MSE: $4.291\times 10^{-3}$ & MSE: $4.281\times 10^{-3}$ & MSE: $0.966\times 10^{-3}$
\\
%%%%%%%%%%%%%%%%%%%%%%%%%%%%
%%%%%%%%%%%%%%%%%%%%%%%%%%%%
%%%%%%%%%%%%%%%%%%%%%%%%%%%%
\rotatebox{90}{\qquad\qquad 5D ($N=1024$)}
&
\begin{tikzpicture}
  \node[anchor=south west,inner sep=0] (image) at (0,0)
  {
    \pdfliteral{ 1 w}\includegraphics[width=1.75in,page=1]{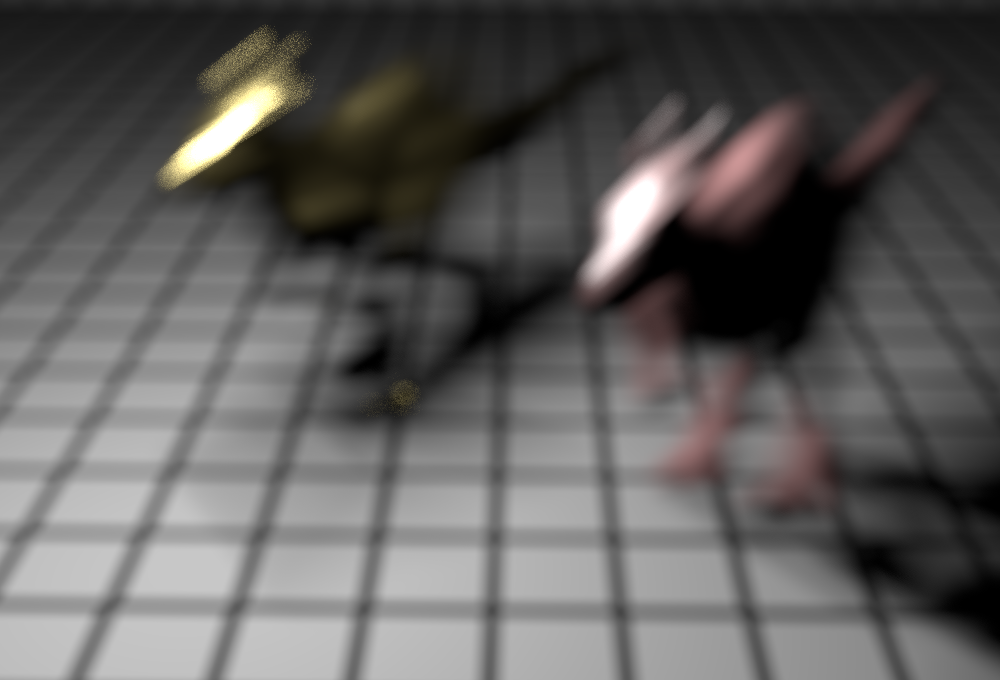}
  };
\end{tikzpicture} 
&
\begin{tikzpicture}
  \node[anchor=south west,inner sep=0] (image) at (0,0)
  {
    \pdfliteral{ 1 w}\includegraphics[width=1.2in,page=1]{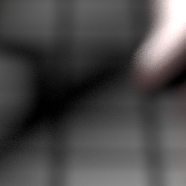}
  };
\end{tikzpicture} 
&
\begin{tikzpicture}
  \node[anchor=south west,inner sep=0] (image) at (0,0)
  {
    \pdfliteral{ 1 w}\includegraphics[width=1.2in,page=1]{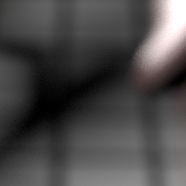}
    };
\end{tikzpicture} 
&
\begin{tikzpicture}
  \node[anchor=south west,inner sep=0] (image) at (0,0)
  {
    \pdfliteral{ 1 w}\includegraphics[width=1.2in,page=1]{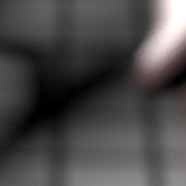}
    };
\end{tikzpicture} 
&
\begin{tikzpicture}
  \node[anchor=south west,inner sep=0] (image) at (0,0)
  {
    \pdfliteral{ 1 w}\includegraphics[width=1.2in,page=1]{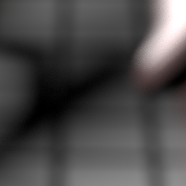}
    };
\end{tikzpicture}
\\
& & MSE: $3.404\times 10^{-4}$ & MSE: $3.182\times 10^{-4}$ & MSE: $3.413\times 10^{-4}$ & MSE: $1.856\times 10^{-4}$
\\
\end{tabular}%
\caption{
\label{fig:renderings}
We render different scenes using the PBRT renderer~\cite{Pharr:2016:PBRT}. We use $N=64,256,1024$ samples for the respective scenes from top to bottom and compare 
our jittered and blue noise (BNOT target) samples with Halton and naive jittered sampling over 3D (\textbf{top row:} pixel and motion blur), 4D (\textbf{middle row:} pixel and depth of field) and 5D (\textbf{bottom row:} pixel, motion blur and depth of field). All scenes are rendered with a point light source. Reference is rendered with $N=4096$ Halton samples. 
%which is used to compute the MSE of the full image for respective sampler. 
} 
\end{figure*}